\pdfoutput=1
\documentclass[universe,article,accept,moreauthors]{Definitions/mdpi} 

\RequirePackage[T1]{fontenc}

\usepackage{pgfplots} 
\usepackage{tikz,amsmath}
\usetikzlibrary{decorations.markings}
\usetikzlibrary{arrows.meta}
\RequirePackage{graphicx}
\usepackage{amstext}
\usepackage[utf8]{inputenc}
\usepackage{amssymb,mathtools}
\RequirePackage{flushend}

\usepackage{environ}

\NewEnviron{myequation}{%
\begin{equation}
\scalebox{0.86}{$\BODY$}
\end{equation}}

\firstpage{1} 
\pubvolume{xx}
\issuenum{1}
\articlenumber{5}
\pubyear{2019}
\copyrightyear{2019}
\history{Received: 05 September 2019; Accepted: 15 December 2019; Published: date}
\updates{yes}

\Title{A Modified Dynamical Model of Cosmology\\ I. Theory}

\Author{Hossein Shenavar *\orcidA{} 
 and Kurosh Javidan *\orcidB{}}

\AuthorNames{Hossein Shenavar, Kurosh Javidan}

\address[1]{Department of Physics,  Faculty of Science, Ferdowsi University of Mashhad, P.O. Box 1436, Mashhad, Iran  }

\corres{\hangafter=1 \hangindent=1.05em \hspace{-0.82em} Correspondence: Javidan@um.ac.ir (K.J.); h.shenavar@mail.um.ac.ir (H.S.)}

\abstract{
Wheeler (1964)  had formulated Mach's principle as the boundary condition for general relativistic field equations. Here, we use this idea and develop a modified dynamical model of cosmology based on imposing  Neumann boundary condition  on cosmological perturbation equations.  Then, it is shown that a new term  appears in the equation of motion, which leads to a modified Poisson equation. In addition, a modified Hubble parameter is derived due to the presence of the new term. Moreover, it is proved that, without a cosmological constant, such a model has a late time-accelerated expansion with an equation of state converging to $w < -1$.   Also, the luminosity distance in the present model is shown to differ from that of the  $\Lambda CDM$ model at high redshifts. Furthermore, it is found that the adiabatic sound speed squared is positive in radiation-dominated era and then converges to zero  at later times.    Theoretical implications of the Neumann boundary condition have been discussed, and it is shown that, by fixing the value of the conjugate momentum (under certain conditions), one could derive a similar  version of modified dynamics. In a future work, we will  confine the free parameters of the Neumann model  based on hype Ia Supernovae, Hubble parameter data, and   the age of the oldest stars.}


\begin{document}

\section{Introduction}
\label{introduction}
The question of the existence of dark matter and dark energy and their implications at different  cosmic scales has been the dominant subject  of the contemporary cosmology. The~main argument in favor of  these two components is that their presence would solve some otherwise baffling discrepancies between observations and predictions of general relativity (GR). See References~\cite{Peebles:1994xt,Peebles:2002gy}. These predictions are mostly based on  the zero- and first-order perturbation equations of GR field equations. See References~\cite{Bardeen:1980kt,Mukhanov:1990me,Bruni:1996im,Malik:2008im,Weinberg:2008zzc} for reviews on perturbation method in general~relativity. 

For example, the~presence of the dark matter particles could explain most notably, among~other phenomena, the~growth of structures in the universe, the~stability of gravitating systems, the~rotation curves of spiral galaxies, and the missing mass problem in gravitational lensing. See References~\cite{Peebles:1994xt,Sanders:2010cle,Jain:2010ka} for  excellent reviews on the essence of these problems. However, it should be noted that, despite  tremendous efforts to detect the most popular candidates of dark matter, such as sterile neutrinos, WIMPs (weakly interacting massive particles), and axions, a~conclusive evidence for the detection  has not yet been reported~\cite{Bertone:2018xtm}.

On the other hand, the~existence of the dark energy is presumed to explain the large-scale structure and  evolution of the cosmos~\cite{Peebles:2002gy,Amendola:2015ksp}. The~main concerns here are accelerated expansion of the universe (proved by References~\cite{Riess:1998cb,Perlmutter:1998np}  based on supernova observations), baryon acoustic oscillations, temperature anisotropies in the cosmic microwave background (CMB), and the age of the universe. See Reference~\cite{Amendola:2015ksp} for a thorough review on theoretical implications and observational evidence of  dark energy. Also, see Reference~\cite{Bamba:2012cp}, which reviews dark energy models from different~perspectives.

To solve GR field equations, the~initial/boundary conditions are usually considered to be trivial. See Referece~\cite{Peebles:1994xt} pages 361--363 for a discussion on this issue.  The~question of initial/boundary condition in GR is closely related to the problem of the boundary term in GR action. At~least from a mathematical point of view, it is well understood that, to derive the Einstein equations, one has to add a surface term to the Einstein--Hilbert action~\cite{Chakraborty:2016yna}. This ensures a well-posed action, meaning that the imposed boundary condition  on the action is compatible with the derived field equations.     The~main issue in adding a surface term to the Einstein--Hilbert action   is the quantity which should be fixed on the boundary. On~the other hand,   Wheeler's~\cite{wheeler1964mach} interpretation of Mach's principle suggests that,  in addition to the energy momentum tensor, one needs  also to specify the boundary condition to derive the structure of spacetime. See also Reference~\cite{1981RPPh...44.1151R} for a review on this idea.  This view toward solving Einstein field equations is mostly based on the necessity of  mathematical and physical well posedness of the theory rather than philosophical  grounds of Mach principle,  which are well discussed by Reference~\cite{Barbour:1995iu}.  In~fact, this interpretation of Mach's principle is closely related to the Cauchy problem of GR~\cite{1980grg1.conf...99C,ChoquetBruhat:1969cb}.   There are various formulations of Mach's principle in the literature.  For~example,   Bondi and Samuel~\cite{Bondi:1996md}  distinguish   eleven statements, Mach0 to Mach10, of~Mach's principle; however, Wheeler's formulation is not included. We will discuss some of these statements below. Also, References~\cite{Sciama:1970yk,1981RPPh...44.1151R} propose similar ideas to Wheeler's in their attempt to find generally covariant integral formulation of~GR.

In their work on the foundations of  gravitational theories, Thorne,  Lee, and Lightman~\cite{1973PhRvD...7.3563T} define boundary conditions as those laws which involve only confined variables. As~these authors have illustrated,  ``if one augments the theory by cosmological demands that $\phi$ and $\psi$  $[$the scalar potentials$]$  go to zero at spatial infinity, those demands are boundary conditions''. We will discuss this particular case in the~following.

Including boundary dynamics while solving GR field equations could lead to interesting results even at the quantum level~\cite{Park:2018xtt,Park:2019lkh}.  For~example,  Park~\cite{Park:2017wiw} considers the issue of ``changing black hole mass through Hawking radiation'' and finds that, by imposing Neumann boundary condition, one could solve the problem.   See also a recent work by Witten~\cite{Witten:2018lgb},  who shows  that the usual Dirichlet boundary condition is indeed  not elliptic. Thus,  in~general, this boundary condition  does not provide a well-defined perturbation theory. On~the other hand, Witten shows that conformal boundary conditions always lead to a sensible perturbation expansion. Therefore, when we solve GR field equations, it is important to check different possibilities of boundary conditions both for the sake of mathematical well posedness of the model and its compatibility with physical~expectations.

In this work, we will impose Neumann boundary condition (B.C.)  on GR perturbation equations.  Also,   the~modifications in trajectories of massive and massless particles are discussed.   The~new term in the equation of motion of this model is found to be proportional to $cH_{0}= 6.59 \times 10^{-10}$ m/s$^{2}$   in which $c$ is the speed of light and $H_{0}$ is the Hubble constant~\cite{Shenavar:2016bnk,Shenavar:2016xcp}. This parameter is also known as  the de Sitter scale of acceleration~\cite{vanPutten:2017lik,vanPutten:2017bqf}.  Thus, the~model is essentially a modified dynamical (MOD) model which, in many features at galactic scales, shows some similarities with  Milgrom's   proposal~\cite{Milgrom:1983ca,Milgrom:1983pn,Milgrom:1983zz} known as MOND (modified Newtonian dynamics). See Reference~\cite{Famaey:2011kh} for a review on MOND. Some particular aspects of galactic dynamics, such as  observed regularities in the properties of dwarf galaxies~\cite{Kroupa:2012qj,Kroupa:2013yd}, are illustrated within the context of MOND quite better than the cold dark matter ( CDM ) 
paradigm. The~main implication of the appearance of such an acceleration scale in the equation of motion is that, probably, the physics of the local universe might be affected by the global expansion of the cosmos~\cite{Sanders:2010cle}. Many works try to explain the phenomenological success of MOND by deriving particular versions of it from action principle. For~example, a~recent article by Vagnozzi~\cite{Vagnozzi:2017ilo} demonstrates that MOND can be recovered in the low-energy limit of  mimetic gravity. The~acceleration scale in this approach  changes with length scale in such a way that the model is able to explain dark matter both on galactic and cluster~scales.

Using our new modified Poisson equation and~the averaging procedure of backreaction cosmology~\cite{Rasanen:2003fy,Hirata:2005ei,Kolb:2004am,Martineau:2005aa}, in~Section~\ref{Modified}, a governing modified Hubble parameter of the present model is derived. Then, in~Section~\ref{Cosmic}, the~main cosmic implications of the new model is surveyed. Among~others, we have derived the evolution of the scale factor, Hubble parameter, deceleration parameter, equation of state, and adiabatic sound speed. In~a future work, we will present a data analysis of the model based on  Supernovae Type Ia (SNIa)  and Hubble parameter data (Shenavar and Javidan, in~preparation). The~basic parameters of the theory could be derived from such~analysis.

Some related explanations and elaborations are postponed to the Discussions section and in the appendices. In~Section~\ref{DisNeumann}, we search for the implications  of Neumann B.C. from an action-principle point of view. Prior works   by Maldacena~\cite{Maldacena:2011mk} and by Anastasiou and Olea~\cite{Anastasiou:2016jix}, who propose  the  equivalence of Einstein and conformal gravity  under Neumann B.C., are discussed. Also, we consider the works of References~\cite{Chakraborty:2016yna,Krishnan:2016mcj} that prove the existence of a well-posed Neumann   boundary condition. The~surface term introduced by References~\cite{Chakraborty:2016yna,Krishnan:2016mcj} has the peculiarity  that it vanishes in a four dimensional spacetime. In~particular, we have shown that, by assuming some conditions on scalar potentials, our choice of Neumann B.C. in Section~\ref{Scalar} could be derived based on the  procedure that References~\cite{Chakraborty:2016yna,Krishnan:2016mcj} define. Discussions regarding the averaging problem and Mach's priciple have been presented in the rest of Section~\ref{Discussion}. In~addition, we have provided detailed analysis of the role of geometrodynamic clocks and  EPS  ( Ehlers,  Pirani and  Schild ~\cite{ehlers2012republication}) theorem in Appendix~\ref{DisMeasurement}. Moreover, a~discussion on the  dimensional analysis of the model is  presented in Appendix~\ref{Disdimensional}. Furthermore, we report a similar  semi-Newtonian model  in Appendix~\ref{DisNewt}, which is built based on some well-motivated theoretical and observational~assumptions.  

In Reference~\cite{Shenavar:2016bnk},  the~existence of a cosmological constant $\Lambda$ was assumed to ensure the existence of a late time  epoch of accelerated expansion. Here, however, the~$\Lambda$ term is abandoned and we will show that  the modified Hubble parameter  derived from imposing a Neumann B.C.---which includes a term proportional to $H$ similar to Veneziano ghost of QCD ( Quantum Chromodynamics ) ~\cite{Cai:2010uf,Cai:2012fq}---would naturally lead to an accelerated  expansion of the universe with an equation of state parameter approaching to $w < -1$ at later cosmic times. The~Neumann model proposes a unification of  dark matter and dark energy.  Such models have been proposed before. For~example, one could see the generalized Chaplygin gas~\cite{Bento:2002ps}, k-essence~\cite{Scherrer:2004au}, and Bose--Einstein condensation model~\cite{Fukuyama:2007sx}. We will compare the results of the Neumann model with these models and MOG (modified gravity of   Moffat~\cite{Moffat:2005si}), which is a model of dark matter (plus $\Lambda$ as dark energy),  in the~following.

\section{A New Model of Modified~Dynamics}
\label{Scalar}

In this section, we use the variational method of Weinberg~\cite{Weinberg:2008zzc}, chapter 5, to~find the Taylor expansion of GR field equations.  One may also see References~\cite{Bardeen:1980kt,Mukhanov:1990me,Bruni:1996im,Malik:2008im} for precise descriptions of the general perturbation theory of Einstein field equations. To~do so, we will first fix the gauge condition to eliminate the spurious gauge degrees of freedom. Then, we will impose the boundary condition on the remaining physical~fields. 

\subsection{Fixing the Gauge and the Boundary~Conditions}

 We assume a flat Friedmann--Lemaitre--Robertson--Walker (FLRW) background metric disturbed by the perturbation of the local universe $h_{\mu \nu}$. Under~spatial rotation,  the~symmetric perturbation metric could be decomposed into a scalar $h_{00}$, a~three-vector $h_{0i}$, and a symmetric two-index spacial tensor $h_{ij}$. The~tensor $h_{ij}$ could itself be decomposed to a trace part and a trace-free part. The~resulting  field equations, as Reference~\cite{Weinberg:2008zzc}  has reported in chapter 5, would be very complicated. Also, they would contain some unphysical scalar and tensor modes which are correlated with the gauge freedom of the field~equations. 

This gauge problem could in principle be solved by working with gauge invariant quantities; however, in~this way, the~issue of imposing a new boundary condition on field equations would not be physically evident. Another method to eliminate the gauge degrees of freedom is by fixing the gauge which we use here. This process could be significantly simplified if the scalar, vector, and tensor modes of  the perturbation metric are treated separately. Consider a general coordinate transformation
\begin{eqnarray}
x^{\mu} \rightarrow  x^{\prime  \mu} = x^{\mu} + \xi^{\mu}
\end{eqnarray}
in which $\xi^{\mu}$ is of the same order of the perturbation. To~distinguish the behaviour of different modes under this transformation, we decompose the spatial part of $\xi^{\mu}$ as
\begin{eqnarray}
\xi_{i} = \partial_{i} \xi^{S}+ \xi^{V}_{i}
\end{eqnarray}
with the condition that the $\xi^{V}_{i}$ is a divergenceless vector, i.e.,~$\partial_{i}\xi^{V}_{i}=0$. As~Weinberg~\cite{Weinberg:2008zzc},  pages 235--238,  shows  that the tensor quantities in perturbation metric and perturbation energy-momentum tensor are gauge invariant, there is no need to gauge-fix these modes. On~the other hand, the~vector $\xi^{V}_{i}$ could be chosen so that some vector degrees of freedom vanish. Lastly, to~gauge-fix the scalar modes, there are various possibilities; however, here, we use the Newtonian gauge because, in this gauge, it is more  convenient to adopt a new boundary condition. Of~course, conversion to other gauges could be done as usual. In~Newtonian gauge, two scalar degrees of freedom could be eliminated by choosing $ \xi^{S} $ and $\xi_{0}$. Thus, the~remaining scalars lead to the next metric (Weinberg~\cite{Weinberg:2008zzc} page 239)
\begin{eqnarray}  \label{metric}
ds^{2} = -(1+2\Phi)dt^{2} + R^{2}(t)(1-2\Psi)\delta_{i j}dx^{i}dx^{j}
\end{eqnarray} 
in which $R(t)$ is the scale factor and the potentials  $\Phi $ and $ \Psi $ are functions of spacetime, i.e.,~$\Phi = \Phi (t,x,y,z) $ and $ \Psi = \Psi (t,x,y,z) $.  Our notation is very close to   Weinberg's~\cite{Weinberg:2008zzc}, however, as~a deviation from Weinberg's notation, we keep $a(t)$ for the dimensionless scale factor $a(t)\equiv R(t)/R_0$ in the~following. 

One should note that the two scalar potentials are, in~principle, distinct degrees of freedom.  In~fact, from a~physical point of view, the~potential $\Phi$  is a  generalization of the Newtonian potential because  it specifies the particle acceleration while the potential $\Psi$ is the 3-curvature perturbation  of a constant-time surface. While  many works consider the two potentials to be the same, the~key assumption of this work is that these potentials  could be essentially~different.

The resulting field equations, as~we will see below, would contain second-order spatial derivatives which certainly need boundary conditions to provide a unique solution.  Indeed, by~imposing a suitable boundary condition, as~Wheeler's interpretation of Mach's principle demands, one could uniquely determine the solution to the field equations. The~aim of the present work is to provide a new solution for the perturbation equations based on imposing Neumann boundary~condition.

Using the metric in Equation \eqref{metric}, it is now straightforward to derive the local and background terms of Christoffel symbols, Riemann, Ricci, and Einstein tensors.   These quantities have already been  reported by many authors  for the full perturbation metric (which includes vector and tensor terms in addition to scalar potentials)    \cite{Bardeen:1980kt,Mukhanov:1990me,Weinberg:2008zzc,Malik:2008im}. For~the sake of brevity, here, we only report Christoffel terms and elements of Einstein tensor which will be needed in~following.

The  Christoffel symbols, up~to first-order terms, are
\begin{eqnarray}    \label{Christoffel1}
\Gamma^{0}_{00} &=&  \dot{\Phi}   \\ \nonumber
\Gamma^{i}_{00} &=&  \frac{\partial_{ i}\Phi }{R^{2}}  \\ \nonumber
\Gamma^{0}_{i0} &=&  \partial_{ i}\Phi    \\ \nonumber
\Gamma^{0}_{ij} &=&  \delta_{ij} \left(R \dot{R} -   (2 R \dot{R} \Phi  + \dot{\Psi} ) \right)    \\ \nonumber
\Gamma^{i}_{j0} &=& \delta_{ij} \left(\frac{\dot{R}}{R} +   \frac{1}{R^{2}} (2 H \Psi -  \dot{\Psi}) \right)  \\ \nonumber
\Gamma^{i}_{jk} &=& \frac{1}{R^{2}} \left(\delta_{jk} \partial_{ i}\Psi -\delta_{ij}\partial_{ k}\Psi  - \delta_{ik} \partial_{ j}\Psi \right)       \\ \nonumber
\end{eqnarray}
while the components of Einstein tensor are as follows
\begin{eqnarray}  \label{Einstein}
G^{0}_{~0} &=& -3 (\frac{\dot{R}}{R})^{2} + \frac{2}{R^{2}} \left(3\Phi \dot{R}^{2} - \nabla^{2} \Psi +3R\dot{R}\dot{\Psi}   \right) \\  \nonumber
G^{i}_{~0} &=&   \frac{2  \partial_{i}}{R^{3}} \left(\dot{R} \Phi + R  \dot{\Psi} \right)  \\  \nonumber
G^{i}_{~j} &=& \frac{\partial_{i}\partial_{j}}{R^{2}}(\Psi -\Phi) ~~~~~~~~~~~~j \neq i  \\  \nonumber
G^{i}_{~i} &=& - \frac{ \dot{R}^{2} + 2R\ddot{R}}{R^{2}}    
+ \frac{1}{R^2} (2\dot{R}^{2} \Phi +4R\ddot{R}\Phi
+(\nabla^{2}-\partial^{2}_{i})(\Phi -\Psi)   
+ 2R\dot{R}\dot{\Phi} +6R\dot{R}\dot{\Psi} +2R^{2}\ddot{\Psi}).     \\  \nonumber
\end{eqnarray}
 All terms  higher than the first order have been dropped in Equations~(\ref{Christoffel1}) and \eqref{Einstein}.

On the other hand, the~right hand side of the Einstein field equation, i.e.,~the energy-momentum tensor $T_{\mu \nu}$, could  be decomposed  to pure time-dependent zero-order terms $\rho$  and $P$ and~also spacetime-dependent first-order perturbations as follows
\begin{eqnarray}  \label{EMtensor}
T^{0}_{~0} &=& -\rho -\delta \rho \\
T^{i}_{~0} &=&   -\frac{\rho + P}{R^{2}} \delta u_{i}  \\
T^{i}_{~j } &=& \left(P +\delta P \right)\delta_{ij} + \partial_{i}\partial_{j} \pi^{S}   
\end{eqnarray}
where  $\delta \rho = \delta \rho (t,x,y,z)$,  $ \delta P = \delta P  (t,x,y,z)$, $\delta u_{i} = \delta u_{i} (t,x,y,z) $, and $\pi^{S} = \pi^{S} (t,x,y,z)$  are respectively the local density, pressure, velocity, and scalar anisotropic perturbations~\cite{Weinberg:2008zzc}. This last term, i.e.,~$ \pi^{S} $, displays the deviation from the perfect fluid model. In~this work, the~vector and tensor parts of the anisotropic stress tensor will be neglected.  Also,  for~the sake of brevity, the~dependency to spacetime parameters $(t,x,y,z)$ is not shown. It should be noted that  the velocity perturbation could be decoupled into scalar and vector parts, i.e.,~$ \delta u_{i} =  \partial_{i} \delta u + \delta u^{V}_{i}  $; however, here, we only consider the scalar share in the  velocity perturbation $ \delta u $. 

Now, one could use the energy-momentum conservation equation, i.e.,
$$ \nabla_{\mu} T^{\mu \nu} =0 $$
to derive the next up-to first-order conditions
\begin{myequation}  \begin{array}{cll}  
0 =& \dot{\rho} +3\frac{\dot{R}}{R}(\rho + P) + \frac{(\rho + P)}{R^{2}}\nabla^{2}(\delta u) +3\frac{\dot{R}}{R}(\delta \rho + \delta P) -6\frac{\dot{R}}{R}(\rho + P)\Phi + \frac{\dot{R}}{R}\nabla^{2} \pi^{S}   -2\Phi \dot{\rho} + \dot{\delta \rho} -3 (\rho + P) \dot{\Psi}  \\
0 =& \partial_{i} \left(3 \frac{\dot{R}}{R} \delta u (\rho + P) +(\dot{\rho} + \dot{P}) \delta u + \delta P +(\rho + P) \Phi  + \nabla^{2} \pi^{S} + (\rho + P) \dot{\delta u} \right) \end{array}
\end{myequation}
which could also be derived, order by order, from~Einstein field~equations.

It is possible now to find the Einstein field equation, i.e.,
\begin{eqnarray} 
G^{\mu}_{~\nu} = \kappa T^{\mu}_{~\nu}
\end{eqnarray} 
in which 
$  \kappa = 8 \pi G$ 
is a constant,  by~using Equations~(\ref{Einstein}) and \eqref{EMtensor}. For~example, by~using $~^{i}_{j}$ components of the Einstein equation, one could find
\begin{eqnarray}    \label{Laplacian}
\partial_{i} \partial_{j} \left(\Phi -\Psi + \kappa R^{2} \pi^{S}   \right) = 0
\end{eqnarray}  
which is reported in many classical literature on cosmic perturbation theory. See, for~instance,  Bardeen~\cite{Bardeen:1980kt} and Mukhanov~et~al.~\cite{Mukhanov:1990me}. Equation \eqref{Laplacian} manifests that  the difference  between tidal forces due to $\Phi $ and $\Psi$ is due to the anisotropic stress $\pi^{S}$. In~particular, if~$\pi^{S}=0$, then the tidal forces are equal $\partial_{i} \partial_{j} \Phi = \partial_{i} \partial_{j} \Psi$.

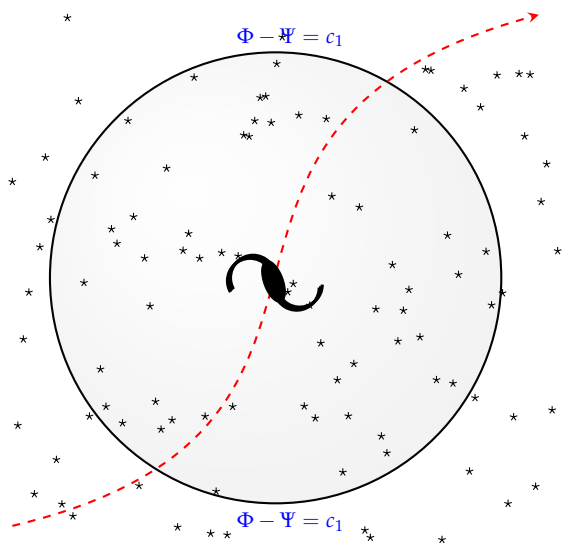
\begin{figure}[H]
\begin{center}
\begin{tikzpicture}[font=\LARGE]
\draw [thick] (2,2) circle (3cm);

\shade[ball color = gray!40, opacity = 0.07] (2,2) circle (3cm);
\draw [->,>=stealth,thick, red, dashed] (-1.5,-1.3)   .. controls  (4,0) and (0,4) .. (5.5,5.5);
\draw plot [only marks, mark=star, mark size=1.5, domain=-1.5:5.75, samples=100] (\x,{rnd*7.0-1.5});
\node [text=blue] at (2.2,-1.25) {\footnotesize $\Phi - \Psi= c_{1}$};
\node [text=blue] at (2.2,5.2) {\footnotesize $\Phi - \Psi= c_{1}$};
\draw[black,fill=black,rotate around={110:(1.99,1.99)},] (1.96,2.02) ellipse (0.29cm and 0.13cm);
\draw [very thick] (2,2) arc (0:210:0.3cm);
\draw [very thick] (1.96,1.96) arc (0:210:0.3cm);
\draw [very thick] (1.98,1.98) arc (0:210:0.3cm);
\draw [very thick] (2.02,2.02) arc (-210:0:0.3cm);
\draw [very thick] (2.04,2.04) arc (-210:0:0.3cm);
\draw [very thick] (2.06,2.06) arc (-210:0:0.3cm);
\end{tikzpicture}
\end{center}
\caption{A particle (dashed curve) is falling into the potential well of a massive object (a galaxy) in a homogeneous and isotropic background. The~sphere includes all the particles which have already been in causal contact with the central mass under consideration; thus, the~radius of the spherical boundary is indeed the size of the particle horizon.  The~presence of distant objects produces a nontrivial  boundary condition  $\Phi - \Psi= c_{1}$ on the boundary surface, which we will consider~here. }
  \label{fig:boundary}
\end{figure}

However, the~field equations are at hand at this point, the~appropriate boundary condition is yet to be imposed~\cite{wheeler1964mach}. See Figure~\ref{fig:boundary}, which shows a typical gravitational source (here a galaxy) in a smooth background and its related boundary. The~size of the boundary is assumed to be determined by the particle horizon because we assume that the boundary in Figure~\ref{fig:boundary} includes all the objects which have already been in causal contact with the central galaxy which is under consideration~\cite{Ellis:2015wdi}. The~matter distribution on large scale is assumed to be homogeneous and isotropic; thus, the  surface of the boundary, which lies at very large distance, must have the same condition too. The~boundary condition will be fixed on this surface in what~follows.

The most general solution to Equation~(\ref{Laplacian}) could be written as $\Phi -\Psi + \kappa R^{2} \pi^{S}  =A_{1}(t,x)+A_{2}(t,y)+A_{3}(t,z) + c_{1}(t)$. Similar to other partial differential equations, here too,  one must impose an appropriate boundary condition to find the unique solution to Equation~(\ref{Laplacian}). In~particular, the~fact that the presumed boundary in Figure \ref{fig:boundary} is assumed to be homogeneous and isotropic everywhere sets the first three functions to zero, i.e.,~$A_{1}(t,x)=A_{2}(t,y)=A_{3}(t,z)=0$. This result could also be obtained by arguing that the perturbations must be statistically homogeneous and isotropic.  On~the other hand, assuming $\pi^{S}=0$, there is no a priori reason for putting $c_{1}(t)=0$; thus,  the~gravitational potential  $\Phi $ and the 3-curvature  perturbation $\Psi$ could in principle be different. In~fact,  it could be shown that a pure time-dependent function $c_{1}(t)$ does not change the homogeneity and isotropy of the system under consideration (see Section~\ref{DisAverage} for the details).

In general, because~the dynamics of the universe at its largest scale is time dependent, the~difference between $\Phi $ and $\Psi$ on the boundary could be, in~principle,  a~time-dependent function. However, as~we have shown in Appendix \ref{timedepend}, a~time-dependent function  $c_{1}(t)$ would result in three independent cosmic equations, i.e.,~two modified Friedmann equations plus a modified conservation equation, such a system of equations is mathematically possible; however, its dynamical behaviour could be completely different in early cosmic times from that of the standard~model.

On the other hand, by~assuming $c_{1}$ as a time-independent factor,  one obtains a model for which only two  cosmological equations (out of three) are independent. In~this way, our analysis of cosmic evolution becomes more straightforward. See Appendix  \ref{timedepend} for details.  Also, as~we will show in the following,  the~existence of this term does not alter the ``form'' of the governing Einstein equations because its space and time derivatives are zero. The~appearance of the new solution, however, is due to imposing a new boundary condition. In~the same way, the~solutions to the geodesic equation  will be modified now; however,  the~form of the geodesic equation is still the same as~before. 

  It could be shown that, after averaging the full metric, the~result contains a term proportional to $c_{1}$. However, by~a simple redefinition of the scales, one could  prove  that the averaged metric is the same as Minkowski metric.  See Section~\ref{DisAverage} for a detailed~discussion.

If one assumes $c_1 = 0$, then the usual formalism of the standard cosmology could be derived readily. This  choice could be interpreted as imposing the Dirichlet boundary condition in solving  Equation~(\ref{Laplacian}). From~our perspective, by~imposing the Dirichlet boundary condition on local perturbation equations, one presumes that distant objects enforce no observable  effect on local physics, i.e.,~no detectable result from Mach's principle.  On~the other hand, a~nonzero $c_1 $ could be seen as imposing a Neumann boundary condition~\cite{Shenavar:2016bnk}. In~fact, a~nonvanishing potential at the boundary, i.e.,~$c_{1} \neq 0$ in Figure~\ref{fig:boundary}, represents the effect of the distant stars  on local dynamics. Therefore, in~the present model, the Mach's principle provides observable effects for the motion of local objects; however, to~be precise, this new model is achieved by following Wheeler's work~\cite{wheeler1964mach} that emphasized the role of the boundary condition in solving Einstein's field equation. See also a discussion in Section~\ref{DisMach}. 

From physical point of view, the~Neumann B.C. is quite sensible because it implies that, instead of the difference between the potentials ($\Phi -\Psi$), the~difference between the  tidal forces, i.e.,~$\partial_{i} \partial_{j} \left(\Phi -\Psi \right)$, vanishes on the~boundary. 

Chameleon cosmology too, uses a nonvanishing potential as its boundary condition~\cite{Khoury:2003rn}; however, the~procedure is quite different in that case.  Also, many recent works presume nonzero anisotropic stress, i.e.,~$\eta = \Psi /\Phi \neq 1 $, to~search for  hints of deviations from the $\Lambda CDM$ model. See, for~example, Reference~\cite{DiValentino:2015bja} and references therein. However, it should be clear now that the new effect in the present model does not occur due to a nonvanishing anisotropic stress. The~appearance of $c_1 $  is because of the presence of matter at the~boundary.

Choosing Neumann B. C.
\begin{eqnarray}   \label{Condition}
 \Psi  &=& \Phi + \kappa R^{2} \pi^{S} - c_1
\end{eqnarray}
one could rewrite Einstein's field equations as
\begin{myequation}   \label{Einstein1} \begin{array}{rcl}  
\underbrace{ -3 (\frac{\dot{R}}{R})^{2} }  +6\frac{\Phi \dot{R}^{2}}{R^{2}} -2\frac{ \nabla^{2} \Phi }{R^{2}} + 6  \frac{\dot{\Phi} \dot{R}}{R} &=&  -\kappa ( \underbrace{ \rho} +\delta \rho + 12 \pi^{S}\dot{R}^{2}  -2 \nabla^{2}\pi^{S} +6  R \dot{R}\dot{\pi}^{S}) \\ 
\underbrace{-\frac{\dot{R}^{2}+2R\ddot{R}}{R^{2}} }  +2 \frac{\Phi \dot{R}^{2}}{R^{2}} + 4\frac{\Phi \ddot{R}}{R} +8 \frac{\dot{R}\dot{\Phi}}{R} +2\ddot{\Phi} &=&   \kappa (\underbrace{P} +\delta P  -16 \pi^{S}\dot{R}^{2} -4 R \ddot{R} \pi^{S} + \nabla^{2} \pi^{S} - 14 R \dot{R} \dot{\pi}^{S} -2R^{2}\ddot{\pi}^{S}) \\  
 \frac{2  \partial_{i}}{R^{3}} \left(\dot{R} \Phi + R  \dot{\Psi} \right) &=& -\kappa \frac{\rho + P}{R^{2}}  \partial_{i} \delta u  - \kappa(4\frac{ \partial_{i}\pi^{S} \dot{R}}{R} + 2 \partial_{i}\dot{\pi}^{S})  \end{array}
\end{myequation}
while the components of energy momentum conservation could be found as follows:
\begin{small}
\begin{eqnarray}  \nonumber  
0 =& \underbrace{\dot{\rho} +3\frac{\dot{R}}{R} (\rho + P)} + \frac{(\rho + P)}{R^{2}}\nabla^{2}(\delta u) +3\frac{\dot{R}}{R}(\delta \rho + \delta P) -6\frac{\dot{R}}{R}(\rho + P)\Phi + \frac{\dot{R}}{R}\nabla^{2} \pi^{S}   -2\Phi \dot{\rho} + \dot{\delta \rho} -3 (\rho + P) \dot{\Phi} \\  \nonumber \\  \label{EMtensor1}
0 =& \partial_{i} \left(3 \frac{\dot{R}}{R} \delta u (\rho + P) +(\dot{\rho} + \dot{P}) \delta u + \delta P +(\rho + P) \Phi  + \nabla^{2} \pi^{S} + (\rho + P) \dot{\delta u} \right) 
\end{eqnarray}
\end{small}
in which pure time-dependent terms are distinguished with an underbrace. As~one could see from Equations~(\ref{Einstein1}) and \eqref{EMtensor1}, as~a result of adding $c_1$ to the solution of Equation~(\ref{Laplacian}), there appears no modification in the zeroth-order (pure time-dependent) GR~equations.

In the rest of this section, we will consider the modifications due to Equation~(\ref{Condition}) on the trajectories of massless and massive particles. This is  essential in understanding the present model, and it has been partially discussed before~\cite{Shenavar:2016bnk}; however, here, we will use the modified Poisson equation introduced in References~\cite{Shenavar:2016xcp,shenavar2018local} to build a full cosmological description through finding a modified dynamical term of cosmic energy. After~finding this term which we will name $\rho_{c_1}$, we will derive  the modified Friedmann equation  and survey its implications in~cosmology.

\subsection{Trajectory of~Particles}
To find the trajectory of a massless particle, we will use the method of perturbative geodesic expansion  introduced by Reference~\cite{Pyne:1995ng}. See also Reference~\cite{Pyne:1995bs} for an extension of the method. To~derive the trajectory of a photon, the~tangent vector, i.e.,~$v^{\mu} = dx^{\mu}/d\lambda$, must be null $v_{\mu} v^{\mu}=0 $ and geodesic $$\frac{dv^{\mu}}{d\lambda}=-\Gamma^{\mu}_{\alpha \beta}v^{\alpha}v^{\beta} $$ at any order. Assuming $k^{\mu}$ and $l^{\mu}$ to be the tangent vectors at zeroth and first orders,  respectively, the~null condition  would provide us with
\begin{eqnarray} \label{Nullzero}
0 &=& -(k^{0})^{2} +R^{2}|\vec{k}|^{2} \\   \label{Nullfirst} 
0 &=& c_{1}R^{2}|\vec{k}|^{2} - k^{0} (l^{0} +k^{0} \Phi) + R^{2}(\vec{k}.\vec{l}-|\vec{k}|^{2} \Phi) 
\end{eqnarray}
while the time and spatial components of the geodesic equation of the photon (at first order) would become
\begin{eqnarray} \label{Geodesictime}
\frac{dl^{0}}{d\lambda} &=& -2R \left(c_{1} |\vec{k}|^{2}  +  \vec{k}.\vec{l} -2  |\vec{k}|^{2} \Phi \right)\dot{R} + R^{2} |\vec{k}|^{2} \dot{\Phi} -k^{0} \left(2 \vec{k}.\vec{\nabla}\Phi   +k^{0}\dot{\Phi} \right) \\   \label{Geodesicspace}
\frac{d\vec{l}}{d\lambda} &=&  -2 \frac{\dot{R}}{R} \left(l^{0}\vec{k} + k^{0}\vec{l}  \right) + 2\vec{k} (\vec{k}.\vec{\nabla} \Phi) - |\vec{k}|^{2}\vec{\nabla} \Phi -\frac{(k^{0})^{2} }{R^{2}} \vec{\nabla}\Phi +2k^{0}\vec{k}\dot{\Phi}
\end{eqnarray} 
in which we have used the Christoffel symbols of Equation \eqref{Christoffel1}. Here, the~zero-order terms of the geodesic equation have been neglected. Also,  spatial vectors have been shown with an arrow $\vec{k}$.

The evolution  of $ l^{0} $, i.e.,~Equation~(\ref{Geodesictime}), essentially illustrates the change in the redshift of photons when they pass through a gravitational well. This equation is the basis for the  discussion of the Sachs--Wolfe effect in CMB physics. The~changes that the presence of a Neumann constant $c_{1}$ introduces to the CMB analysis will be discussed  in more details in a future~work.

The solution to Equation~(\ref{Geodesicspace}), on~the other hand, describes the deviation in photon trajectory while  passing through a gravitational potential which results in   the  lensing equation. In~fact, using Equations~(\ref{Nullzero}) and \eqref{Nullfirst} to replace $k^{0}$ and $l^{0}$, respectively,  in Equation~(\ref{Geodesicspace}) and~neglecting terms like $H\Phi$ and $H\vec{l}$ because of their minute role at local physics, one can  finally derive the governing equation of the  spatial components as
\begin{eqnarray}  \label{trajectory}
\frac{d\vec{l}}{d\lambda} = - 2 |\vec{k}|^{2} \vec{\nabla}_{\perp} \Phi -2 \left(c_{1} \frac{\dot{R}}{R} \right) R  |\vec{k}|^{2}   \hat{k}
\end{eqnarray} 
in which the gradient transverse to the trajectory $ \vec{ \nabla}_{\perp} $ is defined as follows 
$$ \vec{ \nabla}_{\perp} \Phi \equiv \vec{ \nabla} \Phi  - \frac{\vec{k}}{|\vec{k}|^{2}}  (\vec{k}.\vec{\nabla} \Phi).  $$ The deflection angle $\alpha \equiv - \frac{\Delta \vec{l}}{k^{0}}$, in~which $\Delta \vec{l} = \int \frac{d\vec{l}}{d\lambda}d\lambda $, could be simply derived as reported in Reference~\cite{Shenavar:2016bnk}. 

We stress the fact that the above modified trajectory is derived based on the method of Reference~\cite{Pyne:1995ng}. However, if~one hastily applies the usual lens formula of $\alpha \propto \int \vec{\nabla}_{\perp}  (\Phi + \Psi)d\lambda$ and replaces $\Psi$ from Equation~(\ref{Condition}), then there remains no modified term proportional to $c_{1}$. On~the other hand, by~first deriving the deflection value $ d\vec{l}/d\lambda $ at any point of the trajectory as Equation~(\ref{trajectory}) and by~then integrating over photon geodesic $\int \frac{d\vec{l}}{d\lambda}d\lambda $, one could retain the role of the boundary condition $c_1$, i.e.,~the effect of the distant stars is preserved. The~difference between the two methods is that, based on the  method of Reference~\cite{Pyne:1995ng}, which we used here, one has to use Equation~(\ref{Nullfirst}) to  replace the value of $l^{0}$ into Equation~(\ref{Geodesicspace}),  which is a more appropriate~approach. 
 
  The appearance of the acceleration term $ c_{1} c \frac{\dot{R}}{R} $ in Equation~(\ref{trajectory})  is the result of the coupling between the background and the perturbed metrics. Here, we have recovered the velocity of light $c$ to estimate the order of magnitude of this term.  From~physical point of view, this term indicates that the evolution of very far objects affects the local physics of gravitating systems.  The~magnitude of this acceleration is of the order of $10^{-10}$~m/s$^{2}$ to $10^{-11}$~m/s$^{2}$   depending on the value of the Neumann parameter  $c_{1}$.   A~similar quantity has been repeatedly observed in MOND phenomenology~\cite{Milgrom:1983ca,Famaey:2011kh}.

The aforementioned modified lens equation has  been tested against a sample of ten strong lensing systems,  and it has been    shown that the estimated masses of these systems are  within  the observed values  except for the case of the Q0142-100 lensing system~\cite{Shenavar:2016bnk}. The~evaluated mass of this system shows about $ 7.5 \% $ deviation compared to the lower observed bound. This case is not   in strong contradiction with the model regarding the possible sources of uncertainties, such as  error in the position of the image, the~impact parameter, and approximating the lenses as spherical~systems.

The above method also changes the geodesics of massive particles. The~key point here is that, in GR, one has to use geometrodynamic clocks to measure spacetime intervals and physical observables such as velocity, acceleration etc.~\cite{Shenavar:2016bnk}.  Geometrodynamic clocks use particles in free fall and light signals to measure space and time. See Reference~\cite{ohaniangravitation}, chapter 5, for~a review on spacetime measurement through geometrodynamic clocks. Thus, a~modification in the  light's geodesics would consequently result in a change in measurement's outcome.    See Appendix \ref{DisMeasurement} for a related discussion on geometrodynamic clocks and the implications of the  key EPS theorem~\cite{ehlers2012republication}.     Also, it is possible to generalize the equation of motion of a single particle   to a system of particles. See ``Section 3- Systems of Particles'' of Reference~\cite{Shenavar:2016xcp} for the details. Then, the~appropriate potential of MOD is derived, and it has been shown that this model is governed by the next modified Poisson equation:
\begin{eqnarray}   \label{Second}
 c^{2} \frac{1}{R^{2}}\nabla^{2}\Phi &=& 4 \pi G  (\delta \rho_{m}+ \delta \rho_{c_1})
\end{eqnarray}
where
\begin{eqnarray}  \label{dens1}
 \delta \rho_{c_1} &=& \frac{c_{1}  cH(t) }{\pi G M} \int \frac{\delta \rho_{m}(t, \vec{x^{\prime}})d^{3}\vec{x^{\prime}}}{| \vec{x^{\prime}} - \vec{x}|}
\end{eqnarray} 
plays the role of the missing mass at galactic scales.   Appendix \ref{DisMeasurement} provides a thorough derivation of Equation~(\ref{Second}).   The~parameter $c$ in Equations~(\ref{Second}) and \eqref{dens1} is recovered to manifestly show that $\delta \rho_{c_1}$ and $\delta \rho_{m}$  have the same dimension.  At~galactic scales,   the~time evolution of $H(t)$ and $\delta  \rho_{m}(t,\vec{x^{\prime}}) $ is usually considered negligible because the time period that one considers a typical galaxy is much shorter than the cosmic time for which $H(t)$,   and~$ \delta \rho_{m}(t, \vec{x^{\prime}}) $ changes significantly.  Of~course, the~problem of galactic evolution is an exception.  Also, in~this work,  we deal with cosmic evolution; thus, it is important to include the time evolution of these~factors.

We note that $\delta \rho_{c1}$ is in fact determined by $ \delta \rho_{m}$ and that it is actually a new energy between particles;   therefore, $\delta \rho_{c1}$ has  no independent existence apart from the  matter distribution. Strictly speaking, $\delta \rho_{c1}$ does not clump; only matter density $ \delta \rho_{m}$ clumps. Therefore, no interaction between $ \delta \rho_{c1}$  and $ \delta \rho_{m}$ would exist.  However, we will continue to show this energy as $\delta \rho_{c1}$ because, in this way, it is more convenient to incorporate it within the field equations and to derive  the   modified Friedmann~equations. 

  One could evaluate $\delta  \rho_{c_1} $ by integrating Equation~(\ref{dens1}) over the space in which the matter exists. This is a nontrivial work; however, it could be done at least numerically.   The~important point is that the integral in Equation~(\ref{dens1}) has the dimension of mass per scale length $ \propto M/R_{p}$ in which $R_{p}$ is the particle horizon and $M$ is the mass within  it. See Figure~\ref{fig:boundary}. Thus, the~mass $M$ could be  simplified in Equation~(\ref{dens1}).  On~the other hand, if~we show the current size of the particle horizon as $L_{0}$ and its time evolution as the dimensionless function $l(t)$, i.e.,~$R_{p}=L_{0}l(t)$, then the modified density $\delta  \rho_{c_1} $ could be written as
 $$ \delta  \rho_{c_1} = \frac{c_{1}cH(t)  }{\pi G L_{0} l(t)}  $$
which represents the effect of distant cosmic objects on local physics.  We assume for the sake of simplicity  that  the constant dimensionless parameter which emerges from   the  integration in Equation~(\ref{dens1}) has been absorbed in the parameter $L_0$. The~logic behind this is the fact that the parameter $L_{0}$ will disappear when we write the equations based on dimensionless~densities. 

 If the time evolution  of the scale factor was known at this stage, then we could determine the size of the particle horizon from $R_{p} = L_{0} l(t) \equiv  c \int da/a\dot{a}$. However, we still have some clues about the past epochs of the cosmic evolution. For~example, when the universe is in the phase of radiation domination (or matter domination),  the~share of the new term ought to be negligible. Thus,  as~we will show in Section~\ref{Cosmic},  the~evolution of the scale factor could be found as $a(t) \propto t^{1/2}$ and $a(t) \propto t^{2/3}$ for radiation and matter-dominated universes, respectively. Therefore,  the~particle horizon could be calculated as $ R_{p} = 2ct$ for the former  and $ R_{p} = 3ct$ for the~latter. 
 
Although we could not determine the parameter $l(t)$ in recent cosmic era,  the~important point regarding $l(t)$ is that this is a function of time. For~an expanding universe which starts with a singularity and has no contraction, it could be easily understood that the scale factor is a reversible function of the cosmic time. Therefore, the~parameter $l(t)$ could be written as a function of the dimensionless scale factor $a$ instead of $t$. In~this work, we model  the ratio $ \zeta \equiv 1/l(t)$ in $\delta  \rho_{c_1}$ based on the next power-law function  of the scale factor
\begin{eqnarray}    \label{zeta}
 \zeta = a^{\varepsilon}. 
\end{eqnarray} 
in which the value of $\varepsilon$ would be determined through data analysis.  This toy model proves to be useful especially for low-redshift observations.  A~preliminary data analysis of the model, with~the above assumption, has been reported in the first Arxiv version of this work.  See ``Section~5- Data Analysis'', especially Tables~1 and 2 and Figures~9 to 14,  of~arXiv:1810.05001v1 for the detailed discussion.     However, we should mention that, for a problem with a broader range of redshift, for~example CMB physics, one might need a more general function of the scale factor to describe $ \zeta $. 

Using the above definition of $\zeta$, the~modified energy  density of this model is as follows:
\begin{eqnarray}  \label{dens2}
\delta  \rho_{c_1} &=& \frac{c_{1}cH(t)a^{\varepsilon}}{\pi G L_{0}}  
\end{eqnarray}   
 In a future work and based on the Hubble parameter and SNIa data, we will report  that the  value of $\varepsilon$ is most probably about $0.15$ (Shenavar and Javidan, in~preparation). The~effect of changing $\varepsilon$ on data fittings will be also discussed.   On~the other hand, as~mentioned above,  the~exact value of $L_{0}$ will not be needed in building a cosmological model because the~dimensionless density parameter related to $ \delta  \rho_{c_1}  $ is independent of the scale length $L_{0}$.

In classical hydrodynamics of fluids, within~the context of Newtonian mechanics, one does not expect that adding a constant to the gravitational potential result in any change in the final physics. The~reason is that the gravitational effects are introduced to the governing equations through the force (not the potential). On~the other hand, in~general relativity, the~two scalar potentials are basically independent. Regarding this point, the~coupling between zero-order and first-order terms of metric, i.e.,~$H$ and $c_{1}$, respectively, introduces a  nontrivial term of the order of de Sitter scale of acceleration $cH_{0}$ as we saw above. Finally, it is easy to observe that the $MOD$ model reduces to old results in the limit of Dirichlet B.C. $c_{1} = 0$; however, the~vice~versa is neither correct nor necessary according to  ``correspondence principle''.  Although~adding a constant to the potential in Newtonian gravity would provide no observable effect, it is still possible to build a semi-Newtonian model of modified dynamics  based on some simple   and  well-motivated assumptions. See Appendix \ref{DisNewt} for the details. The~result would be almost the same; however, the~main difference is that, in this latter approach, the value of the new acceleration should be determined from~observations. 

\section{Modified Hubble~Parameter}
\label{Modified}
In this section, we want to find the average of the perturbations to derive the governing  equations at cosmic scales, i.e.,~the Friedmann equations. Literature related to backreaction cosmology have produced a rigorous mathematical framework to study different perturbative orders and their averaging process. This is expected because backreaction cosmology  tries to find the effects of inhomogeneities on background metric to explain the accelerated expansion of the universe~\cite{Rasanen:2003fy,Hirata:2005ei,Kolb:2004am,Martineau:2005aa}. Thus, here, we will use the averaging method usually presumed in backreaction cosmology.  In~the following, the~local and background quantities are displayed by $l$ and $b$ indices, respectively. Assuming this,  the~local perturbation equations can be written as
\begin{eqnarray}
~^{l}G_{\mu \nu} &=& \kappa (~^{l}T_{\mu \nu} +~^{b}T_{\mu \nu}) -~^{b}G_{\mu \nu}
\end{eqnarray}
which, by~considering the negligibility of background quantities such as $\rho$, $P$, $H^{2}$, $\ddot{R}$, compared to the local density of galaxies and clusters, could be rewritten as
\begin{eqnarray}
~^{l}G_{\mu \nu} &\simeq & \kappa ~^{l}T_{\mu \nu}. 
\end{eqnarray}
in local universe. See Table~5 of Reference~\cite{jacobs1992obtaining} for order of magnitude estimations of the ratios of local-to-global density and pressure at different~scales.  

On the other hand, the~time evolution of the cosmos, at~its largest scales, could be derived by finding the spatial average of the inhomogeneities. However, the~process of spatial averaging of Einstein field equations  is still debatable~\cite{Zalaletdinov:1992cg,Zalaletdinov:1992cf}, in~the literature related to cosmology  (especially backreaction cosmology), it is customary to use the next definition to find the spatial mean of any inhomogeneities $A (t,  x)$~\cite{Bagheri:2014gwa}:
$$ \langle A \rangle (t,  x) \equiv \frac{\int A(t,x) \sqrt{h(t,x)} d^{3}x}{\int  \sqrt{h(t,x)} d^{3}x}  $$
where $h$ is the determinant of the perturbed metric on hypersurface of constant time. In~this way, the~background Einstein equation is as follows
\begin{eqnarray}
\langle~^{b}G_{\mu \nu} \rangle &=& \kappa~\langle~\left(~^{l}T_{\mu \nu} +~^{b}T_{\mu \nu} \right)~  \rangle - \langle~^{l}G_{\mu \nu} \rangle
\end{eqnarray}
by which one could see that because the spatial average of all space-dependent components of the Einstein field equations---including $\Phi$, $ \delta \rho_{m} $, $\delta P$, $\delta u$, and $ \pi^{S}$---would vanish;  only pure time-dependent terms would survive.  These terms are shown with underbrace in Equations~(\ref{Einstein1}) and \eqref{EMtensor1}. Also, the~term $\delta \rho_{c_1}$ would survive because, as~one could see from Equation~(\ref{dens2}), this term is pure time dependent. We will show the average of $\delta \rho_{c_1}$ as $ \rho_{c_1}$, i.e.,~$ \rho_{c_1} \equiv \langle \delta \rho_{c_1} \rangle$.  Using this process of spatial averaging, one could derive the following  from Equation~(\ref{Einstein1}) and \eqref{EMtensor1}:
\begin{eqnarray}  \label{Friedmann1}
3 \frac{\dot{R}^{2}}{R^{2}} &=& 8 \pi G\rho  \\   \nonumber
\frac{\dot{R}^{2}+2R\ddot{R} }{R^{2}} &=& -8\pi G P  \\   \nonumber
\dot{\rho} +3(\rho +P)\frac{\dot{R}}{R}  &=& 0
\end{eqnarray}
in which now the total energy of the universe consists of radiation $\rho_{r}$, baryonic matter $\rho_{m}$,  and the term due to our modification $ \rho_{c_{1}}$ which is defined in Equation~(\ref{dens2}). In~other words, we have $ \rho = \rho_{m} + \rho_{r} +\rho_{c_1} $. 

In the present work, we neglect the cosmological constant $\Lambda$ in contrast to the procedure that was used in Reference~\cite{Shenavar:2016bnk}. The~reason is that, as~we will discuss below, a term proportional to $H$ in the Friedmann equation could provide a deceleration parameter about $-1$ in recent cosmic time. See Section~\ref{EoS} below.

 The first two equations in Equation  \eqref{Friedmann1} can be rewritten as
\begin{eqnarray}   \label{Friedmann}
H^{2} &=& \frac{8\pi G}{3}\rho -\frac{ k}{R^{2}}  \\   \nonumber
\ddot{R} &=& -\frac{4\pi G}{3}(\rho + 3P)R  
\end{eqnarray}
in which we have resumed the role of the curvature constant $k$ for the sake of completeness and future references. Here, the~parameter $k$ is different from the zero-order wave vector  $k^{\mu}$ (and $\vec{k}$) in the previous section. The~values of $k=1, 0, -1$ refer to closed, flat, and open universes, respectively.  These equations have essentially the form of Friedmann equations because, as~we saw above, the~form of Einstein equations remains unchanged in our approach. However, the~energy content of the universe is now changed due to the presence of $\rho_{c_{1}}$, i.e.,~the modified dynamical energy. In~addition, one can rearrange the conservation equation, i.e.,~the third equation of Equation \eqref{Friedmann1},  as~\begin{eqnarray}   \label{Conserved1}
\frac{d}{dR}(\rho R^{3}) &=& -3PR^{2}
\end{eqnarray}
which  could be dealt with more easily because the time dependency of the parameters is now~implicit. 

In this work, we assume that the energy density of the  universe is only due to  radiation $ \rho_{r} $,  baryonic matter $ \rho_{m} $, and Neuamnn term $ \rho_{c_1} $. We will find that, in this modified cosmological model,  a~large portion of the energy content of the universes is due to this last term. By~including the effect of Neumann B.C. through $ \rho_{c_1}$,   one has the total density as  $ \rho = \rho_{m} + \rho_{r} +\rho_{c_1} $.  The~cosmological energy components are considered to be noninteracting; thus, the~conservation equations, i.e.,~Equation~(\ref{Conserved1}), holds separately for each component. Also, one could assume an equation of state of the form $P_{i} =w_i \rho_i$ in which we have used indices $i=m,~r,~c_1$ for matter, radiation, and $c_1$ components respectively. Then, it is possible to show that, for  dust with $P=0$, one has $\rho \propto R^{-3}$ while, for radiation with $P=1/3 \rho$, one derives $\rho \propto R^{-4}$. The~total equation of state parameter, i.e.,~$w$, is time dependent.    Using the above results and~the expression of $ \rho_{c_1} = \frac{c_{1}c}{\pi G L_{0}} H\zeta $, one could find the evolution of the total density as
$$ \rho = \rho_{0,m} (\frac{R_{0}}{R})^{3}+\rho_{0,r} (\frac{R_{0}}{R})^{4}+\frac{c_{1}c}{\pi G L_{0}}H\zeta  $$
in which $\rho_{0,m}$ and $\rho_{0,r}$ are respectively the matter and radiation densities at the present time.  Knowing that $\zeta$ is dimensionless,  one could easily check that the last term on the rhs has the dimension of~density.

It is worth noticing that there is a fundamental difference between $ \rho_{r} $ and  $ \rho_{m} $ on one side and~$ \rho_{c_1} $ on the other. The~point is that, although $ \rho_{r} $ and  $ \rho_{m} $ are related to physical substances, namely radiation and matter, the~term $ \rho_{c_1} $ is due to the modification of dynamics. In~fact, as~one can see from the definition of $ \rho_{c_1} $, this term is related to Hubble parameter $H$ and scale factor; thus, from~conceptual point of view,  it might be even more appropriate to put it on the left-hand side of the Friedmann Equation~(\ref{Friedmann}) because it is more similar to $H^{2}$ (In this respect, the~present model is within the larger class of dynamical dark energy models.).  However, since we want to compare our results with those of $\Lambda CDM$, we treat this term as  a density and keep it on the left-hand side of Equation  \eqref{Friedmann}. Of~course, the~results of the two approaches would be identical because they are  mathematically~equivalent.

In cosmology, it is well-known that working with dimensionless densities, i.e.,~the density parameters $\Omega$, is easier than working with the physical densities $\rho_i$. Defining the density parameters as
\begin{eqnarray}   \label{densparas}
\Omega_{i}(t) &\equiv &  \frac{8\pi G \rho_{i}(t)}{3H^{2}(t)}  \\   \nonumber
\Omega_{k}(t) &\equiv &   -\frac{k}{H^{2}(t)R^{2}(t)}
\end{eqnarray}
one could rewrite the first Friedmann equation  in Equation~(\ref{Friedmann}) as follows:
\begin{eqnarray}   \label{Hubble1}
H^{2} (t) &=& H_{0}^{2}\left(\frac{\Omega_{0,m}}{a^{3}} +  \frac{\Omega_{0,r}}{a^{4}} +  \frac{\Omega_{0,k}}{a^{2}} + \Omega_{0,c_{1}} \frac{H}{H_0} \zeta (a)  \right)
\end{eqnarray}
in which $ \Omega_{0,i} $ is the $i$th density parameter at present time $t_0$ while  $a(t)\equiv R(t)/R_0=(1+z)^{-1}$ is the normalized scale factor. In~definition of $a(t)$, the~parameter $z$ represents the redshift. From~Equation~(\ref{Hubble1}), one could see that, for the curvature density parameter, we have $\Omega_{k}(t) = 1-\sum_{i} \Omega_{i}(t)$. While other density parameters are positive by definition, the~curvature density parameter could be positive, negative, or zero depending to the the value of $ \sum_{i} \Omega_{i}(t)$.

 Equation~(\ref{Hubble1}) could be simply solved to derive the Hubble parameter $H(t)$. By~doing so, one finds
\begin{eqnarray} \label{Hubble2}
H_{\pm}(t) =\frac{H_0}{2} \left( \Omega_{0,c_{1}} \zeta (a)   \pm  \sqrt{\Omega^{2}_{0,c_{1}}\zeta^{2} (a) +4 ( \frac{\Omega_{0,k}}{a^{2}} + \frac{\Omega_{0,m}}{a^{3}} +  \frac{\Omega_{0,r}}{a^{4}})}      \right)  
\end{eqnarray} 
which is the key formula for our following discussions. Imagine that $ \Omega_{0,k} \geqslant 0 $, i.e.,~$k \leqslant 0$. Then, the~plus sign in Equation~(\ref{Hubble2}) represents an expanding universe, i.e.,~$  H > 0$, while the minus sign shows a contracting cosmos with  $  H < 0$.  On~the other hand, when $ \Omega_{0,k} < 0 $, i.e.,~$k > 0$, then at some normalized scale factor $a$, the value under the square root in Equation~(\ref{Hubble2}) could be negative, which makes the Hubble parameter imaginary and so the real part of the scale factor would be  oscillatory.  However, when this value is nonnegative, then the Hubble parameter could be positive or negative, depending to the sign of $ \pm $ in Equation~(\ref{Hubble2}), and~so we might have an expanding or contracting  universe in this case.  As~we will see in the next section, $ \Omega_{k}$ can be neglected in our~universe. 

It is insightful to compare the behavior of the current model with that of the $\Lambda CDM$ model. The~Hubble parameter in $\Lambda CDM$ cosmology is as follows
\begin{eqnarray}  \label{Hlcdm}
H^{2}_{\Lambda CDM} (t) &=& H_{0}^{2} \left(\frac{\Omega_{0,k}}{a^{2}} + \frac{\Omega_{0,m}}{a^{3}} +  \frac{\Omega_{0,r}}{a^{4}} +   \Omega_{0,\Lambda}  \right)
\end{eqnarray}
in which $ \Omega_{0,\Lambda} \equiv \Lambda/ 3H_{0}^{2} $ is the density parameter of the cosmological constant while the definitions of the other parameters have been presented above. Again, it is possible to introduce an expanding (contracting) universe with $H_{ \Lambda CDM} > 0$  ($H_{  \Lambda CDM} < 0$). Of~course, we will consider an expanding~universe.

As stated above, in~Neumann cosmology, the term responsible for the accelerated expansion changes with time. Models of the universe with variable cosmological terms, i.e.,~dynamical dark energy models, have been repeatedly proposed before.  See Reference~\cite{Overduin:1998zv}, especially Table I,  for~a review on a variety of these models. In~fact, there have been previous papers suggesting models with a dark component of which its  energy density is proportional to Hubble parameter, i.e.,~$\Omega \propto H(t)$. The~behavior of such models are somehow similar to the present model, though their origin is far different. These models are proposed based on Veneziano ghost of QCD. This ghost is considered as unphysical in the usual Minkowski spacetime. However, it could lead to interesting cosmological consequences in dynamical spacetimes or spacetimes with nontrivial topology. See References~\cite{Cai:2010uf,Cai:2012fq} and references therein for more details. Assuming $ \Lambda_{QCD} \approx 100 MeV$ as the mass scale of QCD, one could show that $ \Lambda^{3}_{QCD} H $ is of the order of observed dark energy density. This coincidence makes the Veneziano ghost model particularly interesting. See also a relevant discussion in Appendix \ref{Disdimensional}.

Cai~et~al.~\cite{Cai:2010uf} have shown   that Veneziano ghost  leads to a late time de Sitter phase of the universe and its   accelerated  expansion at $z \approx 0.6$. They have used observational data of big bang nucleosynthesis, baryon acoustic oscillation, type Ia supernovea,  Hubble parameter data, and  cosmic microwave background to estimate the free parameters of the model.  The~best-fit values of the parameters of Veneziano dark energy model lead to a minimum of the probability distributions $\chi^{2}$,  which is about $10 \%$ larger than the minimum of  $\chi^{2}$ derived from $\Lambda CDM$. In~both models. the existence of the dark matter is presumed. Their study has  been expanded by Reference~\cite{Cai:2012fq}  to include a more general model which includes two terms proportional to $H$ and $H^{2}$.  This latter model too approaches a de Sitter phase  while it begins to accelerate at   $z \approx 0.75$. Also, according to Reference~\cite{Cai:2012fq}, the~data analysis of  the growth factor based on this latter model shows a fit as well as those of standard~model.

It is worth mentioning that, by imposing Dirichlet B.C., i.e.,~simply putting $\Omega_{0,c_{1}} =0$ in our equations,  Equations~(\ref{Hubble1}) and \eqref{Hubble2} reproduce the  Hubble parameter of the conventional cosmology without a cosmological constant. Therefore, our model includes  $\Lambda CDM$ without $\Lambda$. However, as~we will see in the following, this model could provide an   accelerated  expansion without assuming the cosmological~constant.

\section{Cosmic Implications of the New~Model}
\label{Cosmic}

In this section, we will outline the main cosmological consequences of the present model. Throughout this section, we will assume that the universe is flat, i.e.,~$\Omega_{0,k}=0$. Also, in~all plots of this section, the~matter density of the $MOD$ model is assumed to be $\Omega_{0,m}=0.15$ and  $\Omega_{0,m}=0.25$. These values are the lower and upper limits of matter density according to our data analysis in Shenavar and Javidan (in preparation). Also, we assume $\varepsilon = 0.15$. 

For the sake of comparison, the~results of $\Lambda CDM$ model is also provided  for which we have assumed the total mass density, the~baryonic plus CDM densities, as~$\Omega_{0,b+cdm}=0.3$.  Here, for both models, the~present radiation density is considered as $\Omega_{0,r}=5 \times 10^{-5}$.

We first start by discussing  the evolution of the scale factor. Then, the~behavior of the Hubble parameter is surveyed and it is proved that the deceleration parameter converges to $< -1$ for a flat universe. This behavior  ensures an accelerating expansion at later cosmic times.  Moreover,  we will derive  the  equation of state parameter and  the  sound speed. Furthermore, it is shown that our model reproduces a  viable sequence of cosmic eras. Also, the~luminosity and angular diameter distances are found to be significantly different at $z \gtrsim 2$, making a good opportunity to contrast the model with $\Lambda CDM$ at high redshifts in future~works.

\subsection{Evolution of the Scale~Factor}
\begin{figure}[H]
\centering
\includegraphics[height=5cm,width=8cm]{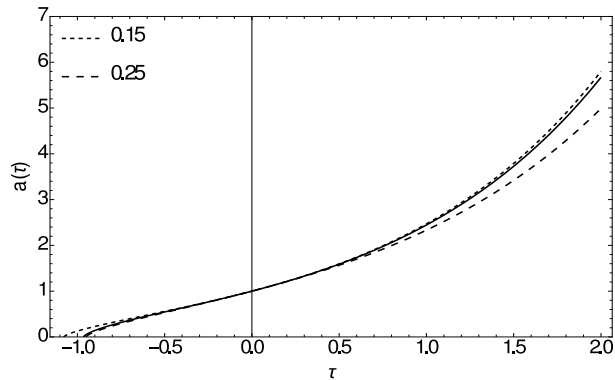}  
\caption{The evolution of scale factor $a$ as a function of dimensionless cosmic time $\tau \equiv H_{0} t$  for  $\Omega_{0,r} =5 \times 10^{-5}$, two values of baryonic density namely $\Omega_{0,m} = 0.15 $ and $ 0.25$, and~$\Omega_{0,c_1} = 1-\Omega_{0,r}-\Omega_{0,m}$: In~addition, we have assumed $\varepsilon = 0.15$.   It could be proved that, by increasing $\varepsilon$, the~scale factor in the past epochs remains almost unchanged while the scale factor in future cosmic time increases.  The~evolution of  the scale factor $a$ in the $\Lambda CDM$ model is also shown by a solid~line. \label{scalefactorplot}}
\end{figure}

In a flat expanding universe, one could see from Equation~(\ref{Hubble2}) that  the evolution of the dimensionless scale factor is  governed by
\begin{eqnarray}   \label{scale1}
\frac{d a(\tau)}{d\tau} = \frac{1}{2} \left(\Omega_{0,c_{1}}a^{1+\varepsilon}(\tau) 
+ \sqrt{\Omega^{2}_{0,c_{1}}a^{2+2\varepsilon}(\tau) +4 (\frac{\Omega_{0,m}}{a(\tau)} +  \frac{\Omega_{0,r}}{a^{2}(\tau)})}  \right)
\end{eqnarray} 
 in which $\tau \equiv H_{0} t $ is the dimensionless time. This differential equation is solved numerically for two values of $\Omega_{0,m}$ and the initial value of $a(0)=1$, and the results are reported in Figure~\ref{scalefactorplot}.  The~evolution of the scale factor in the $\Lambda CDM$ model is also shown by a solid line which closely follows a Neumann cosmology with $0.15 \lesssim \Omega_{0,m} \lesssim 0.25 $. 

Apart from this numerical solution, one could find the time dependency of the scale factor in some special cases. For~example,  in~early universe, the share of the radiation dominates the right hand side of Equation~(\ref{scale1}). In~this case, the scale factor could be derived as $a(\tau) \propto \tau^{1/2}$. Also, in~later times, the matter term dominates the rhs of Equation~(\ref{scale1}) and so we derive $a(\tau) \propto \tau^{2/3}$. Finally, at~later epochs, the terms corresponding to matter and radiation vanish and  $ \Omega_{c_{1}} $ start to govern the cosmic evolution.    Here, if~we have $\varepsilon = 0$,  then  the solution to Equation~(\ref{scale1}) would be exponential, i.e.,~$a(\tau) \propto \exp (\Omega_{0,c_{1}} \tau)$.  These last three statements could be easily derived from Equation~(\ref{scale1}). On~the other hand,   in~the case of a nonzero   $\varepsilon$,  we need to numerically solve Equation~(\ref{scale1}) to derive the scale~factor.   

\subsection{Hubble and Deceleration~Parameter}
\begin{figure}[H]
\centering
\includegraphics[height=5cm,width=8cm]{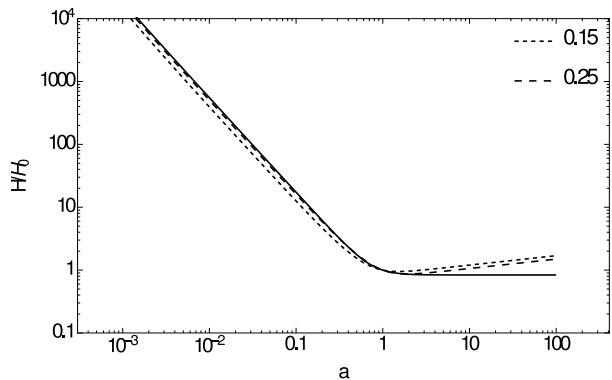} 
\caption{Hubble parameter as a function of scale factor: See Figure~\ref{scalefactorplot} for the values of the basic parameters.   It could be shown that, by increasing $\varepsilon$, the~Hubble parameter in the past epochs remains unchanged while it increases  in future cosmic~time.  \label{hubbleplot}}
\end{figure}

From Equation~(\ref{Hubble2}), one can readily plot the Hubble parameter as a function of the scale factor as you may see in Figure~\ref{hubbleplot}. In~early cosmic time, the~behavior of Hubble parameter in $MOD$ is quite similar to that of the $\Lambda CDM$ model, though~the value of Hubble parameter in the former is generally smaller than the latter for $a<1$. However, the~behavior of $H$ for $a>1$ is completely dependent to the future evolution of $\zeta$ (or $\varepsilon$). The~curves in Figure~\ref{hubbleplot} are graphed for  $\varepsilon = 0.15$; however,    in~the case of  $\varepsilon = 0$,  we would have a flat   Hubble  parameter  for $ a > 1$.

\begin{figure}[H]
\centering
\includegraphics[height=5cm,width=8cm]{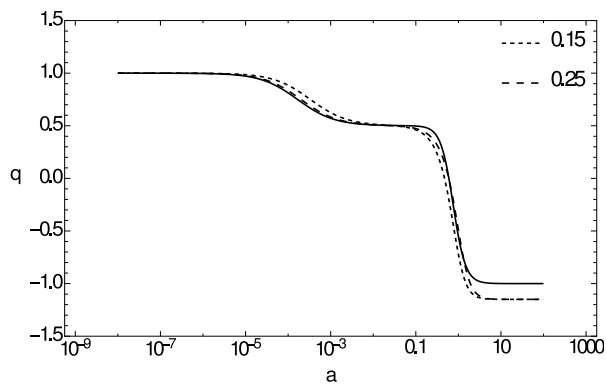} 
\caption{The evolution of the deceleration parameter $q$ as a function of the scale factor:  See Figure~\ref{scalefactorplot} for the values of the basic parameters.   It could be verified that, by increasing $\varepsilon$, the~$q$ parameter in the past epochs remains almost unchanged while it decreases  in future cosmic time as $q = -1-\varepsilon$.  \label{q2plot}}
\end{figure}

The deceleration parameter $q$ is defined as
\begin{eqnarray}  \label{deceleration}
q=-a\ddot{a}/\dot{a}^{2}  &=& -1-\dot{H}/H^{2}
\end{eqnarray}
or equivalently $q= \frac{4\pi G}{H^{2}(t)}\sum_{i}\rho_{i}(1+3w_i) $. The~Friedmann equations  have been used to derive this last expression.   The~evolution of $q$ is graphed in Figure~\ref{q2plot} in which one observes $q=1$ in  radiation-dominated era, then  $q=1/2$ in  matter-dominated epoch, and  finally $q = -1-\varepsilon$ in future cosmic times.  Also, to~plot $q$ as a function of the normalized scale factor, we have  used the next identity:
\begin{eqnarray}   \label{conversion}
\frac{d}{dt} &=& a H(a)\frac{d}{da}
\end{eqnarray}
which converts time derivative of any quantity to its derivative with respect to $a$. The~negativity of the deceleration parameter at  present cosmic time is indeed needed to fit observational data to the model as is discussed by Shenavar $and$ Javidan  (in preparation).  It is also worth noting that, for a model with $\varepsilon = 0$ (similar to QCD ghost model), one would have a similar behavior except that, in future cosmic time, one derives $q = -1$. This is not plotted in Figure~\ref{q2plot}.

The negativity of the deceleration parameter in late cosmic time has an important effect on the arguments related to the value of the current cosmic curvature. In~fact, using the definition of $\Omega_{k} $, one could derive the time derivative of the curvature density parameter as $\dot{\Omega}_{k}=2\Omega_{k}Hq$. Then, if~$\Omega_{k}$ has some nonzero value at a very early universe, the~late-time negativity of $q$ ensures that the value of  $\Omega_{k}$ becomes negligible at present. However, by~assuming  $c_{1} =0 $,   i.e.,~imposing the Dirichlet boundary condition, the~parameter $q$ would always be positive  and the $\Omega_{k}$ would not be negligible. This is one of the reasons that one needs cosmological constant $\Lambda$ in the standard~model.

\subsection{Evolution of the Equation of State Parameter and the Sound~Speed }
\label{EoS}
\begin{figure}[H]
\centering
\includegraphics[height=5cm,width=8cm]{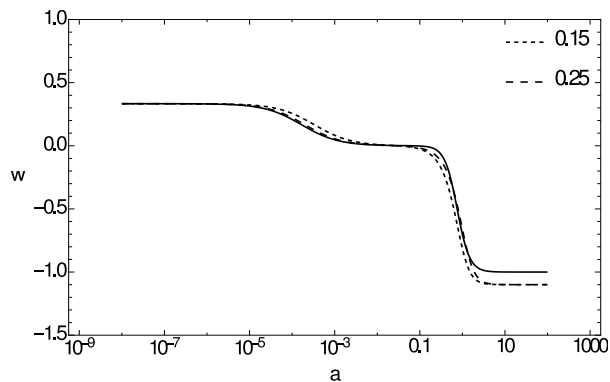}
\caption{Evolution of the equation of state $w$ as a function of scale factor $a$ for  $\Omega_{0,r} =5 \times 10^{-5}$  for two different values of $\Omega_{0,m} $:   One could prove  that, by increasing $\varepsilon$, the~equation of state in the past epochs remains almost unchanged while it decreases  in future cosmic time as $w = -1-2\varepsilon /3$.  The~evolution of  $w$ in the$\Lambda CDM$ model is also shown by a solid~line.   \label{w2plot}}
\end{figure}

In a flat universe, the~total equation of state parameter $ w_t $ could be derived from Friedmann equations as follows:
\begin{eqnarray}
w_t &=& -1-\frac{2\dot{H}}{3H^{2}}.
\end{eqnarray}
  Now one could simply plot $ w_t (a) $ as shown in Figure~\ref{w2plot} by using the conversion formula of Equation \eqref{conversion}.   As~this graph shows, the~value of $  w_t $  is initially very close to $1/3$ because the universe starts from a radiation dominated era.  Then, in~a matter-dominated era, $  w_t $ approaches very rapidly to zero.   Finally, at~later times, $ w_t (a) $ converges (from above) to $w = -1-2\varepsilon /3$, which results in an accelerating expansion of the universe.  Therefore, in~a completely $\Omega_{ c_1}$-dominated universe, the~equation of state must be less than $-1$ for $\varepsilon > 0$. In~this regard,  the~present  model is similar to phantom models of dark energy; however, the~existence of a big rip in future, which is the characteristic sign of a phantom scenario, crucially depends on the value of $\varepsilon$ at that time. In~fact, it is more likely that big rip never happens in Neumann cosmology because all forces of this model are attractive. See the modified Poisson Equation~\eqref{Second} above.  In~contrast, in~a phantom model, the big rip happens when the repulsive force of  dark energy rips apart all structures down to subatomic particles~\cite{Caldwell:2003vq}. See also Reference~\cite{Bamba:2012cp} for a thorough review. The~issue of the behavior of $\varepsilon$ and,~thus, the possibility of a doomsday scenario in future must be resolved with a  more careful study to determine the behavior of   $\zeta(t) = 1/ l(t)$. 

See   Vagnozzi~et~al.~\cite{Vagnozzi:2018jhn}, who compare  dynamical dark energy models with $w \geq -1$ and $w < -1$. The~authors show that, unlike models which allow  values of $w < -1$,  a~dark energy component with $w \geq -1$ is unable to reduce the tension between  observables of high redshift and direct measurements of $H_{0}$. We will investigate this matter in Shenavar and Javidan (in preparation). It is worthy to note that the curves with smaller $\Omega_{0,m}$ tend to $w < -1$   more quickly because the share of $\Omega_{0,c_1} $ is larger in these cases. Also, we point out that the parameter  $  w_t $ of the $\Lambda CDM$ model coincides (approximately) with $0.15 \lesssim \Omega_{0,m} \lesssim 0.25 $ of the present model. See Figure~\ref{w2plot}.
 
\begin{figure}[H]
\centering
\includegraphics[height=5cm,width=8cm]{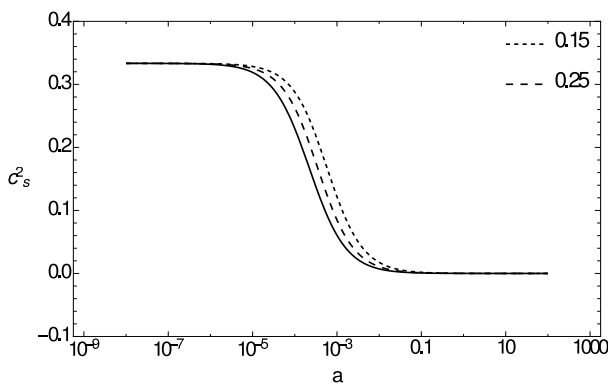}
\caption{Evolution of the adiabatic sound speed $c^{2}_{s(a)} $ as a function of scale factor $a$:    It is straightforward to show  that, by changing $\varepsilon$, the~adiabatic sound speed  remains almost unchanged   in all cosmic~time.  \label{cs2plot}}
\end{figure}

Another important quantity to be compared with its $\Lambda CDM$ counterpart is the sound speed. For~a barotropic fluid, in~which the pressure is only dependent to the density $\rho$, the~sound speed is defined as $c^{2}_{s}= \dot{P} / \dot{\rho}$. This quantity is sometimes called adiabatic sound speed because, in such medium, the entropy per particle is assumed to be constant. As~mentioned above, the~only physical quantities in $MOD$ model, as~in $\Lambda CDM$, are radiation and (baryonic) matter. Thus, the~sound speed could be found from Friedmann equations as $c^{2}_{s}= (\dot{P}_{r} + \dot{P}_{m}) /  (\dot{\rho}_{r}+\dot{\rho}_{m})$. Using this equation and~changing the time derivatives to derivatives with respect to scale factor Equation~(\ref{conversion}), one could plot the sound speed as Figure~\ref{cs2plot}. The~sound speed squared in a radiation-dominated era is about $1/3$ while, in a matter-dominated era and~beyond, tends to zero. The~important point here is the positivity of the sound speed of $MOD$ which results in a bounded (non-exponential) rate of the structure formation. Also, from~$c^{2}_{s} $, one could find the angular size of the sound horizon of $MOD$ and compare it with that of the $\Lambda CDM$ model. This  will be done in~future.

\unskip

\subsection{Sequence of the Cosmological~Epochs}
\begin{figure}[H] 
\centering
\includegraphics[height=5cm,width=8cm]{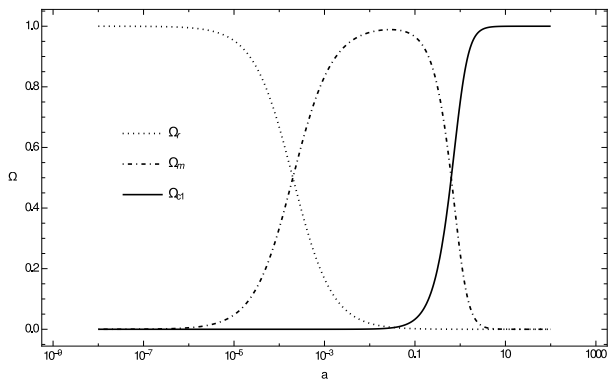}  
\caption{The evolution of the dimensionless density~parameters.     \label{densities}}
\end{figure}

It is easy to check that the present model starts from an unstable radiation-dominated mode, then proceeds toward a matter dominated era, which is also unstable, and then ends its evolution at a stable epoch of $\Omega_{c_1}$ domination. See Figure~\ref{densities}. In~this plot, we have assumed that $\Omega_{0,r} = 5 \times 10^{-5}$, $\Omega_{0,m} = 0.25$, and $\Omega_{0,c_1} = 1- \Omega_{0,m} - \Omega_{0,r}$. One could easily check that,  by choosing other fractions of $\Omega_{0,c_1} $ and $\Omega_{0,m} $ and by keeping  the radiation density at the order of $\Omega_{0,r}  \approx  10^{-5}$, similar behaviors would~emerge.  

In some alternative cosmologies, the~sequence of cosmological epochs is not necessarily as described above.   If~this happens, one might   encounter some  problems in dealing  with the formation of structures in a matter-dominated era. For~example,   Amendola~et~al.~\cite{Amendola:2006we} have  proved that $f(R)$ models of the forms $f(R)=\alpha R^{-n}$ and $f(R)=R + \alpha R^{-n}$ do not show  a viable matter-dominated epoch prior to a late-time acceleration for  any $n > 0$ and $n < -1$.  In~some cases of $f(R)$ theories, the~matter dominated epoch is replaced by an era of  cosmic expansion in which the scale factor varies as $ a \propto t^{1/2}$, which is cosmologically~unacceptable.

Moreover, in~the case of scalar-tensor-vector theory of Moffat~\cite{Moffat:2005si},   Jamali  and Roshan~\cite{Jamali:2016zww} have  shown that there are two radiation-dominated eras, two matter-dominated epochs, and two late time-accelerated phases. In~the matter-dominated phases, the~growth of the scale factor is found to be as $ a(t) \propto t^{0.46} $ and  $ a(t) \propto t^{0.52} $ slower than the growth in standard model and the present model reported above as  $ a(t) \propto t^{2/3} $. However, it should be mentioned that, as~\cite{Jamali:2017zrh} has reported, the~standard MOG possesses a valid sequence of standard cosmological~eras. 

Because the dimensionless density parameter of the Neumann term is inversely proportional to Hubble parameter,  i.e.,~$\Omega_{c_1} \propto 1/H(t)$,  it plays no significant role in an early universe which is dominated by radiation and then matter. A~similar situation exists in the context of the standard model of cosmology, in~which  $\Omega_{\Lambda}$ is negligible until recent time. However, in~the case of $\Lambda CDM$,  one deals with the cosmological constant which is   very tiny $\Lambda \approx 10^{-52} m^{-2}$, while the present model   is built   on introducing a dimensionless parameter with a value much closer to unity $c_1 =0.065$. Also, the~density of the cosmological constant, $\rho_{\Lambda} = \Lambda c^{2} / 8\pi G$ is always a constant while the density of $\rho_{c_1}$ varies with time according to Equation~(\ref{dens2}).

\subsection{Luminosity and Angular Diameter~Distances}

\begin{figure}[H]
\includegraphics[width=7.5cm]{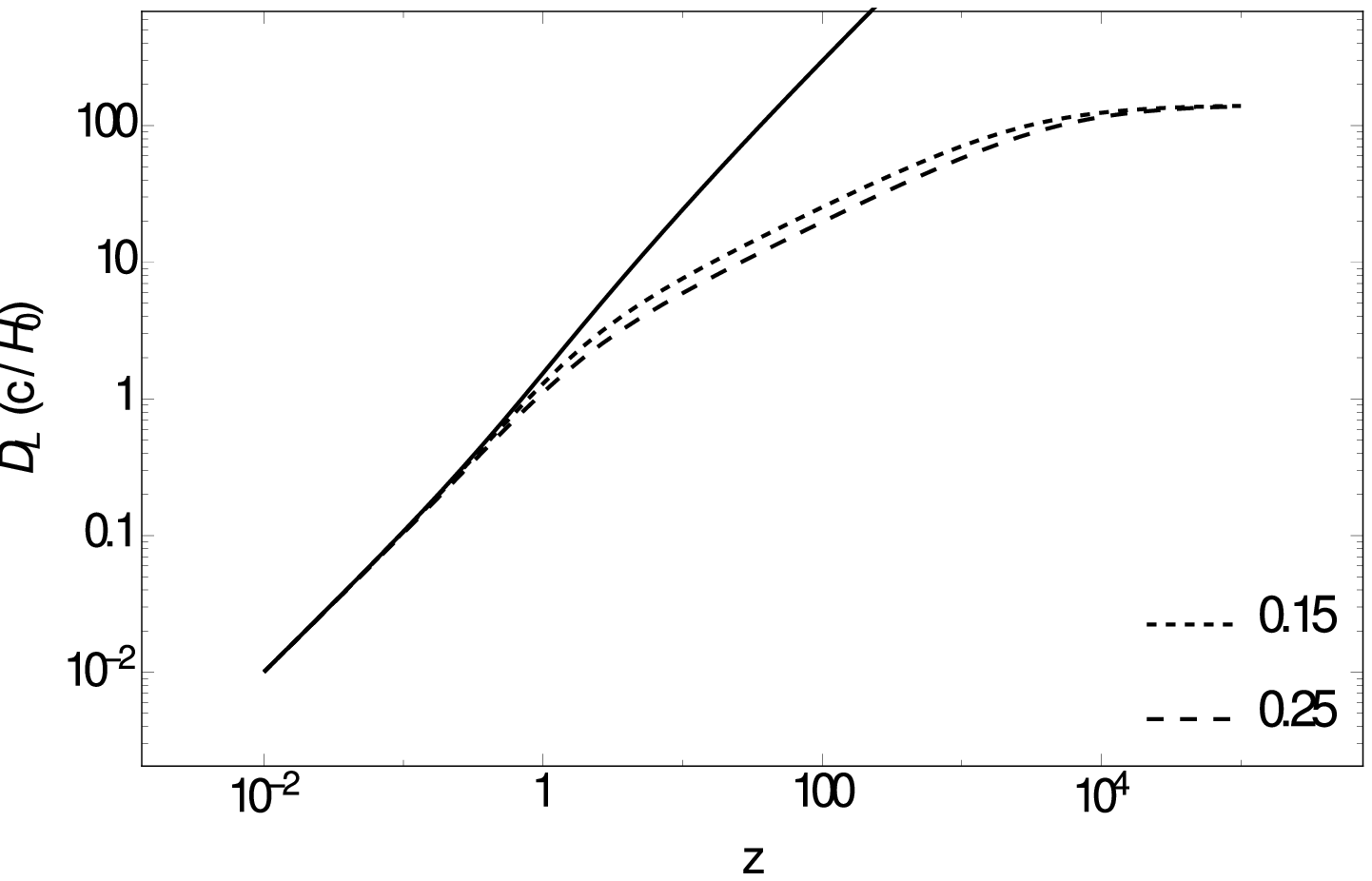}\hfil
\includegraphics[width=7.5cm]{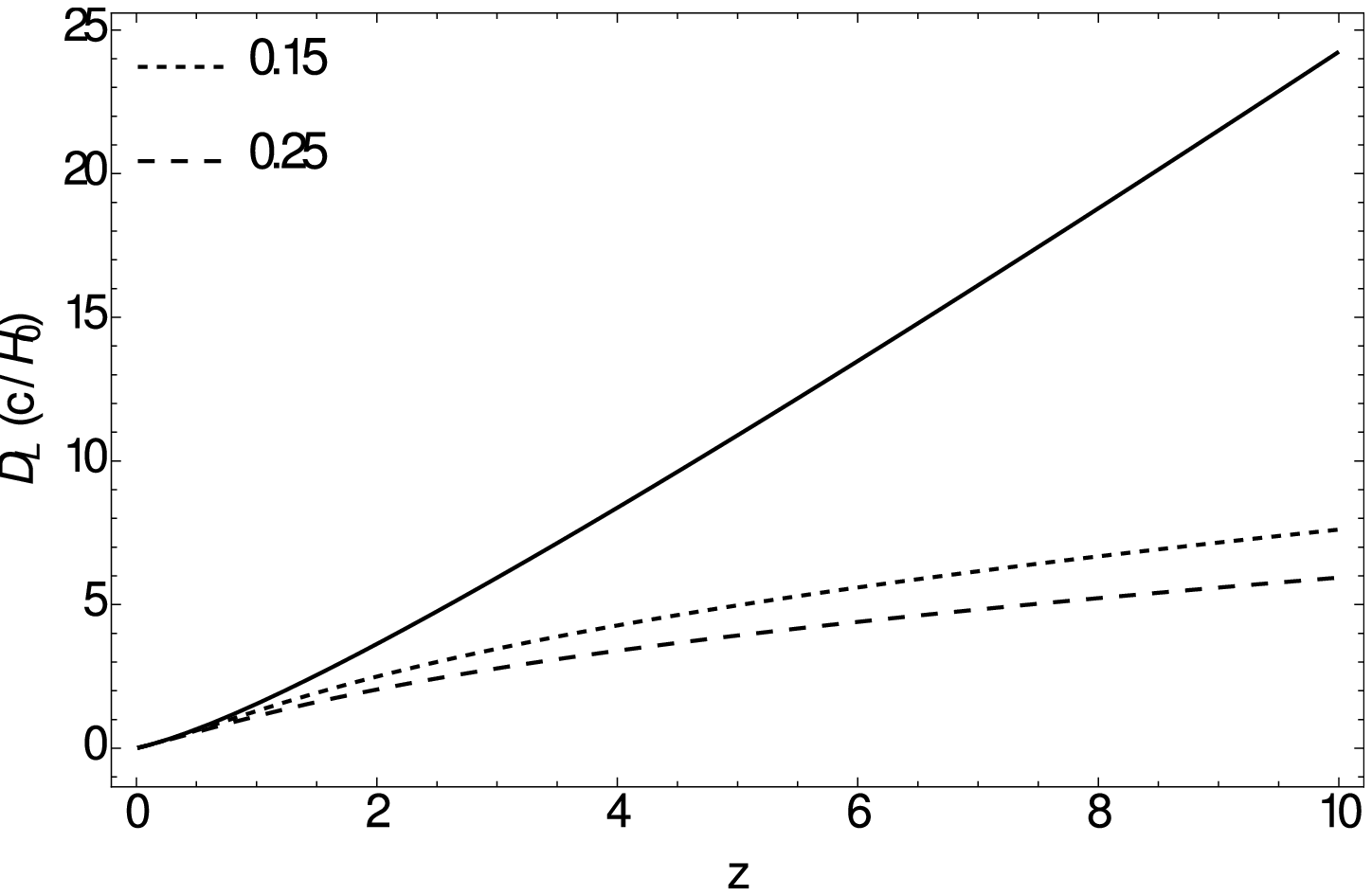}  
\caption{(\textbf{Left}) Luminosity distance $D_L$ for $\Omega_{0,r}=5\times 10^{-5}$ and different values of $\Omega_{0,m}$ in a flat universe: Both vertical and horizontal axes are logarithmic. The~luminosity distance in the present model converges to a fixed value at very high redshifts while that of the $\Lambda~CDM$ model continues to grow. (\textbf{Right}) The behavior of $D_L$  in relatively small redshifts: One could see that, for most values of $\Omega_{0,m}$, the difference between the two models is not negligible even  at $z \approx 2.0$.   One could prove that, by changing $\varepsilon$, the~luminosity distance  remains almost unchanged   in all cosmic~time.  \label{DLplot}}
\end{figure}

\begin{figure}[H] 
\centering
\includegraphics[height=5cm,width=8cm]{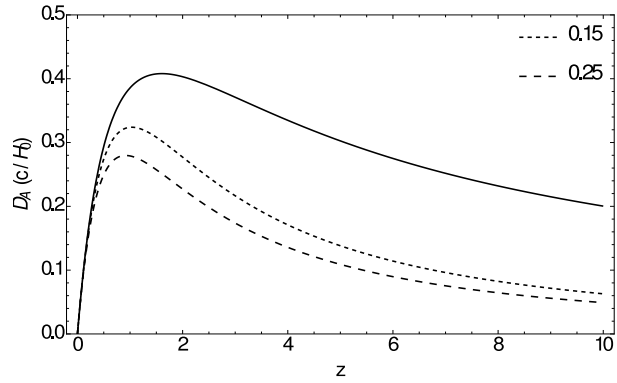}
\caption{The angular diameter distance $D_A$ for $\Omega_{0,r}=5\times 10^{-5}$ and different values of $\Omega_{0,m}$ in a flat universe:   It could be verified that, by changing $\varepsilon$, the~angular diameter distance slightly~changes.  \label{DAplot}}
\end{figure}

To compare our model with observational data, we need to measure distances in the universe. Shenavar and Javidan (in preparation) study the behavior of  luminosity distance
\begin{eqnarray}
D_{L} &=& c (1+z)   \int^{z}_{0} \frac{dz^{\prime}}{H(z^{\prime})}
\end{eqnarray}
while the angular diameter distance
\begin{eqnarray}
D_{A} &=& \frac{c}{1+z}\int^{z}_{0} \frac{dz^{\prime}}{H(z^{\prime})}
\end{eqnarray}
is  studied by Reference~\cite{2019MNRAS.488.3876B}. Both distances are defined for a flat universe. The~plot of luminosity distance $ D_{L} $ and angular diameter distance $ D_{A} $  for two values of $\Omega_{0,m}$ are shown in Figures~\ref{DLplot} and \ref{DAplot}, respectively. The~analogous curves for $\Lambda CDM$ model too are presented by a solid line in both~figures. 
\unskip

As it is shown in the left panel of Figure~\ref{DLplot},  the~luminosity distance of the present model converges to a constant value at $ z\gtrsim  10^{4}$ while the same curve for the $\Lambda CDM$ model varies linearly. Please note that the left panel is logarithmic on both axes. Fortunately however, to~compare the present model with the standard one, it is not needed to observe objects at  very high redshifts. The~reason is that, as~it is shown in the right panel of Figure~\ref{DLplot}, the~luminosity distance in the present model begins to differ significantly from that of the standard model at $z \approx 4$. If~the baryonic content of the universe at the present time  is larger than a minimal value, say $\Omega_{0,m} \gtrsim 0.15$, then the deviation between the two models might be detectable with a precise data analysis at even $z \approx 2.0$. Therefore, it seems that the luminosity distance provides one of the best possibilities to check the viability of the present model. In~conclusion, we should say that, although  the luminosity distance of the present model is rising with increasing $z$, it is always smaller than $ D_{L} $ of $\Lambda CDM$ at the same redshift. In~fact, $ D_{L} $ of $\Lambda CDM$ acts as an upper boundary of $ D_{L} $ predicted by the present~model.

One of the most important consequences of changing the value of $D_{L}$ is that now the absolute visual magnitude $M_V$ of the objects observed at high redshifts has to be modified (compared to the values derived from standard model). Assuming the apparent visual magnitude as $m_V$, the~relation between these two magnitudes is
\begin{eqnarray}  \nonumber
M_V-m_V &=& 5 \log_{10} (\frac{D_{L,0}}{D_{L}})
\end{eqnarray}
in which $D_{L,0}$ is some constant distance, say $ 10~pc $ for near objects or $1~Mpc$ in cosmological measurements. Then, one could simply prove that
\begin{eqnarray}
M_{V,~MOD} -M_{V,~\Lambda CDM} &=& -5 \log_{10}(\frac{D_{L}|_{MOD}}{D_{L}|_{\Lambda CDM}})
\end{eqnarray}
in which $M_{V, MOD}$ is the absolute visual magnitude in our modified model,  $ M_{V, \Lambda CDM}  $ is the absolute visual magnitude in the standard model, $ D_{L}|_{MOD} $ is the luminosity distance in the present model, and finally $D_{L}|_{\Lambda CDM}$ is the  luminosity distance of the standard model. As~it is clear from above equation, the~fact that the luminosity distance in the present model is approximately equal to or smaller than that of the standard model leads to the fact that the absolute visual magnitude of the objects would be approximately equal to or larger than the absolute visual magnitude of the standard model.  This point must  be included especially when one deals with objects at higher~redshifts.

In addition, the~change in $ D_{L} $ could affect the mass estimations in lensing problem if they lie at relatively high redshifts.  In~the lensing sample which was used in Reference~\cite{Shenavar:2016bnk}, the~lenses lie at redshifts $0.2 \leq z  \leq  0.86$, which are relatively low redshifts. Depending to the values of $\Omega_{0,m}$ and $\Omega_{0,c_1}$, the~mass estimation of  lenses at higher redshifts might be~affected. 

The angular diameter distance of the present model too is (almost) always smaller than $ D_{A} $ of $\Lambda CDM$, as shown in Figure~\ref{DAplot}. The~value of $ D_{A} $ for both models decreases  when $z\gtrsim 2$, though~the decline of $ D_{A} $ in the present model occurs more rapidly. The~angular diameter distance provides a good opportunity to confront different cosmological models through the famous ``angular size--redshift'' problem. See Reference~\cite{Gurvits:1998hs} for ane xample. A~comparison  between $\Lambda CDM$ model and some other cosmologies, including the present one, based on angular size--redshift data  is provided by Reference~\cite{2019MNRAS.488.3876B}.

\section{Discussion}
\label{Discussion} 
\vspace{-6pt}
\subsection{Discussion on the Boundary Condition of~GR}
\label{DisNeumann}
Assuming Neumann boundary conditions to solve Equation~(\ref{Laplacian}), which results in the solution of Equation~\eqref{Condition}, is valid as a mathematical possibility. However, this choice of boundary condition would be more motivated if we could enforce it based on some physical backgrounds, i.e.,~action principle, for~example. In~fact, the~debate on  the proper boundary condition of Einstein--Hilbert action dates back to early days of the introduction of general relativity. See Reference~\cite{Realdi:2009zz} for a review on this subject.  Especially, one could mention the discussion between de Sitter and Einstein on this matter, which could be considered as the root of relativistic cosmology. In~the heart of the debate lies the possibility of a static universe, presumed by Einstein before Hubble's discovery of the expansion of cosmos~\cite{Peebles:1994xt}. To~achieve  the relativity of inertia, Einstein proposed a metric of the form $g_{0 i} \equiv \infty$ at spatial infinity while other components are zero. However,  de Sitter criticized this point of view because it leads to the notion of Newtonian ``absolute time'' and some invisible masses. In~the context of GR, this could be considered as  the first mentioning  of ``dark matter''.  In~fact, through trying to solve  the problem of boundary conditions at spatial infinity, de Sitter and Einstein introduced two distinct cosmic models, the~first two models of relativistic cosmology~\cite{Realdi:2009zz}. 

From a pure mathematical point of view too, the~issue of boundary condition seems to be quite controversial in GR. In~fact, since Einstein's field equations contain second-order derivatives of the metric tensor, it could be shown that  the Einstein--Hilbert action  is not well posed, i.e.,~the boundary condition of this action is not compatible with the obtained field  equations~\cite{Chakraborty:2016yna}.  To~cure this problem, however, one could eliminate the surface terms   by adding a boundary term to Einstein--Hilbert action. For~example,   York~\cite{York:1972sj} and Gibbons and Hawking~\cite{Gibbons:1976ue} proposed  the next action with a surface term
\begin{eqnarray}  \label{YGH}
S_{D} &=& \frac{1}{2\kappa}\int_{\mathcal{M}} d^{D}x \sqrt{-g}R + \frac{1}{\kappa} \int_{\partial \mathcal{M}} d^{D-1}y \sqrt{|h|} \gamma K
\end{eqnarray}
which is invariant under diffeomorphism. In~this action, known as YGH action,  $R$ is the Ricci scalar, $h$ is the induced metric on the boundary  $ \partial \mathcal{M} $ with the coordinates $ y^{i} $,   $ K$ is the extrinsic curvature of the boundary, and  $g$ is the determinant of the metric tensor. Also, we have  $\gamma = +1$ for time-like and $\gamma = -1$ for space-like boundaries.  This action presumes Dirichlet boundary condition by killing all the normal derivatives of the metric tensor on the surface. See Reference~\cite{Chakraborty:2016yna,Krishnan:2016mcj} for a review. It is also worth mentioning that, according to Reference~\cite{Charap:1982kn},   there could be infinitely many boundary terms to make the above action well posed. Therefore, it is not uniquely~determined.

On the other hand, by assuming Neumann B.C.,   Chakraborty~\cite{Chakraborty:2016yna} and Krishnan and Raju~\cite{Krishnan:2016mcj}  have shown that the action takes the following form:
\begin{eqnarray}    \label{NeumannBC}
S_{N}= \frac{1}{2\kappa}\int_{\mathcal{M}} d^{D}x \sqrt{-g}R
+ \frac{D-4}{2 \kappa} \int_{\partial \mathcal{M}} d^{D-1}y \sqrt{|h|} \gamma K
\end{eqnarray} 
with the unique property that the surface term vanishes in $D=4 $ dimension. Krishnan and Raju~\cite{Krishnan:2016mcj}  suggest that this property of $ S_{N}$ could be interpreted as follows: \textit{``Standard Einstein--Hilbert gravity in four dimensions, without~boundary terms, has an interpretation as a Neumann problem''.}  In addition, Equation~(\ref{NeumannBC}) is interesting from the point of view of  gravitational theories in higher dimensions since it singles out the $D=4 $ dimension. In~other words,  our four dimensional spacetime is not merely one of the possibilities; i.e.,~$D=4 $ is the  dimension that one has the elegant Einstein--Hilbert action (without any surface term) if one imposes the Neumann boundary condition. Krishnan and Raju~\cite{Krishnan:2016mcj} have also shown that, in a three-dimensional space, the Neumann action $S_N$ becomes  the  Chern--Simons action, which is another interesting feature of this~action. 

In addition, Maldacena~\cite{Maldacena:2011mk} has proved that, by using Neumann B.C., one can derive the semiclassical (or tree level) wavefunction of the universe in 4-D asymptotically de-Sitter or Euclidean anti-de Sitter spacetimes. Since, conformal gravity has many solutions,   Neumann B.C. seems to  select the Einstein solution out of all possible choices. A~more elaborate derivation of the equivalence between Einstein theory and conformal gravity  by assuming Neumann B.C. is also provided by Reference~\cite{Anastasiou:2016jix}. 

Chakraborty~\cite{Chakraborty:2016yna} shows that, to impose the Neumann B.C.,  one has to fix the momentum conjugate on the boundary:
\begin{eqnarray}   \label{Mconjugate}
 \Pi^{ij} &=& \sqrt{h} (K^{ij} - K h^{ij}) 
\end{eqnarray}

In~what follows in this subsection, we will impose the Neumann B.C. on cosmic perturbation  equations and then derive the modifications due to this new boundary condition on cosmic equations. To~do so,   assume a  metric of the form $$ds^{2}=-(1+2\phi(t,\vec{x}))dt^{2}+(R^{2}(t)-2\psi(t, \vec{x}))d\vec{x}^{2}.$$ Please note that, unlike metric (Equation \eqref{metric}), the~scale factor $R(t)$ is not included in the potential $\psi$. See Reference~\cite{jacobs1992obtaining} for a discussion on the difference between the two metrics. For~this metric, the momentum conjugate could be derived as
\begin{eqnarray} \nonumber
 \Pi^{ij} &=& -2\dot{R} + 2\frac{\dot{R}}{R^{2}}(R^{2}\phi - \psi +\frac{\dot{\psi}}{H})
\end{eqnarray}
in which the first term on the rhs is   of zeroth order  while the rest represent the  perturbed terms. Neumann boundary condition must be imposed on physical degrees of freedom, i.e.,~the potentials $\phi$ and $\psi$. Also, presuming that the time derivative of the 3-curvature perturbation $\psi$ is negligible, i.e.,~$\dot{\psi} / H  \approx 0$, we would have
\begin{eqnarray} \nonumber
~^{l}\Pi^{ij} &=&   2\frac{\dot{R}}{R^{2}}(R^{2}\phi - \psi)
\end{eqnarray}
for the local conjugate momentum. Remembering that $\psi = R^{2}\Psi$ and assuming $~^{l}\Pi^{ij} = 2\dot{R}c_{1}$ at any time, one would derive $\Phi - \Psi = c_{1}$ as suggested in Equation~(\ref{Condition}) (neglecting the anisotropic stress). We note that  the Dirichlet B.C. could be reproduced simply by replacing $c_1 = 0$ in Equation~(\ref{Condition}). 

In deriving the above condition from the Neumann boundary of Equation \eqref{Mconjugate}, we assumed the negligibility of the time variation of $\psi$. The~validity of this presumption  at different scales needs to be studied more carefully and we do not consider it here. Also, another critical assumption is that we only imposed Neumann B.C. on potentials (perturbations) and not the background term, i.e.,~$-2\dot{R}$, which is related to the geometry of the space. This is justified    because we prefer to maintain the underlying geometry intact; otherwise, the~whole method of perturbation theory which we  used here needs to be revised. In~conclusion, the~above method provides a possible interpretation for the Neumann boundary condition from an action-principle point of view. The~other possible interpretation which was mentioned above and is based on the equivalence between Einstein and conformal gravity would not be discussed here. We should point out that  boundary conditions, such as mixed boundary condition, are also possible. See, for~instance,   Peebles~et~al.~\cite{Peebles:2011kv} and Peebles~\cite{peebles2017dynamics}   impose mixed boundary condition to study the dynamics of the local~group.

It is interesting that the  possibility of a Neumann B.C. to solve GR field equations has been rarely discussed before. One of the reasons for this shortcoming is that proving the well posedness of GR field equations under a particular boundary condition has been found to be a very difficult task. The~well posedness, i.e.,~the  existence and uniqueness of the solution in a small neighborhood, of~GR under Cauchy B.C. has  been proved by Choquet-Bruhat for the first time and   after a long debate. A~review on this proof which is based on harmonic coordinates could be found in Reference~\cite{1980grg1.conf...99C}. Then,   Choquet-Bruhat and Geroch~\cite{ChoquetBruhat:1969cb} proved the theorem of the global existence and uniqueness of GR.  See also Reference~\cite{gourgoulhon20123+} for a good introduction to this matter. As~the authors of Reference~\cite{Arnowitt:1962hi} have discussed too, the~Cauchy problem starts with assuming $(h_{ij} , \Pi^{ij})$ as a complete set of Cauchy data (initial data). It is  not yet clear that, if we change  the boundary condition, the~proof of ``the  existence and uniqueness of solutions'' would remain intact or not.  However,  a~systematic treatment is needed to prove (or disprove)  the well posedness of GR with a Neumann B.C., which is beyond the scopes of the present~work.

Another reason for the usual discarding of Neumann B.C. is that  the meaning of this boundary condition in a four-dimensional spacetime is not necessarily evident. For~example, in~the field of numerical relativity, there have been some attempts to impose Neumann B.C. to  prevent the violation of  constraints~\cite{Kidder:2004rw} or to investigate the numerical stability of Cauchy evolution~\cite{Szilagyi:1999nu} under the new boundary condition. To~impose Neumann B.C., Kidder~et~al.~\cite{Kidder:2004rw} placed restrictions on the normal derivatives of some characteristic fields while \citep{Szilagyi:1999nu} determined $\partial_{z} \Phi$, $z $  being a specific direction in their simulations, at~$z=0$.  Thus, there had been different conceptions in dealing with Neumann boundary condition. However, now, by the method that References~\cite{Chakraborty:2016yna,Krishnan:2016mcj} present, there is a well-motivated mathematical definition of imposing Neumann B.C. as determining the value of $\Pi^{ab}$ on the boundary. Therefore, there is at last a clear definition of ``imposing Neumann B.C.'' which could guide us through the complexity of GR~formalism.

As it is discussed before~\cite{Shenavar:2016bnk,Shenavar:2016xcp,shenavar2018local}, assuming Neumann B.C. could lead to three different possibilities. First, by~comparing our results with  observations, we might realize that  the derived value  of $c_1$ is very tiny. Of~course, in~this case, we will conclude that  Neumann B.C. is not reliable, i.e.,~Dirichlet B.C. triumphs. Second, one might derive   different values for $c_{1}$ from various observations, i.e.,~observations based on  lensing, rotation curves of galaxies, CMB,  etc. report contradictory results for $c_1$. Then,   we would   conclude that our fundamental assumption, i.e.,~Neumann B.C., is not self-consistent and thus excluded. The~last possibility is that the value of the Neumann constants $c_1$ which is found from the results of different observations is nonzero and compatible at various scales.  In~this case,  the significance of the  Neumann B.C. should not be~underestimated.

\subsection{Discussion on Mach's Principle as Boundary~Condition} 
\label{DisMach}

Using the metric in Equation \eqref{metric}, one could derive the field and conservation equations. These are reported in Section~\ref{Scalar}. Then, one could immediately see from  Equation~(\ref{Laplacian}) that the scalar fields  are not physically independent. Here,  one usually concludes that, if the anisotropic stress is negligible, then the two scalars are equal. In~other words, it is presumed that $\partial_{i}\partial_{j}(\Phi - \Psi)=0$ leads automatically to $\Phi = \Psi$. However, simple investigation of this partial differential equation ( PDE ) reveals  that  a presumption is implicitly made about the boundary condition. Namely, it is assumed that $\Phi - \Psi$ vanishes at the boundary, i.e.,~Dirichlet B.C on the boundary is presumed. On~the other hand, by~imposing a new boundary condition, e.g.,~Neumann B.C., the~solution to the field equations could substantially change. In~conclusion,  unless~particular boundary conditions are imposed,  the~gauge conditions do not uniquely fix the potentials. See Reference~\cite{1973ASSL...38..127B}, page 144, for~a~discussion. 

Wheeler~\cite{wheeler1964mach}, too, argues that Einstein field equations, as~a system of PDEs, do not suffice to define a solution. These equations must also be supplemented by well-defined boundary conditions which he recognizes as Mach's principle. Other formulations of Mach's principle  are also possible which might be considered as mathematically vague. See References~\cite{wheeler1964mach,1981RPPh...44.1151R,Bondi:1996md,Barbour:1995iu} for more details.  According to Wheeler, the~role of Mach's principle, i.e.,~boundary condition, is to select the solutions of Einstein's equations based on the physics of the system.  This method of selecting solutions is familiar in solving Poisson equations (electrostatics or Newtonian gravitation) which  generally goes without a specific name. In~fact, as~the boundary condition,  the~Poisson equation  is typically supplemented by  the statement that the potential decreases as $1/r$. Then, one could see that the distribution of   matter (or electric charge) uniquely determines the general form of the potential. If, on~the other hand, another boundary condition is imposed, the~behaviour of the potential could be entirely different~\cite{Jackson:1998nia}.  In~the case of general relativity too, the~boundary condition is used to select the physically allowable solutions. For~example, when one seeks a stellar interior solution, e.g.,~Tolman--Oppenheimer--Volkoff solution, one might insist that, as the boundary condition, the~pressure goes to zero and the solution matches the exterior solution, i.e.,~the Schwarzschild metric,  at~the boundary of the star. In~the present work, we implement Neumann B.C. on cosmological perturbation equations to pursue a modified model which is more in accord with the observable universe and, thus,  free of dark matter and dark energy. As~discussed in Section~\ref{DisAverage}, by~using Neumann B.C., the homogeneity and isotropy of the background universe is still~preserved.   

However, we do not consider the details here,  we only mention that, by imposing Neumann B.C., the~super-horizon solutions would also be modified.  Moreover,  if~the parameter $c_{1}$ is time dependent, then the results  could be substantially different from that of References~\cite{Weinberg:2003sw,Weinberg:2008zzc} depending on the rate of change in $c_{1}$. In~this case, the~Weinberg theorem should be modified. See Reference~\cite{Akhshik:2015rwa} for other models which violate the Weinberg~theorem.

\subsection{Discussion on the Averaging~Procedure}
\label{DisAverage}
We  should note that the averaging procedure defined in Section~\ref{Modified} is carried out on a hypersurface of  constant time $t$. In~other words, it is presumed that  the averaging integrals are performed  instantaneously on the whole of the universe. See   Amendola and Tsujikawa~\cite{Amendola:2015ksp}, page 294, for~a discussion on this matter. Of~course, one could assume that, in the absence of walls and other huge inhomogeneities, the~structures in the cosmos form quite similarly everywhere; thus, the~presumption of an instantaneous average is somehow justified. However, as~an alternative, the~averaging procedure could be done by calculating integrals over the ``light-cone'' which is a more subtle method.  See Reference~\cite{Gasperini:2011us} for a complete description of this procedure. The~integration in this method is carried out on a section of spacetime which is ``causally'' connected with the observer (us). Basically, this type of averaging could lead to quite different consequences (at least for some redshifts and some particular mass distributions) compared to the procedure that we used in Section~\ref{Modified}; however,   a~survey on the difference between the two approaches is beyond the scopes of the present~work.

The problem of  averaging  perturbations at large distances needs to be examined more carefully. In~this regard, one might argue that either $i)$ the perturbed metric, when averaged out on large scales,  must result  in an isotropic and homogeneous background or $ii)$ the spacial average of the perturbations needs to vanish on large scales. This latter condition presumes that the perturbed metric provides no contribution to the background metric of the universe at large scale. From an~observational point of view, the~condition $i$ must always be satisfied. We will call $i$ and $ii$ the weak and strong smoothness conditions,   respectively, and it is the intention of this part to show that a constant $c_{1}$ satisfies the strong smoothness condition while a time-dependent Neumann parameter $c_{1}(t)$ is compatible with weak smoothness~condition.  

To see this, one could readily check that, by using the definition of $ \langle A \rangle $ in Section~\ref{Modified}, the  averaged  metric could be found as
\begin{eqnarray}  \label{averaged}
\langle ds^{2} \rangle = -dt^{2} + R^{2}(t) \left(1+2c_{1}(t) \right)  \delta_{i j}dx^{i}dx^{j}
\end{eqnarray} 
in which we have recovered the time dependence    of $ c_{1} $ for the sake of completeness of the discussion. The~elements of this  averaged metric only depend on time while the  governing cosmic equations, which are reported in Appendix \ref{timedepend}, are modified now. Therefore, a~time-dependent Neumann parameter changes the form of the Friedmann equations by providing a share into the background equations.   However,  a~suitable rescaling of the scale factor as $R^{\prime}(t) = R(t) \left(1+2c_{1}(t) \right)^{1/2}$ shows that the geometry of the metric \eqref{averaged} is equivalent to FLRW geometry at any time. In~general, we conclude that a time-dependent Neumann parameter satisfies condition $i$. Moreover, regarding the pure time dependence of the averaged metric in Equation \eqref{averaged} and the assumption that the anisotropic scalar is negligible, i.e.,~$\pi^{S}=0$, one concludes that the background universe is essentially a perfect fluid. On~the other hand, if~$ c_{1} $ is a constant, then the form of the Friedmann equations remains intact as reported in Equation \eqref{Friedmann1}. In~this case, which is the main focus of this work, the~strong smoothness condition is satisfied. Of~course,  the~result in this latter case is again a perfect~fluid. 

The appearance of  the new term, i.e.,~$c_{1}(t)$, in~the metric only changes large-scale systems. The~reason is that we have only changed the boundary condition of the gravitational field. In~particular, in~the same way that the expansion of the universe does not change the laws governing the nongravitational forces, i.e.,~electromagnetic, etc., these laws remain intact under appearance of $c_{1}(t)$ in Equation \eqref{averaged}. For~example, a~typical scattering between subatomic particles happens in a very short time compared to the rate which $R(t)$ or $c_{1}(t)$ change significantly. Therefore, in~calculating a cross section, it is safe to neglect the time evolution of the~universe.

\section{Conclusions}
\label{Conclusion}

In this study, we built   a model based on the idea that local scale physics might be affected by the global expansion of the universe through a term which is related to the de Sitter scale of acceleration $cH_{0}$. To~do so,  we used Neumann B.C. instead of the usual Dirichlet B.C to solve  Einstein perturbation equations.  This new boundary condition is mathematically well motivated, however, its mathematical well posedness is yet to be~surveyed. 

The outcome of the model is  reminiscent of Mach's principle. The~Mach's principle argues that  the condition of the  ``distant objects of the cosmos'' somehow enters into the laws of local mechanics. There are various interpretations and formulations of this principle which are thoroughly presented in Reference~\cite{Barbour:1995iu}. See also References~\cite{Bondi:1996md} or~\cite{Misner:1974qy}, pages 543--549. However, the~present model is  based on the idea  that  ``the expansion of the universe'' is essential and that the new term  enters the equation of motion through presuming Neumann B.C. in this expanding universe   (Wheeler's formulation of Mach's principle~\cite{wheeler1964mach}). By~imposing a boundary condition, according to  Wheeler, one essentially  ``selects'' a specific solution. The~Neumann boundary condition which is used here presumes a nonnegligible effect of the distant objects on local dynamics.  Another possible path to the present model might be through evaluating the surface term of the generally covariant integral formulation of GR~\cite{Sciama:1970yk} by imposing a Neumann boundary~condition. 

In the standard model of cosmology, the~accelerating expansion of the universe is provided by assuming a constant density of energy due to  $\Lambda $. This term provides a repulsive force at large scales  in $\Lambda CDM$ while  presuming that a dark halo provides a more attractive force (of course through Newtonian force of gravity)  at galactic scales. On~the other hand,  the~new term in the present theory, i.e.,~the term proportional to $c_1cH(t)$ in the Poisson equation (Equation~(\ref{Second})), provides more attractive forces at galactic scales while still produces a negative deceleration parameter at later times, i.e.,~see $q<0$ in Figure~\ref{q2plot}. Also, the~total equation of state parameter $w_{t}$ tends to $w_{t} < -1$ at later times, as was shown in Figure~\ref{w2plot}. The~key to understanding this seemingly  odd  behavior, i.e.,~making attraction and repulsion at different scales, is that, unlike the $\Lambda CDM$ model with its static $\Lambda$, the~new term in our model changes with time because it is proportional to Hubble parameter. As~is shown by References~\cite{Cai:2010uf,Cai:2012fq} and here, the~existence of a term proportional  to $H(t)$ in Friedmann equations could be considered as a model of dark~energy. 

Unified models of dark matter and dark energy, which rely on ``dark fields'' instead of dark matter or $\Lambda$ term, have been proposed before.  See~\cite{Amendola:2015ksp}, chapter 8 and references therein, for~a brief review on these models.  Among~these models, one could mention generalized Chaplygin gas model~\cite{Bento:2002ps}, k-essence model~\cite{Scherrer:2004au}, and models based on Bose--Einstein condensation~\cite{Fukuyama:2007sx}. However, these models might encounter problems in dealing with observations at early or late times. For~example, the~sound speed in a generalized Chaplygin gas model is small at early cosmic times while it shows a growth  at later epochs. Compared with   the behavior of $c^{2}_{s}$ in the $MOD$ model and $\Lambda CDM$ as reported in Figure~\ref{cs2plot}, this evolution of sound speed in generalized Chaplygin gas results in incompatibilities with observations of large-scale structure~\cite{Amendola:2015ksp}. 

However, the~observational success of the standard  model seems to mostly rely on the fact that $\Lambda CDM$ needs only a few parameters to fit the cosmological observations~\cite{Jain:2010ka}. Beside the parameters which are usually derived by fitting cosmological data to $\Lambda CDM$ predictions; there are some other parameters presumed at galactic scales. For~instance, a  NFW  profile (introduced by Navarro,  Frenk,   and White~\cite{Navarro:1995iw})   relies on two free parameters, i.e.,~the scale length of the halo and its central density. Furthermore,  one also needs more parameters to build a successful particle theory of dark matter, e.g.,~ supersymmetry ( SUSY ), which we will not discuss here.  In~this respect, the~present model could be considered quite economical because at galactic scales it relies only on $c_1$. See  Equation~(\ref{Second}).  At~cosmic scales too, the~Neumann model is dependent to  $c_1$, $\varepsilon$ and $\Omega_{c_1}$, which seems quite parsimonious. Also, it attributes the puzzling phenomena of dark matter and dark energy  to the boundary conditions of the field equations (the state of matter distribution at the boundary) or to the presence of new substances or other dark~fields.

Admittedly, the~question of boundary/initial conditions in cosmology has not been thoroughly studied so far. The~same is true about a possible link between  local and global dynamics. However, some exceptions could be found in the works of the founders of the standard model. For~example,     Dicke and Peebles~\cite{dicke1964evolution} criticize  the proposal  of Pachner~\cite{Pachner:1963zz,pachner1965problem}, who suggested the existence of a connection  between  local and global dynamics based on Mach's principle. In~fact, Dicke and Peebles  argue that the apparent connection proposed by Pachner is only formal, i.e.,~such effects would be unobservable. In~addition, Peebles~\cite{Peebles:1994xt} (part III,  pages 361--363) argues that the notion of initial condition is not necessary for a cosmological model  unless probably for the fine-tuning problem of very early universe.   However, these works do not include a systematic imposition of a new boundary condition and its local effects and they mostly argue, on~general grounds, that such modifications would be unnecessary or~unobservable.

The degree of reliability of Neumann model needs to be further studied in galactic, cluster, and cosmic scales. At~galactic scales, the~local and global stability of the systems have to be surveyed. Specifically, by~developing the work of Reference~\cite{shenavar2018local}, one could derive the local stability of a gas + stellar system and compare the results with star formation rate. Moreover, the~scaling rules of the $MOD$ model at galactic scales show interesting properties which will be reported in the future. Another challenge is the problem of dwarf galaxies which is well established within the MOND paradigm  but seems quite problematic in $\Lambda CDM$ \cite{Kroupa:2012qj}. 

However, the~main issue at cosmic scales remains to be the CMB analysis of MOD. Regarding the analysis which we presented in Section~\ref{Scalar}, one could see that the $MOD$ model works based on a modified Poisson equation which is indeed of fourth order. See Equation~(3) of Reference~\cite{shenavar2018local} for the derivation.  The~homogeneous form of this equation is usually discussed in the theory of linear elasticity and has been named \textit{ biharmonic equation}. The~solutions to  biharmonic equation include the solutions of Laplace equation. By~implementing the solutions of the modified Poisson equation into perturbation equations, one can derive  CMB spectrum. This is a delicate issue which would be reported in a following work. However, as~the authors of Reference~\cite{DiValentino:2015bja} have discussed, there is already a powerful technique to systematically search for a deviation between the two scalar potentials at cosmic scales, i.e.,~the so-called anisotropic stress function $\eta = \Psi /\Phi$, which hints to possible need for a modification in gravity (at 95$\%$  confidence~level). 

\vspace{6pt} 

\authorcontributions{Conceptualization, H.S. and K.J.; formal analysis, H.S.; supervision, K.J.; writing---original draft, H.S.; writing---review and editing, K.J.}

\funding{Hossein Shenavar is supported by Ferdowsi University of Mashhad under grant No. 3/41795 (13/07/1395).}

\acknowledgments{ H.S. is   grateful to Pavel Kroupa for his support during a visit at Bonn University and for also suggesting some  tests to confront the present model with. H.S. is also thankful to Vasanth B. Subramani, Hossein Afsharnia, and Mehdi Ebrahimi for various discussions.  The~authors also appreciate help from  David Merritt. It is a pleasure to thank Mahmood Roshan and Neda Ghafourian, who kindly reviewed the draft and provided insightful comments. Many thanks to Yoshiaki Sofue for providing his galactic data freely.  We also acknowledge the anonymous reviewer whose comments helped us to  clarify the  presentation of the work. The~appendix on  trajectory of massive particles is added due to reviewer's suggestion. This research has made use of NASA Astrophysics Data~System.}

\conflictsofinterest{The authors declare no conflict of~interest. } 

\abbreviations{The following abbreviations are used in this manuscript:\\

\noindent 
\begin{tabular}{@{}ll}
GR & General Relativity\\
PDE & Partial Differential Equation\\
CMB & Cosmic Microwave Background\\
($\Lambda$)CDM & $\Lambda + $ Cold Dark Matter\\
B.C. & Boundary Condition\\
%
MOND & MOdified Newtonian Dynamics \\
MOD & Modified Dynamics\\
NFW & Navarro,  Frenk,   and White halo profile\\
SUSY & Supersymmetry\\
EPS & Ehlers,  Pirani, and  Schild \\
QCD & Quantum Chromodynamics\\
FLRW & Friedmann--Lemaitre--Robertson--Walker \\
YGH & York, Gibbons, and Hawking \\
WIMP & Weakly Interacting Massive Particles \\
MOG & MOdified Gravity~\cite{Moffat:2005si} \\
SNIa &  Supernovae (type) Ia \\
\end{tabular}}
All abbreviations have also been defined within the~text.

\appendixtitles{yes} 

\appendix

\numberwithin{equation}{section}
\numberwithin{table}{section}
\numberwithin{figure}{section}

\section{Time-Dependent Neumann~Parameter}
\label{timedepend}
Similar to the method that was used in Section~\ref{Scalar}, one can prove that, by assuming a time-dependent Neumann parameter $c_{1}(t)$, the next two pure time-dependent equations could be derived from Einstein field equations:
\begin{eqnarray}  \label{Einsteinc1t}
3(\frac{\dot{R}}{R})^{2}+6\dot{c}_{1}\frac{\dot{R}}{R} = 8 \pi G \rho \\ \nonumber
\frac{\dot{R}^{2} +2R\ddot{R}}{R^{2}}+6\dot{c}_{1}\frac{\dot{R}}{R} + 2\ddot{c}_{1} = - 8 \pi G P
\end{eqnarray} 
while, by using energy-momentum conservation equation, it is possible to achieve
\begin{eqnarray}  \label{consc1t}
\dot{\rho}+3(\rho +P)(\frac{\dot{R}}{R} +\dot{c}_{1})=0.
\end{eqnarray} 
As one could see immediately, these three equations are independent, i.e.,~ the third one  could not be derived from the first two equation unless~one assumes that $\dot{c}_{1} =0$.  This is the key reason for assuming a constant $c_{1}$ in Section~\ref{Scalar}. 

It is worthy to note that the conservation equation could be derived  ``order by order'' from   GR field equations. However, the~key point here is that Equation~(\ref{consc1t}) is independent from Equations~(\ref{Einsteinc1t}) because we have mixed zeroth- and first-order terms in these equations to obtain all pure time-dependent~terms.

On the other hand, if~one drops the assumption that \textit{ there is  necessarily two independent equations}, then Equations~(\ref{Einsteinc1t}) and \eqref{consc1t} are still mathematically self-consistent. The~reason is that one could derive the evolution of radiation (with $P_{r}=1/3$) and baryonic matter (with $P_{m}=0$) from Equation~(\ref{consc1t}) as follows:
\begin{eqnarray}
\rho_{r}(t) = \rho_{0,r}(\frac{R_{0}}{R})^{4}e^{-4c_{1}(t)} \\  \nonumber
\rho_{m}(t) = \rho_{0,m}(\frac{R_{0}}{R})^{3}e^{-3c_{1}(t)}
\end{eqnarray}
One can then put the results in Equation~(\ref{Einsteinc1t}) to obtain a system of two independent equations of two unknown functions $R(t)$ and $c_{1}(t)$. This system could be solved numerically by presuming suitable initial conditions, i.e.,~determining $a(0)$, $\dot{a}(0)$, $c_{1}(0)$, and $\dot{c}_{1}(0)$. The~results are very similar to the present model at small $z$ as one expects. However, the~dynamical system of such a model is quite complicated and we will not consider it~here.

\section{Discussion on  the Trajectory of Massive Particles, GR Measurement Process, and the EPS~Theorem}
\label{DisMeasurement}
Light signals, alongside  free falling particles, form the basis of measurement process in general theory of relativity~\cite{marzke1964gravitation,ehlers2012republication,ohaniangravitation}. Therefore,  if~the trajectory of the light is modified by Equation~(\ref{trajectory}), the~measured  values for physical quantities  would change too.  It is the aim of this appendix to show how the change in the trajectory of photons would lead to  a modification in the trajectory of massive particles. However, before~we analyse this issue, we need to review the measurement process of spacetime intervals, which lies in the center of a constructive-axiomatic approach to GR~\cite{ehlers2012republication,ehlers1973survey}. 

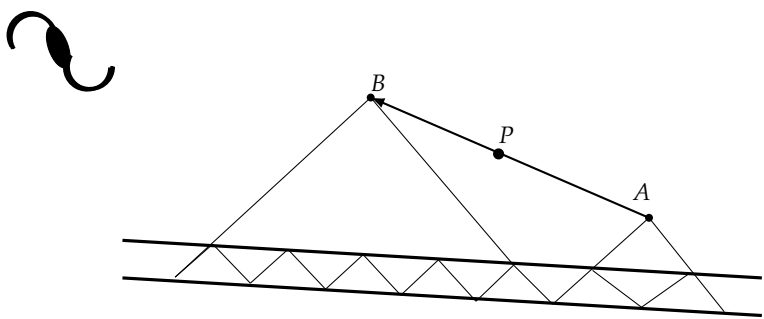
\begin{figure}[H]
\begin{center}
\begin{tikzpicture}[font=\LARGE]
\draw[very thick] (0.5,0) -- (-8,0.5);
\draw[very thick] (0.5,0.5) -- (-8,1);
\draw (0,0.05)-- (-1.,1.3); 
\draw (-1,1.3)-- (-2.28,0.15);
\draw (-0.45,0.57)-- (-1.1,0.11);
\draw  (-1.1,0.11) -- (-1.75,0.6);
\draw (-2.28,0.15) -- (-2.8,0.67);
\draw (-2.8,0.67) -- (-3.3,0.19);
\draw (-3.3,0.19) -- (-3.8,0.74);
\draw (-3.8,0.74) -- (-4.3,0.27);
\draw (-4.3,0.27) -- (-4.8,0.81);
\draw (-4.8,0.81) -- (-5.3,0.35);
\draw (-5.3,0.35) -- (-5.8,0.88);
\draw (-5.8,0.88) -- (-6.3,0.43);
\draw (-6.3,0.43) -- (-6.8,0.95);
\draw (-6.8,0.95) -- (-7.3,0.51);
\draw (-2.8,0.67) -- (-4.7,2.9);
\draw (-4.7,2.9) -- (-7.3,0.51);
\draw [-latex, thick] (-1. ,1.3) -- (-4.7,2.9);

\fill (-1. ,1.3)  circle[radius=1.5pt];
\node at (-1.1,1.65) {\small $A$};

\fill (-4.7,2.9)  circle[radius=1.5pt];
\node at (-4.6,3.1) {\small $B$};

\fill (-3.0,2.15)  circle[radius=2.1pt];
\node at (-2.9,2.4) {\small $P$};

\draw[black,fill=black,rotate around={110:(-3.99,-.99)},] (1.96,2.02) ellipse (0.29cm and 0.13cm);
\draw [very thick] (-8.92,3.73) arc (0:210:0.3cm);
\draw [very thick] (-8.88,3.7) arc (0:210:0.3cm);
\draw [very thick] (-8.9,3.7) arc (0:210:0.3cm);
\draw [very thick] (-8.73,3.45) arc (-210:0:0.3cm);
\draw [very thick] (-8.7,3.47) arc (-210:0:0.3cm);
\draw [very thick] (-8.68,3.46) arc (-210:0:0.3cm);
\end{tikzpicture}
\end{center}
\caption{Trajectory of a particle $P$  in the field of a massive object  is   measured by using geometrodynamic clocks. The~light rays of the clock are shown by solid lines. The~clock itself is  affected by the gravitational field of the central mass and the  boundary condition. Accordingly, the~trajectory of the massive test particle too will be modified because the physical quantities such as velocity, acceleration, etc., are measured by geometrodynamic~clocks. }
  \label{fig:measurement}
\end{figure}

The accepted procedure of spacetime measurement in GR is known as the method of geometrodynamic clocks \citep{ohaniangravitation}. This method had been introduced by References~\cite{marzke1964gravitation,kundt1962determination}, and~later modified slightly by Reference~\cite{desloge1989simple}  to eradicate the need for atomic standards which tend to change when placed in gravitational fields. See Reference~\cite{ohaniangravitation}, chapter 5, for~a thorough review on measuring spacetime intervals in curved spaces by using  geometrodynamic clocks. These clocks are made of two mirrors with a constant separation, with~light rays bouncing back and forth  between them while both mirrors are in free fall.  See Figure~\ref{fig:measurement} for a sketch of this clock.  In~a gravitational field, all particles fall at the same rate; thus, geometrodynamic clocks are in fact matter independent. Although~the tidal forces might destroy the parallelism of the mirrors,  one might continually re-parallel them by using extra~clocks.

The key advantage of the method of geometrodynamic clocks lies in using free falling particles and light rays. This is crucial because as Ehlers,  Pirani, and  Schild   have proved~\cite{ehlers2012republication}, the~light rays build a conformal structure  while free falling particles determine a projective structure. This is known as the EPS theorem, which is inspired by previous works of Hermann Weyl on the foundations of differential geometry and their relation to physics \citep{2012GReGr..44.1581T}. The~EPS theorem could be used to derive many interesting results. Most notably, by~applying geometrodynamic clocks, one could  obtain  affine and metric structures of spacetime manifold. To~prove this, the~authors of Reference~\cite{ehlers2012republication} have assumed some well-motivated postulates from which one could define exact operations to measure length and time intervals.   See Reference~\cite{ehlers2012republication} for the exact definitions of conformal, projective, affine, and metric structures. In~addition, Ehlers~\cite{ehlers1973survey}  provides details of a constructive-axiomatic approach to GR  which employs EPS theorem as the cornerstone of the theory.  Einstein field equations are derived in this axiomatic method from a few well-motivated set of postulates. Thus, the~EPS theorem is essential in axiomatic description of~GR.

The notion of geometrodynamic clocks is also important from a physical point of view because it can be proved that the particles in free fall move along the geodesics of a metric $g_{\mu \nu}$ which is measured by geometrodynamic clocks. See Reference~\cite{ohaniangravitation}, page 203, for~a proof of this theorem. Therefore, if~due to the change in boundary conditions the~governing equation of motion of light rays is modified---as it has happened now according to Equation~(\ref{trajectory})---then one could expect that the measured values of time, space, acceleration, etc. would change~accordingly. 

In this appendix, we resume the velocity of light $c$ in our equations to compare the orders of magnitudes more easily.  Assume a massive test particle $P$ which moves from point $A$ to point $B$, with~an infinitesimal distance  $d\vec{x}_{p}$ and within infinitesimal time $dt_{p}$. See Figure~\ref{fig:measurement}. The~proper time for this motion could be written as
\begin{eqnarray} 
 c^{2}d\tau^{2}_{p} = c^{2} (1+2\Phi)dt^{2}_{p} - R^{2}(t)(1-2\Psi)\delta_{i j}dx^{i}_{p}dx^{j}_{p}.
\end{eqnarray}
 The particle is accompanied by a geometrodynamic clock, which is meant to measure  $dt_{p}$ and $d\vec{x}_{p}$.  The~unit of time $ t_{c}$ is considered to be the interval between two ticks  of the clock while the unit of length $ x_{c}$ is  the distance that light goes back and forth in one unit of time, i.e.,~$ x_{c}$  is twice the distance  between the two mirrors in Figure~\ref{fig:measurement}. The~clock measures the  time interval $dt_{p}$ and the distance $d\vec{x}_{p}$ based on these units as explained by References~\cite{ohaniangravitation,desloge1989simple}. 

For the light signals of the clock,  we need not to worry about the deflection from the radial direction because the background metric is homogeneous and isotropic; therefore, any deflection would have to be of first order and its effect in the perturbation metric would be of the second order.   Thus, any first order change in the geodesic of the light signals  would be toward the center of the perturbation. For~any light ray within the horizon (including the signal  of the clock) which is moving toward the perturbation,  one could write
\begin{eqnarray} \label{nullcondition}
\frac{d x^{r}_{c}}{dt_{c}}   = -  \frac{c}{R(t)} (\frac{1+2\Phi}{1-2\Psi})^{1/2} \simeq - \frac{c}{ R(t)} \left(1+\Phi +\Psi    \right) +...
\end{eqnarray} 
in which  $x^{r}_{c}$  stands for the infinitesimal displacement in the radial direction of the signal. We will use this equation in the following to derive the effect of  imposing Neumann boundary conditions on the~clocks.

\unskip

\subsection{Trajectory of Massive~Particles}
In this part, we first prove that  imposing a Neumann boundary condition on perturbation  equations would result  in negligible terms on the right-hand side of the geodesic equation, i.e.,~$-\Gamma^{i}_{\mu \nu}q^{\mu}q^{\nu}$, for~massive particles. Then, we will show that, because  imposing  Neumann B. C. would affect the light signals of the geometrodynamic clock, the~left-hand side of the geodesic equation  of massive particles $dq^{i} / d \tau$ will ultimately change to include a term proportional to $cH$.   Also, we will provide a simple observational evidence for the existence of a term proportional to $cH$ in the equation of motion of massive particle. In~addition, a~proof of Equation~(\ref{Second}) is provided at the end of the results. During~our calculations, we will use Table~\ref{ratia}, which provides the order of magnitude of the perturbation terms, to~systematically neglect the terms with insignificant share in the~equations.

The new term, i.e.,~$2c_{1}cH$, has been introduced into the equation of motion of massive particles by Reference~\cite{Shenavar:2016bnk}  on the general ground that, because the light rays are disturbed by Neumann B. C., its effect must show itself into the equation of motion of massive objects. However, Reference~\cite{Shenavar:2016bnk} provides no conclusive method of how one could include the change in geometrodynamic clocks (due to Neumann boundary condition) into the equation of motion. However, in~that work, a~sample of 101 galactic systems has been surveyed to show that, at the last data points, in which the Newtonian attraction is small, the~ratio of the observed acceleration to $ cH_{0}$ is almost constant. In~addition, MOND literatures have already revealed ample signs of the reliability of $ cH_{0}$ to the galactic data~\cite{Milgrom:1983zz,Milgrom:1983pn,Milgrom:1983ca,Famaey:2011kh}. In~this sense, the~inclusion of  $2c_{1}cH_{0}$ into the equation of motion had been justified through a phenomenological  approach by Reference~\cite{Shenavar:2016bnk}.

\begin{table}[H]   
  \begin{small}
  \centering
 \caption{ Typical magnitudes of perturbations in solar, galactic, and cluster scales. We have assumed  typical rotation velocity of the objects in the solar system (galaxy and cluster of galaxies, respectively) as 30  km/s  (100 km/s and 1000 km/s, respectively), mass of the Sun $M_{\odot}= 1.988 \times 10^{30}$~kg ($10^{12} M_{\odot}$, $10^{14} M_{\odot}$), and~typical distance to the Sun $AU= 1.496 \times 10^{11}~m$ (10~kpc, 10~Mpc). In~general, this table shows that the terms proportional to $v/c$, $\dot{\Phi}$, and $H\Phi$ are negligible in Equation~(\ref{geo2}). Also, solar system is governed by the term $\vec{\nabla} \Phi$ (the effect of $2c_{1}cH$ due to boundary condition is very small in this scale~\cite{Shenavar:2016xcp}) while the effect of $2c_{1}cH$ is considerable at galactic and cluster scales.   In~the last row, we have reported   typical observed accelerations $a_{obs}$ in these~scales. }
 \label{ratia}  
  \begin{tabular}{ c  c c c }   
\toprule 
     \multicolumn{1}{c}{}& \multicolumn{1}{c }{\textbf{Solar System}} &\multicolumn{1}{c }{\textbf{Galaxies}} &\multicolumn{1}{c }{\textbf{Cluster of Galaxies}}   \\\midrule
$v/c \approx $    & $10^{-4}$ & $3 \times 10^{-4}$  & $ 3 \times  10^{-3}$   \\   
$\Phi \approx$   & $10^{-8}$& $5 \times 10^{-6}$ & $5 \times 10^{-7}$   \\  
  $ \frac{c\dot{\Phi}}{a_{0}} \approx \frac{cH \Phi}{a_{0}} \approx$ &  $10^{-8}$& $5 \times 10^{-6}$ & $5 \times 10^{-7}$   \\    
 $ \frac{c^{2} |\vec{\nabla} \Phi |}{a_{0}} \approx  $  & $10^{+8}$ & 2 & $2 \times 10^{-3}$  \\
$ \frac{a_{obs}}{a_{0}}  \approx  $  & $10^{+8}$ & 0.01 to 10 & 0.001 to 0.1 \\\bottomrule
  \end{tabular}

 \end{small}
\end{table} 
 

To find the equation of motion of a massive particle with mass $m$, we write the physical momentum~as 
 $$p^{\mu}=(\frac{E}{c},p^{i})$$
in which $E$ is the physical energy
$$E=(\vec{p}^{2}c^{2}+m^{2}c^{4})^{1/2},$$ 
while  the co-moving momentum $q$, to~first order of approximation, could be found as (Dodelson~\cite{Dodelson:2003ft}, page 102)
$$q^{\mu}=(\frac{E(1-\Phi)}{c},p^{i}\frac{(1+\Psi)}{R})$$
where 
 $$g_{\mu \nu}q^{\mu}q^{\nu}=-m^{2}c^{4}.$$
Please note that $\Phi $ and $\Psi$ in Dodelson's book corresponds to $-\Psi$ and $\Phi$ in our~notation.
 
Then, the~geodesic equations for a massive particle
\begin{eqnarray}
\frac{dq^{0}}{d \tau_{p}} = -\Gamma^{0}_{\mu \nu}q^{\mu}q^{\nu}\\ \nonumber
\frac{dq^{k}}{d \tau_{p}} = -\Gamma^{k}_{\mu \nu}q^{\mu}q^{\nu}
\end{eqnarray}
could be found  as (in first order of approximation)
\begin{eqnarray}
\frac{dq^{0}}{d \tau_{p}}  &=& \frac{1}{mc} \left(2\frac{\dot{R} }{R} \vec{p}^{2} \Phi - 2\frac{E}{R}\vec{p}.\vec{\nabla}\Phi -\frac{E^{2}}{c^{2}}\dot{\Phi} +  \vec{p}^{2} \dot{\Psi} \right)             \\ \nonumber
\frac{d \vec{q}}{d \tau_{p}}  &=&  \frac{1}{mc^{2}R^{2}}   \left( - E^{2} \vec{\nabla} \Phi + 2E\vec{p}(\Phi - \Psi)\dot{R} +2c^{2}\vec{p}\vec{p}.\vec{\nabla} \Psi -c^{2}\vec{p}^{2} \vec{\nabla} \Psi +2ER\vec{p} \dot{\Psi}     \right) 
\end{eqnarray}
by applying the Christoffel symbols of Equation \eqref{Christoffel1}. Now, by~imposing Neumann B.C.  $\Psi= \Phi -c_{1}$ into the last two equations, one could arrive at
\begin{eqnarray}\label{geo1} 
\frac{dq^{0}}{d \tau_{p}}  &=& \frac{1}{mc} \left(2\frac{\dot{R} }{R} \vec{p}^{2} \Phi - 2\frac{E}{R}\vec{p}.\vec{\nabla}\Phi -\frac{E^{2}}{c^{2}}\dot{\Phi} +  \vec{p}^{2} \dot{\Phi} \right)             \\ \nonumber  \\  \label{geo2}
\frac{d \vec{q}}{d \tau_{p}}  &=&  \frac{1}{mc^{2}R^{2}}   \left( - E^{2} \vec{\nabla} \Phi + 2c_{1}E\vec{p}\dot{R} +2c^{2}\vec{p}\vec{p}.\vec{\nabla} \Phi -c^{2}\vec{p}^{2} \vec{\nabla} \Phi +2ER\vec{p} \dot{\Phi}     \right). 
\end{eqnarray}
Here, we need to evaluate the terms on the right-hand side of Equations~(\ref{geo1}) and \eqref{geo2}. First of all, we assume particles with non-relativistic velocities $\mid \vec{p} \mid /mc \ll 1$. Thus, we would have $E \simeq mc^{2}$. Now, by~considering typical values of perturbations from Table~\ref{ratia}, one could readily see that the terms on the rhs of Equation~(\ref{geo1}) are negligible. Thus, we could find $dq^{0} / d \tau_{p}  \simeq 0 $, which means that $E$ is conserved. The~right-hand side of Equation~(\ref{geo2}) too could be estimated in the same way. Comparing  these terms with the values of Table~\ref{ratia}, one could see that the only term which survives on the rhs is the first term, i.e.,~$\propto \vec{ \nabla } \Phi$. Thus, Equation~(\ref{geo2}) could be approximated as 
$$ \frac{d \vec{q}}{d \tau_{p}} \simeq  -mc^{2}\frac{ \vec{ \nabla } \Phi}{R^{2}} $$
Therefore, we find out that the imposition of   Neumann boundary condition   does not provide a significant share on the rhs of Equation~(\ref{geo2}).   The~same result had been reported by Reference~\cite{Shenavar:2016bnk}.

Now, we evaluate the left-hand side of Equation~(\ref{geo2}) to derive the equation of motion of massive particles.   Knowing that the proper time of the particle could be written as follows
$$d \tau_{p} = dt_{p} \left(1+2\Phi -  R^{2}(1-2\Psi)\frac{\vec{v}^{2}}{c^{2}} \right)^{1/2},$$
the left-hand side of the geodesic  equation could be found as
$$ \left(1+2\Phi -  R^{2}(1-2\Psi)\frac{\vec{v}^{2}}{c^{2}} \right)^{-1/2} \frac{d }{d t_{p}} \left(\frac{(1+\Psi)}{R}\vec{p} \right) \simeq  -m c^{2}\frac{ \vec{ \nabla } \Phi}{R^{2}}. $$
Since $v/c$ and $\Phi$ are considered to be small according to Table~\ref{ratia}, the~last equation could be simplified as
$$  \frac{d }{d t_{p}} \left(\frac{(1+\Psi)}{R}\vec{p} \right) \simeq  -m c^{2}\frac{ \vec{ \nabla } \Phi}{R^{2}} $$
or equivalently
$$ \frac{ \dot{\Psi}R-\dot{R} \Psi }{R^{2}}\vec{p} + \frac{(1+\Psi)}{R} \frac{d \vec{p} }{d t_{p}}    \simeq  -m c^{2}\frac{ \vec{ \nabla } \Phi}{R^{2}} $$.
Checking the values of Table~\ref{ratia} again, one could immediately see that the first term on the left-hand side is negligible; thus, we have
$$  (1+\Psi)  \frac{d \vec{p} }{d t_{p}}    \simeq  -mc^{2}\frac{ \vec{ \nabla } \Phi}{R } $$
or simply 
$$  (1+\Psi)  \frac{d^{2} \vec{x}_{p} }{d t^{2}_{p}}    \simeq  - c^{2}\frac{ \vec{ \nabla } \Phi}{R } $$
in which $\vec{x}_{p}$ is the physical coordinate of the particle. If~we presume Dirichlet B. C., i.e.,~$\Phi = \Psi$, the~above equation would readily provide the  equation of motion in Newtonian~limit.

Now, we want to prove that, because the trajectory of light signals is altered, the~effect would adjust acceleration measurements  in geometrodynamic clocks. Suppose that a geometrodynamic clock accompanies a particle, as~in Figure~\ref{fig:measurement}, on~a trajectory which is parametrized by $\lambda$. The~time  $t_{p}$ and displacement  $x_{p}$ of the particle are measured based on the units of the geometrodynamic clock. The~clock measures  time intervals directly $t_{P}=Nt_{c}$, i.e.,~the motion has occurred in $N$ units of time $t_{c}$. However, the~displacement will be a function of both $t_{c}$ and $ x_{c}$, i.e.,~$x_{p}=x_{p}(t_{c}, x_{c})$ because~the units of geometrodynamic clocks are always linked by being null, i.e.,~Equation~(\ref{nullcondition}). We will write all derivatives with respect to $\lambda$. Also, as~mentioned above, since the deflection from the radial direction is negligible for light signals, we will only consider the displacement of light rays toward center $x^{r}_{c}$.  Moreover, we will presume that the observation of the test particle has occurred in a very small time. The~smallest possible time is when $t_{P}= t_{c}$ or when $N=1$.

For the physical velocity of the particle, we have (using the chain rule)
\begin{eqnarray} \label{veloci}
\frac{dx^{i}_{p}}{dt_{p}}=\frac{dx^{i}_{p} / d \lambda}{dt_{p}/d\lambda }=\dfrac{ \displaystyle \frac{\partial x^{i}_{p}}{\partial t_{c}}\frac{dt_{c}}{d\lambda}+ \frac{\partial x^{i}_{p}}{\partial x^{r}_{c}}\frac{dx^{r}_{c}}{d\lambda} }{ \displaystyle   \frac{\partial t_{p}}{\partial t_{c}}\frac{dt_{c}}{d\lambda}+ \frac{\partial t_{p}}{\partial x^{r}_{c}}\frac{dx^{r}_{c}}{d\lambda}} 
\end{eqnarray} 
which is equal to
\begin{eqnarray} 
\frac{dx^{i}_{p}}{dt_{p}}=  \left(\frac{\partial x^{i}_{p}}{\partial t_{c}}\frac{dt_{c}}{d\lambda}+ \frac{\partial x^{i}_{p}}{\partial x^{r}_{c}}\frac{dx^{r}_{c}}{d\lambda} \right)  \left( \frac{dt_{c}}{d\lambda} \right)^{-1}
\end{eqnarray} 
because $\partial t_{p} / \partial x^{r}_{c}=0$. In~the same way
\begin{eqnarray}
 \frac{d^{2} x^{i}_{p}}{dt^{2}_{p}} &=& \dfrac{d \left(\dfrac{dx^{i}_{p}}{dt_{p}} \right) / d \lambda}{dt_{p}/d\lambda}  \\ \nonumber
 &=&   \dfrac{\displaystyle \left[ \frac{d}{d\lambda}(\frac{\partial x^{i}_{p}}{\partial t_{c}}\frac{dt_{c}}{d\lambda})+ \frac{d}{d\lambda}(\frac{\partial x^{i}_{p}}{\partial x^{r}_{c}}\frac{dx^{r}_{c}}{d\lambda}) \right] \left( \frac{dt_{c}}{d\lambda} \right)^{-1} - \left[ \frac{\partial x^{i}_{p}}{\partial t_{c}}\frac{dt_{c}}{d\lambda}+ \frac{\partial x^{i}_{p}}{\partial x^{r}_{c}}\frac{dx^{r}_{c}}{d\lambda} \right] \dfrac{\frac{d^{2}t_{c}}{d\lambda^{2}}}{(\frac{dt_{c}}{d\lambda})^{2} }  }{ \displaystyle   \frac{\partial t_{p}}{\partial t_{c}}\frac{dt_{c}}{d\lambda}+ \frac{\partial t_{p}}{\partial x^{r}_{c}}\frac{dx^{r}_{c}}{d\lambda}} 
 \end{eqnarray}
which results in
\begin{eqnarray}
 \frac{d^{2} x^{i}_{p}}{dt^{2}_{p}} &=&   \left[ \frac{d}{d\lambda}(\frac{\partial x^{i}_{p}}{\partial t_{c}}\frac{dt_{c}}{d\lambda})+ \frac{d}{d\lambda}(\frac{\partial x^{i}_{p}}{\partial x^{r}_{c}}\frac{dx^{r}_{c}}{d\lambda}) \right]    \left( \frac{dt_{c}}{d\lambda} \right)^{-2}  \\ \nonumber
 &-&   \left[ \frac{\partial x^{i}_{p}}{\partial t_{c}}\frac{dt_{c}}{d\lambda}+ \frac{\partial x^{i}_{p}}{\partial x^{r}_{c}}\frac{dx^{r}_{c}}{d\lambda} \right] \dfrac{\frac{d^{2}t_{c}}{d\lambda^{2}}}{(\frac{dt_{c}}{d\lambda})^{3} } 
 \end{eqnarray}
or
\begin{eqnarray} \label{geo0}
 \frac{d^{2} x^{i}_{p}}{dt^{2}_{p}} &=&    \left[ \frac{d^{2} t_{c}}{d\lambda^{2}}(\frac{\partial x^{i}_{p}}{\partial t_{c}})+  \frac{dt_{c}}{d\lambda}  \frac{d}{d\lambda}(\frac{\partial x^{i}_{p}}{\partial t_{c}}) 
 + \frac{d^{2} x^{r}_{c}}{d\lambda^{2}} (\frac{\partial x^{i}_{p}}{\partial x^{r}_{c}})  
  + \frac{dx^{r}_{c}}{d\lambda} \frac{d}{d\lambda}(\frac{\partial x^{i}_{p}}{\partial x^{r}_{c}}) \right]    \left( \frac{dt_{c}}{d\lambda} \right)^{-2}  \\ \nonumber
 &-&  \left[ \frac{\partial x^{i}_{p}}{\partial t_{c}}\frac{dt_{c}}{d\lambda}+ \frac{\partial x^{i}_{p}}{\partial x^{r}_{c}}\frac{dx^{r}_{c}}{d\lambda} \right] \dfrac{\frac{d^{2}t_{c}}{d\lambda^{2}}}{(\frac{dt_{c}}{d\lambda})^{3} } .
 \end{eqnarray}
We just mention that there is no summation on indices $c$ or  $r$ in above equations. These indices only indicate the relation to ``clock'' rays and their ``radial''~direction.

The meaning of  the partial derivatives $ \partial x^{i}_{p} / \partial t_{c} $  and $\partial x^{i}_{p} / \partial x^{r}_{c}$ needs to be discussed. The~derivative  $ \partial x^{i}_{p} /  \partial t_{c} $ provides the rate of the change in the position of the particle when $x_{c}$ is kept constant. In~other words, this term does not include the motion of the light signal. Thus, $ \partial x^{i}_{p} /  \partial t_{c} $ coincides with the classical definition of velocity which is defined independent of light rays. In~the same way, $\partial x^{i}_{p} / \partial x^{r}_{c}$ measures the change in the direction  of $x^{i}_{p}$ with respect to $x^{r}_{c}$ when  $t_{c}$ is kept constant.  The~function $ dx^{i}_{p} / dt_{p} $ in Equation~(\ref{veloci}), on~the other hand, includes the fact that $t_{c}$ and $ x_{c}$ are related through Equation~(\ref{nullcondition}), i.e.,~they are the elements of a null~trajectory.
 
We would also have
 $$\frac{d}{d\lambda}(\frac{\partial x^{i}_{p}}{\partial t_{c}})  = \frac{dt_{c}}{d\lambda} (\frac{\partial^{2} x^{i}_{p}}{\partial t^{2}_{c}}) +\frac{d x^{r}_{c}}{d\lambda} \frac{\partial^{2} x^{i}_{p}}{\partial x^{r}_{c} \partial t_{c}} $$
and 
$$ \frac{d}{d\lambda}(\frac{\partial x^{i}_{p}}{\partial x^{r}_{c}}) = \frac{dt_{c}}{d\lambda}\frac{\partial^{2} x^{i}_{p}}{ \partial t_{c} \partial x^{r}_{c} }+ \frac{d x^{r}_{c}}{d\lambda} \frac{\partial^{2} x^{i}_{p}}{\partial x^{r~2}_{c}} $$
for the second derivatives alongside the path of the particle. In~addition, because~$x^{i}_{p}$ indicates a physical coordinate while $x_{c}$ is a co-moving one, we will have $\partial x^{i}_{p} / \partial x^{r}_{c}  = R \delta^{i}_{r}$. Thus, the~last two equations could be obtained as 
 $$\frac{d}{d\lambda}(\frac{\partial x^{i}_{p}}{\partial t_{c}})  = \frac{dt_{c}}{d\lambda} (\frac{\partial^{2} x^{i}_{p}}{\partial t^{2}_{c}}) +\dot{R}\frac{d x^{r}_{c}}{d\lambda} \delta^{i}_{r} $$
and 
$$ \frac{d}{d\lambda}(\frac{\partial x^{i}_{p}}{\partial x^{r}_{c}}) = \dot{R} \frac{dt_{c}}{d\lambda} \delta^{i}_{r}. $$
Therefore, Equation~(\ref{geo0}) could be written as
\begin{myequation}  \label{acc1}\begin{array}{ccl}
 \frac{d^{2} x^{i}_{p}}{dt^{2}_{p}} &=&     \left[ \frac{d^{2} t_{c}}{d\lambda^{2}}(\frac{\partial x^{i}_{p}}{\partial t_{c}})+  \frac{dt_{c}}{d\lambda}  \left(\frac{dt_{c}}{d\lambda} (\frac{\partial^{2} x^{i}_{p}}{\partial t^{2}_{c}}) +\dot{R}\frac{d x^{r}_{c}}{d\lambda} \delta^{i}_{r} \right) 
 +R \frac{d^{2} x^{r}_{c}}{d\lambda^{2}} \delta^{i}_{r}
  +  \dot{R}  \frac{dx^{r}_{c}}{d\lambda}\frac{dt_{c}}{d\lambda} \delta^{i}_{r}  \right]    \left( \frac{dt_{c}}{d\lambda} \right)^{-2}  \\ 
 &-&   \left[ \frac{\partial x^{i}_{p}}{\partial t_{c}}\frac{dt_{c}}{d\lambda}+ R \delta^{i}_{r} \frac{dx^{r}_{c}}{d\lambda} \right] \dfrac{ \displaystyle \frac{d^{2}t_{c}}{d\lambda^{2}}}{ \displaystyle (\frac{dt_{c}}{d\lambda})^{3} } .\end{array}
\end{myequation}

Now, we need to find $dx^{r}_{c} / d\lambda $ and $d^{2} x^{r}_{c} / d\lambda^{2}$ to determine Equation~(\ref{acc1}). From~the null condition of Equation \eqref{nullcondition}, one would have
\begin{eqnarray}  \label{eq1}
\frac{dx^{r}_{c}}{d\lambda} = - \frac{c}{R} \left[ 1 +\Phi +\Psi +...  \right]\frac{dt_{c}}{d\lambda}
\end{eqnarray}
and
\begin{eqnarray}
\frac{d^{2}x^{r}_{c}}{d\lambda^{2}} =     - \frac{c}{R} \left(1 +\Phi +\Psi +...  \right)  \frac{d^{2}t_{c}}{d\lambda^{2}} 
 - c \frac{dt_{c}}{d\lambda} \left[ \frac{dt_{c}}{d\lambda}  \frac{\partial  }{\partial t_{c}} + \frac{dx^{r}_{c}}{d\lambda}   \frac{\partial  }{\partial x^{r}_{c}} \right]
  \left(\frac{1}{R} (1 +\Phi +\Psi +... ) \right)
\end{eqnarray}
to the first order of approximation. This last equation is equal to
\begin{myequation}\begin{array}{cll}
\frac{d^{2}x^{r}_{c}}{d\lambda^{2}} &=&    - \frac{c}{R} \left(1 +\Phi +\Psi +...  \right)  \frac{d^{2}t_{c}}{d\lambda^{2}} \\ 
&-  &  c (\frac{dt_{c}}{d\lambda})^{2} \left[ -\frac{\dot{R}}{R^{2}} (1 +\Phi +\Psi +... ) +\frac{ \dot{\Phi} + \dot{\Psi} +...   }{R} + \frac{c}{R^{2}} (\partial_{r} \Phi+ \partial_{r} \Psi + ...) \left(1 +\Phi +\Psi +...  \right)  \right].\end{array}
\end{myequation}
Since the measurement takes place in a sufficiently small length and small time,   the~spatial and time variations of  $\Phi$ and $ \Psi $ are negligible. Thus, for~the light signal, we could write
\begin{eqnarray}  \label{eq2}
\frac{d^{2}x^{r}_{c}}{d\lambda^{2}} &=&    - \frac{c}{R} \left(1 +\Phi +\Psi +...  \right)  \frac{d^{2}t_{c}}{d\lambda^{2}} +   c (\frac{dt_{c}}{d\lambda})^{2} \left[ \frac{\dot{R}}{R^{2}} (1 +\Phi +\Psi +... )   \right].
\end{eqnarray}

In Equation~(\ref{acc1}), the~terms proportional to $\partial x^{i}_{p} / \partial t_{c} $ are negligible because they are proportional to the velocity of the moving object. Thus, the~first terms on the right-hand side of the first and the second rows are dismissed. Also, the~third and the fifth terms in the  first row are similar and are equal to
\begin{eqnarray} \label{eq3}
2 \dot{R} \dfrac{ \displaystyle  dx^{r}_{c} / d\lambda  }{ \displaystyle  dt_{c} / d\lambda  } =-2\frac{\dot{R}}{R}c (1 +\Phi +\Psi +... )
\end{eqnarray}
in which we have used Equation~(\ref{eq1}).  In~addition, by~using Equations~(\ref{eq1}) and \eqref{eq2}, the~fourth term on the right-hand side of Equation~(\ref{acc1}) could be found as
\begin{eqnarray} \label{eq4}
 R \dfrac{ \displaystyle  d^{2}x^{r}_{c} / d\lambda^{2}  }{ \displaystyle  (dt_{c} / d\lambda)^{2}  } = -R\frac{c}{R}  (1 +\Phi +\Psi +... )   \dfrac{ \displaystyle  d^{2}t_{c} / d\lambda^{2} }{ \displaystyle (dt_{c} / d\lambda)^{2} } + cR \left(\frac{\dot{R}}{R^{2}}(1 +\Phi +\Psi +... )  \right)
\end{eqnarray}
while the second term on the second row could be found as
\begin{eqnarray} \label{eq5}
R  \frac{dx^{r}_{c}}{d\lambda}  \dfrac{    d^{2} t_{c} / d\lambda^{2} }{   (dt_{c} / d\lambda)^{3}}= -c (1 +\Phi +\Psi +... )\dfrac{    d^{2} t_{c} / d\lambda^{2} }{   (dt_{c} / d\lambda)^{2}}
\end{eqnarray}
 Putting Equations~(\ref{eq3}) to \eqref{eq5} together, one could find
\begin{eqnarray}   
2 \dot{R} \dfrac{ \displaystyle  dx^{r}_{c} / d\lambda  }{ \displaystyle  dt_{c} / d\lambda  } &+&  R \dfrac{ \displaystyle  d^{2}x^{r}_{c} / d\lambda^{2}  }{ \displaystyle  (dt_{c} / d\lambda)^{2}  } - R  \frac{dx^{r}_{c}}{d\lambda}  \dfrac{    d^{2} t_{c} / d\lambda^{2} }{   (dt_{c} / d\lambda)^{3}}  \\ \nonumber
 &=& -2\frac{\dot{R}}{R}c (1 +\Phi +\Psi +... ) - R\frac{c}{R}  (1 +\Phi +\Psi +... )   \dfrac{ \displaystyle  d^{2}t_{c} / d\lambda^{2} }{ \displaystyle (dt_{c} / d\lambda)^{2} }  \\ \nonumber
&+ & cR \left(\frac{\dot{R}}{R^{2}}(1 +\Phi +\Psi +... )  \right) +c (1 +\Phi +\Psi +... )\dfrac{    d^{2} t_{c} / d\lambda^{2} }{   (dt_{c} / d\lambda)^{2}}  \\ \nonumber
&= & -\frac{\dot{R}}{R}c (1 +\Phi +\Psi +... )   \\ \nonumber
&= & -\frac{\dot{R}}{R}c (1 +\Phi + \Phi -c_{1} +... )   \\ \nonumber
&\simeq &  -cH(1-c_{1})
\end{eqnarray}
in which ,in the fifth row, we have imposed Neumann boundary condition $ \Psi= \Phi - c_{1}$ while, in the last row, we have used the fact that $cH\Phi$ is negligible in local scales according to Table~\ref{ratia}. 
 Finally, by~employing this last equation, the~equation of motion could be written as
\begin{eqnarray}
  (1+\Psi)  \frac{d^{2} \vec{x}_{p} }{d t^{2}_{p}}    &\simeq & (1+\Psi) \left( \frac{\partial^{2} x^{i}_{p}}{\partial t^{2}_{c}} +  2 \dot{R} \dfrac{ \displaystyle  dx^{r}_{c} / d\lambda  }{ \displaystyle  dt_{c} / d\lambda  }\delta^{i}_{r} +  R \dfrac{ \displaystyle  d^{2}x^{r}_{c} / d\lambda^{2}  }{ \displaystyle  (dt_{c} / d\lambda)^{2}  }\delta^{i}_{r} - R  \frac{dx^{r}_{c}}{d\lambda}  \dfrac{    d^{2} t_{c} / d\lambda^{2} }{   (dt_{c} / d\lambda)^{3}}\delta^{i}_{r}  \right) \\ \nonumber  
&\simeq &      \frac{\partial^{2} x^{i}_{p}}{\partial t^{2}_{c}} +  2c H c_{1} \hat{r}. 
\end{eqnarray}
in which $\hat{r}$ shows the unit vector in radial direction. The~value $\partial^{2} x^{i}_{p} / \partial t^{2}_{c}$ is the classical definition of acceleration because it measures the rate of change in velocity when $x_{c}$ is constant, i.e., when we do not use light rays in our measurements. For~example, in~the radial direction,  $\partial^{2} x^{i}_{p} / \partial t^{2}_{c}$ is equal to the centripetal acceleration $v^{2}/r$ since, if~$\theta$ represents the azimuthal angle, then 
$$ \vec{v}_{\theta}= \frac{\partial  \vec{x}_{p}}{\partial t_{c}}\mid_{\underset{x_{c} = constant}{r = constant}}~=~r\dot{\theta} \hat{\theta}$$
while, for the centripetal acceleration, we would have
$$ \frac{\partial^{2} \vec{x}_{p}}{\partial t^{2}_{c}} = \frac{\partial  \vec{v}_{\theta}}{\partial t_{c}}\mid_{\underset{x_{c} = constant}{r = constant}}~=~r\dot{\theta}^{2} \hat{r}$$,
which is equal to $v^{2}/r$.

\begin{figure}[H]
\centering
\includegraphics[width=14cm]{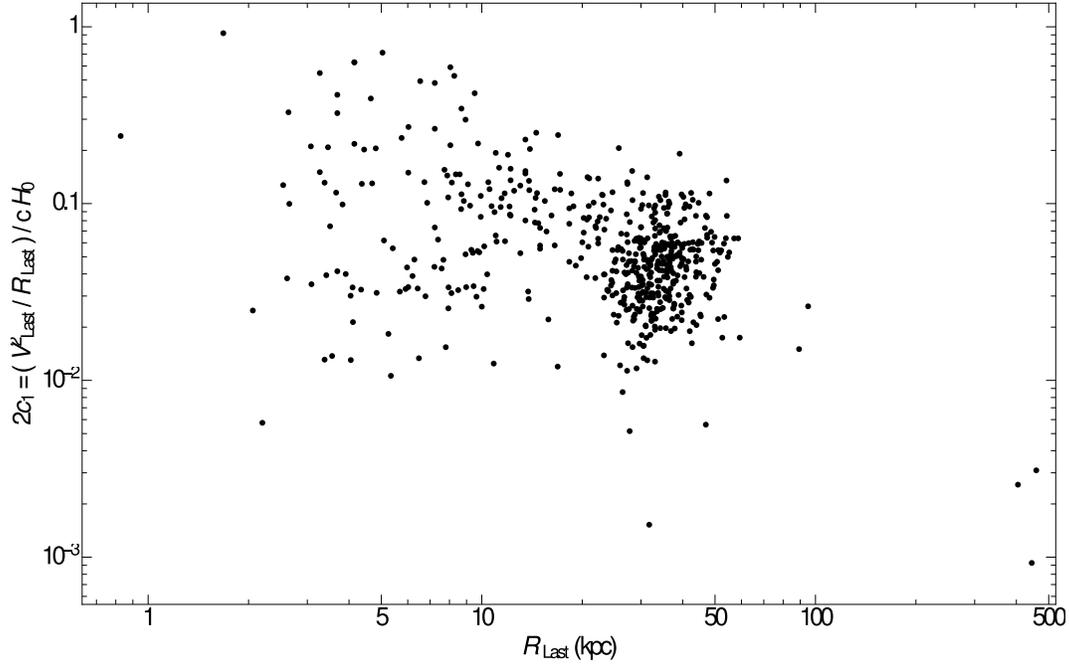} 
\caption{ Estimating $2c_{1}$ based on the last data points of the rotation velocity. Systems with small $R_{last}$ are more probably governed by the Newtonian term; systems with very large $R_{last}$ might be affected by the long-range interaction  of their environment, which is proportional to $\propto 2c_{1}cH_{0}$. On~the other hand, galaxies with extended rotation velocities $20~kpc~<~R_{last}~< 60~kpc$ show a concentration around the value $2c_{1}=0.06$. One could check that about $67 \%$ of the data points lie within the bound of $0.03~<~2c_{1}~<~0.12$.  Data of this plot is extracted from Reference~\cite{2018PASJ...70...31S} and includes three subsamples (containing 551 objects): S-sample provided  by Reference~\cite{2016PASJ...68....2S}, P-sample which contains galaxies with W1-band photometry~\cite{2014AJ....148..134P}, and C-sample (galaxies with r-band photometry and rotation velocities reported by \citet{1996ApJS..103..363C,1997AJ....114.2402C}). \label{lastpoints}}
\end{figure}

We mention that the effect of the constraint Equation~(\ref{nullcondition})  displays itself in the term  $2c_{1}H$. Thus, this term represents the effect of which the imposition of a Neumann boundary condition has on the light rays of geometrodynamic clocks and,~through them, into~the equation of motion of massive particles. Therefore, the~new equation of motion due to a point-like perturbation at the center  becomes
\begin{eqnarray}
\frac{d^{2} \vec{x}_{p} }{dt^{2}} +2c_{1}cH(t) \hat{r} &=& -c^{2} \frac{\vec{\nabla}}{R} \Phi 
\end{eqnarray}
or in  physical coordinates $x_{p}$ as
\begin{eqnarray}
\frac{d^{2} \vec{x}_{p} }{dt^{2}} +2c_{1}cH(t) \hat{r} &=& -c^{2} \vec{\nabla}_{p} \Phi .
\end{eqnarray} 
 For a particle around a galaxy, one expects that the centripetal acceleration due to Newtonian potential decreases while there is still a constant acceleration term due to the boundary condition. Thus, if~we measure the acceleration at  the outermost points of galactic rotation curves, the~Newtonian share would be minimized and we could have an estimation of the parameter $c_{1}$. To~do so, we have use data from Reference~\cite{2018PASJ...70...31S} that contain 551 galaxies (including three subsamples). The~result is provided in Figure~\ref{lastpoints}. In~this plot, $R_{last}$ shows the radius of the last data point (in $kpc$) while $V_{last}$ is its rotational velocity (in $km/s$). We have assumed $H_{0}=70~km/s/Mpc$. For~galaxies with small $R_{last}$  (these are the systems for which the observations of rotational velocity are not extended enough),  the acceleration due to Newtonian potential is still dominant. Thus, the~ratio of $(V^{2}_{last}/R_{last})/cH_{0}$ is high for these systems. The~maximum value of $2c_{1}$ is found to be about $1$. As~we move toward  right of the plot---systems with extended rotation velocities---the   value of $2c_{1}$ becomes more concentrated in the interval $0.01~<~2c_{1}~<~0.13$ (88 $\%$ of the data points). Eight systems (about 1.5 $\%$ of the data) show a value less than $0.01$ for $2c_{1}$. Three galaxies out of eight posses very large  $R_{last}$. These eight systems are probably galaxies with strong interaction  with their neighbouring environments (at least at their outer parts).  See also Tables~3--5 of Reference~\cite{Shenavar:2016bnk} for a similar analysis of 101 galaxies. The~value of the Neumann parameter is therefore estimated to be with the bound of $0.01~<~2c_{1}~<~0.13$. This coincides with some previous estimations~\cite{Shenavar:2016xcp,Shenavar:2016bnk,shenavar2018local}

Since, the~long-range interaction is always present in the MOD model, it is essential to go beyond the simple approximation of point-like perturbation.  To~survey systems with many sources of gravitation,  one should sum  (integrate) over different patches of matter to derive~\cite{Shenavar:2016xcp}
\begin{eqnarray}  
c^{2}\Phi (\vec{x}_{p})=-G \int \frac{ \delta \rho_{m} (\vec{x^{\prime}_{p}})d^{3}\vec{x^{\prime}_{p}}}{| \vec{x^{\prime}_{p}} - \vec{x}_{p}|}  + \frac{2c_{1}cH(t)}{M}\int \delta \rho_{m}(\vec{x^{\prime}_{p}})d^{3}\vec{x^{\prime}_{p}}|\vec{x^{\prime}_{p}} - \vec{x}_{p}|
\end{eqnarray} 
in which $\vec{x}_{p}$ still represents the physical units. The~factor $c^{2}$ on the left-hand side makes $\Phi$ dimensionless as it is assumed in Section~\ref{Scalar}. This potential could lead to the next modified Poisson equation:
\begin{eqnarray}
c^{2} \vec{\nabla}^{2}_{x_{p}} \Phi (\vec{x}_{p}) = 4\pi G \delta \rho_{m} + \frac{4c_{1}  cH(t) }{  M} \int \frac{\delta \rho_{m}(t, \vec{x^{\prime}_{p}})d^{3}\vec{x^{\prime}_{p}}}{| \vec{x^{\prime}_{p}} - \vec{x}_{p}|}
\end{eqnarray} 
In addition, to~write the integrals in the co-moving coordinates, which is used in Section~\ref{Scalar}, we should note that $dx_{p}= R(t)dx$ and $\partial_{x_{p}}=\partial_{x}/R(t)$; thus, the~modified Poisson's equation in the co-moving coordinates could be written as
\begin{eqnarray}
c^{2}\frac{1}{R^{2}}\nabla^{2}\Phi (\vec{x}) = 4\pi G \left(  \delta \rho_{m} + \frac{c_{1}  cH(t) R^{3}(t) }{\pi G MR^{3}(t)} \int \frac{\delta \rho_{m}(t, \vec{x^{\prime}})d^{3}\vec{x^{\prime}}}{| \vec{x^{\prime}} - \vec{x}|}  \right)
\end{eqnarray}
which is equal to Equation~(\ref{Second}).

One should note that, even when the particle is   moving away from the perturbation,   for~which  the radial direction of the light rays should be assigned with a $+$ sign on the right-hand side of the Equation~(\ref{nullcondition}), the~result for the equation of motion would still be the same. The~reason is that, according to  Equation~(\ref{trajectory}), the~imposition of Neumann boundary condition always leads to more gravitation so far as $c_{1}>0$. The~direction of the light signal does not change this point. In~fact, Reference~\cite{Shenavar:2016bnk} presumes $c_{1}>0$ to solve the mass discrepancy problem in lensing~systems. 

Also, we should point out that the effect that a boundary condition introduces to the geometrodynamic clocks could not be removed by readjusting the clocks. The~reason is that, unlike the influence of the tidal forces which could be eradicated by re-paralleling of the mirrors, the~effect of the boundary condition happens even in a single tick of the device. Thus, it must be included within the equations of~motion.

\subsection{ The EPS Theorem and the Structure of~Spacetime }
As we proved in this appendix, because~the velocity of a massive particle at galactic scales is typically much smaller than the speed of light, the~modification in the geodesy of the massive object (due to Neumann B.C.) is far smaller than $cH$. However, as~discussed above, since the trajectory of the light signal includes a term proportional to de Sitter scale of acceleration, the~measured values of the massive particle's acceleration  would include the term $cH$. This procedure is motivated by the observational fact that both massless and massive particles trajectories show dependency to $cH$. A~good example for the former is the observations of strong lenses~\cite{Shenavar:2016bnk}, while for the  latter, we could mention the rotation curve data of galaxies~\cite{Shenavar:2016xcp}, local stability of galactic disks~\cite{shenavar2018local},  and  Figure~\ref{fig:measurement} here. In~this way, the~missing mass discrepancy is united by introducing a new term into the equation of~motion. 

On the other hand, if~one presumes the other procedure of spacetime measurement, e.g.,~atomic standards~\cite{ohaniangravitation}, then the two theorems of EPS~\cite{ehlers2012republication} and Ohanian~\cite{ohaniangravitation} could not be used. Thus, the~unification between the massive and massless particle's trajectories would not be possible and the missing mass discrepancy would remain unsolved for massless particles. To~summarize,  the~inclusion of the notion of geometrodynamic clock  is the second critical assumption of Neumann cosmology (beside Neumann B.C.).

One could even further argue that the viability of Neumann cosmology based on observations  at different scales could be considered as a direct ``empirical evidence'' for the advantage of the Marzke--Wheeler~\cite{marzke1964gravitation} measurement method  (geometrodynamic clocks)  over other methods of spacetime measurement. Especially, this could demonstrate   the correctness of EPS theorem (within the present model), which is a key tool in constructing the geometrical structure of~spacetime. 

\section{Discussion on the Dimensional Analysis of the~Model}
\label{Disdimensional}
  In this part, the~physical constants are recovered in all the formulas.   The~potential of the Neumann model grows linearly with distance. See Figure~1 of~\cite{Shenavar:2016xcp} for a plot of the potential. Thus, for~the centripetal acceleration, we have $v^{2}/r = GM/r^{2}~+~2c_{1}a_{0} $. However, the~rotational velocity could not exceed the speed of light. This point confines the size and the mass of  observable universe for the Neumann model as follows. Putting $v \approx c$ into the centripetal acceleration,  one could see that, to derive a unique cosmic  scale length $r_{s}$,  one must have
\begin{eqnarray} \label{scaling}
r_{s} &\approx& c/(4c_{1}H_{0})\\ \nonumber
M_{s} &\approx& c^{3}/(8 c_{1}GH_{0})
\end{eqnarray}
in which $M_{s}$ is the mass within the boundary in Figure~\ref{fig:boundary}. The~value of $r_{s}$ appears to be of the order of the Hubble distance, which is of course expected. On~the other hand, from~the value of the enclosed mass $M_{s}$, one could see the reason that the mass density $\rho_{m} \propto M_{s}/r^{3}_{s}$ of the observable universe is of the same order of the critical density $\rho_{crit} = 3H^{2}/8\pi G$ and of the same order of the new density   $\rho_{c1}$,  which is supposed to play the role of dark energy at large distances. To~see this, one could simply estimate $\rho_{m}/\rho_{crit}$ and $\rho_{m}/\rho_{c1}$ and prove that both ratios reduce to the second equation in Equation \eqref{scaling}. The~dimensional relation $M_{s} \propto  c^{3}/(c_{1}GH_{0})$ could be interpreted as follows: the gravitational acceleration felt by any particle  from the total cosmic mass within a radius $r_{s}$, i.e.,~$GM_{s}/(c/H_{0})^{2}$, is of the same order of the fundamental acceleration $cH_{0}$.  Also, this relation is Mach8 formulation of the Mach principle discussed by Reference~\cite{Bondi:1996md}.

The dimensional relation $M_{s} \approx c^{3}/(8 c_{1}GH_{0})$ is also important from the point of view of Dirac's hypothesis of large numbers~\cite{Dirac:1937ti,Dirac:1938mt}.    As~Weinberg (1972)   \cite{Weinberg:1972kfs}, chapter 16, has explained, the~pion mass   $m_{\pi} $  could be approximated as
\begin{equation} \label{pion}
    m_{\pi} \approx (\frac{h_{p}^{2}H_{0}}{Gc})^{1/3}
\end{equation}
in which $h_{p}=6.62 \times 10^{-34}$ Js  is  the Planck's constant. As~ Milgrom~\cite{Milgrom:1983ca} has recognized, this second relation could also be interpreted as follows: the gravitational acceleration generated  by  a pion at the scale of its Compton wavelength is of the order of the fundamental acceleration of modified dynamics $cH_{0}$. Of~course, any particle with a mass close to the mass scale of QCD, i.e.,~$\approx 100$~MeV, could play the role of pion as well. By~eliminating $G$ from Equation~(\ref{pion}) and the second equation of \eqref{scaling}, one could find  the next  dimensional relation:
\begin{equation}
\frac{M_{s}}{(c/H_{0})^{2}} \approx \frac{m_{\pi}}{(h_{p}/ m_{\pi}c)^{2}}.
\end{equation}
which is independent of any interaction and only depends on the mass and size of the systems. Also, by~eliminating the Hubble constant between the two dimensional relations, one finds
\begin{equation}
M_{s} \approx M^{4}_{p}/ m^{3}_{\pi}
\end{equation} 
in which $M_{p}=\sqrt{h_{p}c/G}$ is the Planck's mass. From~this relation, for~example, one could estimate the number of baryons within the horizon by calculating $M_{s}/m_{p}$ in which $m_{p}$ is the mass of proton. Moreover, other dimensional relations could be found by employing Planck's length and time. In~one or another form, these dimensional relations have been mentioned in the literature before and, of~course, one might simply argue that these are meaningless numerical coincidences. However, in~light of the present work, these dimensional relations might also be interpreted as evidence that the microphysics is related to the  cosmological dynamics through Mach's~principle.

Knowing that the Hubble parameter in Equation~(\ref{pion}) varies with time, if~one assumes that the relation in Equation \eqref{pion} is indeed fundamental, then at least one of the other constants in this equation must    change with time. The~factor $G$ is most commonly speculated as the time dependent parameter in the literature; otherwise, one must be open to reformulating all atomic physics by introducing $c$, $m_{\pi}$, or $h_{p}$ as time-dependent variables. Many interesting results have been derived in the literature based on $G(t)$ strategy. See, for~example,   Weinberg~\cite{Weinberg:1972kfs}, chapter 16, for an interesting cosmic model based on a time-dependent gravitational ``constant''.  The~assumption that $G$ is time dependent is the core idea in Mach1 formulation    of Bondi and Samuel~\cite{Bondi:1996md}.  

However, the~Neumann model   presents another possibility since,  in all our formulation, the~new acceleration term is introduced as $c_{1}cH_{0}$ in which the parameter $c_{1}$ could be, in~general, time dependent. The~reason is that $c_{1}$ represents the effect of the distant objects, which are themselves  evolving. Thus,   it might be insightful to introduce $c_{1}(t)$ in Equation \eqref{pion} and to investigate its time variability (instead of $G$). However, as~mentioned before, the~Friedmann equations would be modified in this case. See Appendix \ref{timedepend} for the details.  Assuming the Compton wavelength of pion $h_{p}/ m_{\pi}c$, one might rewrite Equation~(\ref{pion}) as
\begin{equation}
\frac{    Gm_{\pi}}{(h_{p}/ m_{\pi}c)^{2}} \approx c c_{1}(t)  H(t).
\end{equation}
If the left-hand side of the above equation is assumed to be independent of time, then we could see that there should be a relation as $c_{1}(t) \propto 1/ H(t)$ to maintain the rhs too as time independent. Therefore, the~dimensionless parameter $c_{1}(t)$  must be very small in early epochs, i.e.,~when $H(t)$ is very large, and~then show its effects in later times.  This provides a new possibility for cosmic evolution; however, for~the sake of brevity, we do not follow this case~here.

\section{ A Semi-Newtonian Model of Gravity: Some Properties and~Issues}
\label{DisNewt}
Having built a modified dynamical model based on imposing Neumann boundary condition on cosmological perturbation equations, a~question naturally arises about the possibility of such a model in Newtonian mechanics. At~first glance, it seems that there is no such possibility in Newtonian dynamics because this theory exploits only one potential, i.e.,~there is $\Phi$ but  no $\Psi$. Therefore, even in a Newtonian model of cosmic expansion, it seems  that one could not derive a term proportional to $a_{0}=cH_{0}$ due to the limited degrees of freedom. It could be also proved  that such an exotic term does not appear in the equation of motion even by imposing Neumann B.C. on Poisson equation. Assuming the classical Poisson's equation with Neumann B.C.,
\begin{eqnarray} \label{NewtNeumann}
\nabla^{2} \Phi (\vec{r}) &=& 4 \pi G \rho (\vec{r})~~~~~\vec{r} \in \mathcal{V} \\ \nonumber
\frac{\partial \Phi (\vec{r})}{\partial \vec{n}} &=& \vec{g} (\vec{r})  ~~~~~~~~~~~ \vec{r} \in \partial \mathcal{V},
\end{eqnarray}
in which $\frac{\partial \Phi (\vec{r})}{\partial \vec{n}} = \vec{g} (\vec{r})$ is the normal derivative at the boundary of the volume $\partial \mathcal{V}$, one could see that  the normal derivative of the potential is related through the divergence theorem to~its Laplacian, i.e.,
\begin{eqnarray} \label{solvable}
4 \pi G  \int_{\mathcal{V}} \rho (\vec{r})d^{3} \vec{r} = \int_{\partial \mathcal{V}} \frac{\partial \Phi (\vec{r})}{\partial \vec{n}}. d\vec{S}
\end{eqnarray}
where $\vec{S}$ is the surface element on the boundary. See Reference~\cite{hassani2013mathematical},  pages 619--620,  for~more details.   In~this section, the~potential has the dimension of Length$^{2}$/Time$^{2}$ as it is usually presumed in Newtonian~mechanics. 

 This last equation is Gauss's law in gravitation, and it imposes  a strict solvability conditions on the Poisson equation with Neumann boundary condition. Thus, one cannot choose quite arbitrarily the value of the force on the boundary  because it is related to the total mass of the bulk, i.e.,~$|\partial \Phi (\vec{r}) / \partial n|=GM_{in}/r^{2}$, in which $M_{in}$ is the total mass enclosed in a sphere of radius $r$. In~other words, no fifth force can appear from imposing Neumann B.C.  in Newtonian gravity. For~the sake of completeness, we mention that  it is always  possible to rewrite the Poisson equation as
$$   \nabla^{2} \Phi (\vec{r}) = 4 \pi G (\rho (\vec{r})- \Bar{\rho}) $$
where $ \Bar{\rho}$ is the average density of the bulk. This equation is always solvable with Neumann B.C.~\cite{hassani2013mathematical}. However, it should be mentioned that this modified Poisson  equation cannot solve the dark matter problem since it  decreases the amount of gravitating source by $\Bar{\rho}$.

However, it is still possible to construct a semi-Newtonian model of local dynamics similar to the model introduced in Section~\ref{Scalar} by considering Mach's principle. To~do so, we will use a well-defined property of classical mechanics and three well-supported observational facts regarding our universe. In~classical mechanics,  $(a)$ Newton's second law is valid only in inertial frames.   The~observational facts used here are  as follows: $(b)$ All astrophysical systems are accelerating; thus, they could only  be considered as an approximation of inertial frames. $(c)$ The typical acceleration in the scale of galaxies and~cluster of galaxies is of the order of de Sitter scale of acceleration $cH_{0}$. $(d)$ The universe is isotropic and homogeneous on very large scales. Assumptions $a,~b, and~d$ are well accepted within the community while  assumption $c$ is supported by an impressive wealth of observations modeled based on the MOND~paradigm. 

Now, imagine a particle moving around a galaxy. See Figure~\ref{fig:boundary}. This galaxy is not, in~principle, an~inertial frame; however, we could assume an inertial frame $i$ far from the galaxy $g$ and could write the equations of motion of a particle $p$, with~mass $m$, in~that inertial frame. Then, the~position of the particle with respect to the inertial frame could be written as $\vec{r}_{pi}=\vec{r}_{pg}+\vec{r}_{gi}$; thus, the~Newton's second law in that frame is as follows:
\begin{equation} \label{ModNewt}
    m(\vec{a}_{pg}+\vec{a}_{gi}) = -m \vec{\nabla} \Phi_{N}
\end{equation}
in which $\Phi_{N}$ must be derived from Poisson equation,  $|\vec{a}_{pg}|=v^{2}/r$ provides the centripetal acceleration of the particle $p$ around the galaxy $g$ and $\vec{a}_{gi}$ is the acceleration of the galaxy with respect to the inertial frame. In~writing Equation \eqref{ModNewt}, we have used the assumptions $a$ and $b$ above.   Also, based on assumption $c$ and~to solve the mass discrepancy problem at galactic scale, we presume that the value of $\vec{a}_{gi}$ is of the same order of de Sitter acceleration. This assumption also helps us to understand how far the inertial frame ought to be; however, the~argument comes from the causal structure of special relativity and as a result of a time-dependent boosts. In~fact, for~an accelerating observer with acceleration $g$, the~local frame attached to the  observer could  be safely used to set coordinates within the Rindler horizon defined as the distance $\mathcal{L}_{R}\equiv c^{2}/g$. For~example, for~typical accelerations on earth, i.e.,~$g~\approx~10$~m/s$^{2}$, one has  $\mathcal{L}_{R} \approx $ light  year  which is quite satisfying. However, at~galactic scale with a typical de Sitter acceleration $cH_{0}$, the~Rindler horizon is of the order of the Hubble distance $\mathcal{L}_{R}\equiv c/H_{0}$. Thus, one infers that the origin of the term $\vec{a}_{gi}$ is  related to the general evolution of the cosmos. Since the cosmic fluid at Hubble distance is  homogeneous and isotropic, i.e.,~assumption $d$, the~only form of  $\vec{a}_{gi}$ compatible with this condition  is when $\vec{a}_{gi}$ is radial. Thus, the~magnitude of $\vec{a}_{gi}$ is of the order of $cH_{0}$ and it is toward the center. Therefore,  the~governing equations in this semi-Newtonian model could be written as
\begin{eqnarray} \label{ModNewt1}
\vec{a}_{pg}+2c_{1}cH_{0}\hat{\vec{r}} &=& - \vec{\nabla} \Phi_{N}  \\  \nonumber
\nabla^{2} \Phi_{N} &=& 4\pi G \rho 
\end{eqnarray}
in which the term $2c_{1}$ is introduced to make this model compatible with the Neumann model of Section~\ref{Scalar}. It should be noted that no horizon, including  Rindler horizon, exists in Newtonian mechanics and that the inertial frames of Newtonian physics are presumed to be satisfied quite globally. The~notion of   Rindler horizon is borrowed from special relativity, hence, the name semi-Newtonian. It is also worthy to note  that the result derived in this part fulfils Mach3 formulation of   Bondi and Samuel~\cite{Bondi:1996md}, who imply  that the ``local inertial frames are affected by the cosmic motion and distribution of matter''; however, Mach3 is usually discussed in connection with rotating frames. In~conclusion, the~inertial frame $i$ is not attached to a ``single object'' very far away; $i$ is linked to the homogeneous and isotropic matter which is distributed at cosmic scale. In~this way,  the~symmetry of the model is~preserved.

The appearance of the new term on the left-hand side of Equation~(\ref{ModNewt1}) stresses the fact that this model is indeed a modified dynamical model. However, to~use the full capacity of potential theory, it is more appropriate to transfer this term to the rhs of the equation of motion and to rewrite Equation~(\ref{ModNewt1}) for a system of particles as
\begin{eqnarray} \label{ModNewt2}
\vec{a}_{pg}&=& - \vec{\nabla} \Phi  \\  \nonumber
\nabla^{2} \Phi &=& 4 \pi G  \rho + \frac{4c_{1}  a_{0}}{M}  \int \frac{\rho(\vec{x^{\prime}})d^{3}\vec{x^{\prime}}}{| \vec{x^{\prime}} - \vec{x}|}
\end{eqnarray}
in which $\Phi$ is the modified potential which must be derived from the second equation, i.e.,~the modified Poisson equation. The~method to derive this modified Poisson equation is exactly the same as the method used in the case of Neumann model in Reference~\cite{Shenavar:2016xcp}. 

It is important  to note that this semi-Newtonian model shares  all transformation properties of the Newtonian theory. To~see this point,  assume the next extended Galilean group of transformations
\begin{eqnarray}
x^{\prime}_{i}&=&R_{ij}x_{j}+d_{i} \\ \nonumber 
t^{\prime}&=& t + \tau
\end{eqnarray}
in which $R_{ij}$ and $d_{i}$ are time-dependent rotation matrix and vector, respectively. See Reference~\cite{1967rta1.book..218S} and~\cite{ehlers1973survey} for more details. Then, for~the Galiliean group $\mathcal{G}$, we have $\dot{R}_{ij}=0$ and $\ddot{d}_{i}=0$ while, for the Newtonian group  $\mathcal{N}$, we  have $\dot{R}_{ij}=0$, thus, $\mathcal{G} \subset \mathcal{N}$. As References~\cite{1967rta1.book..218S,ehlers1973survey}   have explained, under~a Newtonian transformation, the gravitational potential transforms as \begin{equation} \label{NewtTrans}
\Phi (\vec{r},t) \rightarrow \Phi (\vec{r}^{\prime},t^{\prime})= \Phi (\vec{r},t) - \vec{r}^{\prime}.\vec{A}
\end{equation}
in which $\vec{A}$ is some constant acceleration. This transformation explains in an elegant mathematical way that the field strength is frame dependent, i.e.,~it changes by an amount of $-\vec{A}$ under a Newtonian transformation (the basis of Einstein's elevator argument). On~the other hand, when the matter distribution is given, the~tidal forces could be  solely~determined.  

The semi-Newtonian potential of Equation \eqref{ModNewt2} preserves  the transformation in Equation \eqref{NewtTrans} as one may check easily. As~a result, the~homogeneity of the universe will be preserved under this model. This is a crucial requirement  for models of modified dynamics. As~Benisty and Guendelman~\cite{Benisty:2019wpm}  have argued, when Newton's second law is modified  by a nonlinear function of the acceleration, as the~MOND model proposes,  the~concept of relative acceleration would be lost. As~a result,  it is probably not possible   to maintain the homogeneity principle~anymore.

Although the semi-Newtonian model seems to be a  mathematically consistent model, we have clear reasons to favor the general relativistic view of Section~\ref{Scalar}. First of all, the~appearance of the new acceleration term in Equation~(\ref{ModNewt})  had to be ``introduced'' based on four assumptions while such term is ``derived''  in Section~\ref{Scalar} by imposing Neumann boundary conditions and as a result of background-perturbed metrics coupling. (Of course, to~find a modified equation of motion at a large scale which goes beyond Newtonian mechanics, one must be open to introduce some aspects of the phenomenology of large-scale systems into the model~\cite{Pachner:1963zz}.)  Moreover, even if we attribute $2c_{1}cH_{0}$ in Equation~(\ref{ModNewt2}) to  distant objects (Mach's principle), the~notion   remains conceptually  vague in contrast to Wheeler's idea, which is mathematically well defined. Furthermore, we had to use the the concept of Rindler horizon to determine the domain of reliability of inertial frames in Newtonian mechanics. Regarding this last issue, the~basis of the semi-Newtonian model could be strengthened by considering accelerated frames in  special relativity. In~this way, the~Rindler horizon would naturally emerge. See Reference~\cite{vanPutten:2017lik} for a relativistic method of introducing de Sitter acceleration into the equation of motion and its consequences at galactic and extra-galactic scales~\cite{vanPutten:2017bqf}.  

 The second reason to favor the GR version of modified dynamics is the clear   triumph of the Einstein equation  in explaining many gravitational phenomena specially in the case of  strong gravitational fields. This could not be incorporated in a semi-Newtonian model of gravity. Third, precise experiments of high-energy collisions, time dilation effect, spin-statistics theorem, etc. ensure us the superiority of the Lorentz group of special relativity over the Galilean group of Newtonian mechanics. Thus, as~a natural extension of special relativity, GR provides a better chance toward unification of~physics.

\reftitle{References}

\begin{thebibliography}{999}
\providecommand{\natexlab}[1]{#1}

\bibitem[Peebles(1994)]{Peebles:1994xt}
Peebles, P.J.E.
\newblock {\em {Principles of Physical Cosmology}}; Princeton University Press; Princeton, New Jersey, USA, 1994. 


\bibitem[Peebles and Ratra(2003)]{Peebles:2002gy}
Peebles, P.J.E.; Ratra, B.
\newblock {The Cosmological constant and dark energy}.
\newblock {\em Rev. Mod. Phys.} {\bf 2003}, {\em 75},~559--606,
  doi:{\changeurlcolor{black}\href{https://doi.org/10.1103/RevModPhys.75.559}{\detokenize{10.1103/RevModPhys.75.559}}}.

\bibitem[Bardeen(1980)]{Bardeen:1980kt}
Bardeen, J.M.
\newblock {Gauge Invariant Cosmological Perturbations}.
\newblock {\em Phys. Rev.} {\bf 1980}, {\em D22},~1882--1905, doi:10.1103/PhysRevD.22.1882.

\bibitem[Mukhanov \em{et~al.}(1992)Mukhanov, Feldman, and
  Brandenberger]{Mukhanov:1990me}
Mukhanov, V.F.; Feldman, H.A.; Brandenberger, R.H.
\newblock {Theory of cosmological perturbations. Part 1. Classical
  perturbations. Part 2. Quantum theory of perturbations. Part 3. Extensions}.
\newblock {\em Phys. Rept.} {\bf 1992}, {\em 215},~203--333, doi:10.1016/0370-1573(92)90044-Z.

\bibitem[Bruni \em{et~al.}(1997)Bruni, Matarrese, Mollerach, and
  Sonego]{Bruni:1996im}
Bruni, M.; Matarrese, S.; Mollerach, S.; Sonego, S.
\newblock {Perturbations of space-time: Gauge transformations and gauge
  invariance at second order and beyond}.
\newblock {\em Class. Quant. Grav.} {\bf 1997}, {\em 14},~2585--2606,
doi:10.1088/0264-9381/14/9/014.

\bibitem[Malik and Wands(2009)]{Malik:2008im}
Malik, K.A.; Wands, D.
\newblock {Cosmological perturbations}.
\newblock {\em Phys. Rept.} {\bf 2009}, {\em 475},~1--51,
 doi:10.1016/j.physrep.2009.03.001.

\bibitem[Weinberg(2008)]{Weinberg:2008zzc}
Weinberg, S.
\newblock {\em {Cosmology}}; Oxford University Press; Oxford, UK, 2008. 


\bibitem[Sanders(2010)]{Sanders:2010cle}
Sanders, R.H.
\newblock {\em {The Dark Matter Problem}}; Cambridge University Press: Cambridge, UK,   {2010.} 


\bibitem[Jain and Khoury(2010)]{Jain:2010ka}
Jain, B.; Khoury, J.
\newblock {Cosmological Tests of Gravity}.
\newblock {\em Ann. Phys.} {\bf 2010}, {\em 325},~1479--1516,
 doi:10.1016/j.aop.2010.04.002.

\bibitem[Bertone and Tait(2018)]{Bertone:2018xtm}
Bertone, G.; Tait, T.M.P.
\newblock {A new era in the search for dark matter}.
\newblock {\em Nature} {\bf 2018}, {\em 562},~51--56,
 doi:10.1038/s41586-018-0542-z.

\bibitem[Amendola and Tsujikawa(2015)]{Amendola:2015ksp}
Amendola, L.; Tsujikawa, S.
\newblock {\em {Dark Energy}}; Cambridge University Press: Cambridge, UK,  2015.

\bibitem[Riess \em{et~al.}(1998)Riess et~al.]{Riess:1998cb}
Riess, A.G.; Filippenko, A.V.; Challis, P.; Clocchiatti, A.; Diercks, A.; Garnavich, P.M.; Gillil, R.L.; Hogan,~C.J.; Jha, S.; Kirshner, R.P.; et al.
\newblock {Observational evidence from supernovae for an accelerating universe
  and a cosmological constant}.
\newblock {\em Astron. J.} {\bf 1998}, {\em 116},~1009--1038,
doi:10.1086/300499.

\bibitem[Perlmutter \em{et~al.}(1999)Perlmutter et~al.]{Perlmutter:1998np}
Perlmutter, S.; Aldering, G.; Goldhaber, G.; Knop, R.A.; Nugent, P.; Castro, P.G.; Deustua, S.; Fabbro, S.; Goobar, A.; Groom, D.E.; et al.
\newblock {Measurements of Omega and Lambda from 42 high redshift supernovae}.
\newblock {\em Astrophys. J.} {\bf 1999}, {\em 517},~565--586,
doi:10.1086/307221.

\bibitem[Bamba \em{et~al.}(2012)Bamba, Capozziello, Nojiri, and
  Odintsov]{Bamba:2012cp}
Bamba, K.; Capozziello, S.; Nojiri, S.; Odintsov, S.D.
\newblock {Dark energy cosmology: The equivalent description via different
  theoretical models and cosmography tests}.
\newblock {\em Astrophys. Space Sci.} {\bf 2012}, {\em 342},~155--228,
 doi:10.1007/s10509-012-1181-8.

\bibitem[Chakraborty(2017)]{Chakraborty:2016yna}
Chakraborty, S.
\newblock {Boundary Terms of the Einstein–Hilbert Action}.
\newblock {\em Fundam. Theor. Phys.} {\bf 2017}, {\em 187},~43--59,
doi:10.1007/978-3-319-51700-1\_5.

\bibitem[Wheeler(1964)]{wheeler1964mach}
Wheeler, J.A.
\newblock Mach's principle as boundary condition for Einstein's equations.
In  {\em Gravitation and Relativity}; NASA Goddard Institute for Space Studies,  Vol. 2; New York, USA,  {\bf 1966} 303--349. 

\bibitem[{Raine}(1981)]{1981RPPh...44.1151R}
{Raine}, D.J.
\newblock {Mach's principle and space-time structure}.
\newblock {\em Rep. Prog. Phys.} {\bf 1981}, {\em
  44},~1151--1195, doi:10.1088/0034-4885/44/11/001.

\bibitem[Barbour and Pfister(1995)]{Barbour:1995iu}
Barbour, J.B.; Pfister, H. (Eds.)
\newblock {\em Mach's Principle: From Newton's Bucket to Quantum Gravity};  Springer Science \& Business Media: Berlin/Heidelberg, Germany,  
 1995.

\bibitem[{Choquet-Bruhat} and {York}(1980)]{1980grg1.conf...99C}
{Choquet-Bruhat}, Y.; {York}, J.W., Jr.
\newblock {The Cauchy Problem}. In
\newblock  \emph{General Relativity and Gravitation. Volume 1---One Hundred Years after
  the Birth of Albert Einstein}; Held, A., Ed.;  Plenum Press: New York, NY, USA, 
 1980;  p.~99.

\bibitem[Choquet-Bruhat and Geroch(1969)]{ChoquetBruhat:1969cb}
Choquet-Bruhat, Y.; Geroch, R.P.
\newblock {Global aspects of the Cauchy problem in general relativity}.
\newblock {\em Commun. Math. Phys.} {\bf 1969}, {\em 14},~329--335, doi:10.1007/BF01645389.

\bibitem[Bondi and Samuel(1997)]{Bondi:1996md}
Bondi, H.; Samuel, J.
\newblock {The Lense-Thirring effect and Mach's principle}.
\newblock {\em Phys. Lett.} {\bf 1997}, {\em A228},~121,
doi:10.1016/S0375-9601(97)00117-5.

\bibitem[Sciama \em{et~al.}(1969)Sciama, Waylen, and Gilman]{Sciama:1970yk}
Sciama, D.W.; Waylen, P.C.; Gilman, R.C.
\newblock {Generally covariant integral formulation of einstein's field
  equation}.
\newblock {\em Phys. Rev.} {\bf 1969}, {\em 187},~1762--1766, doi:10.1103/PhysRev.187.1762.

\bibitem[Thorne \em{et~al.}(1973)Thorne, Lee, and
  Lightman]{1973PhRvD...7.3563T}
Thorne, K.S.; Lee, D.L.; Lightman, A.P.
\newblock {Foundations for a Theory of Gravitation Theories}.
\newblock {\em  Phys. Rev. D} {\bf 1973}, {\em 7},~3563--3578, doi:10.1103/PhysRevD.7.3563.

\bibitem[Park(2018)]{Park:2018xtt}
Park, I.Y.
\newblock {Boundary dynamics in gravitational theories}.
\newblock {\em arXiv} {\bf 2018}, arXiv:1811.03688.

\bibitem[Park(2019)]{Park:2019lkh}
Park, I.
\newblock {Foliation-Based Approach to Quantum Gravity and Applications to
  Astrophysics}.
\newblock {\em Universe} {\bf 2019}, {\em 5},~71,
doi:10.3390/universe5030071.

\bibitem[Park(2017)]{Park:2017wiw}
Park, I.Y.
\newblock {Foliation-based quantization and black hole information}.
\newblock {\em Class. Quant. Grav.} {\bf 2017}, {\em 34},~245005,
 doi:10.1088/1361-6382/aa9602.

\bibitem[Witten(2018)]{Witten:2018lgb}
Witten, E.
\newblock {A Note On Boundary Conditions In Euclidean Gravity}.
\newblock {\em arXiv} {\bf 2018}, arXiv:1805.11559.


\bibitem[Shenavar(2016{\natexlab{a}})]{Shenavar:2016bnk}
Shenavar, H.
\newblock {Imposing Neumann boundary condition on cosmological perturbation
  equations and trajectories of particles}.
\newblock {\em Astrophys. Space Sci.} {\bf 2016}, {\em 361},~93, doi:10.1007/s10509-016-2676-5.

\bibitem[Shenavar(2016{\natexlab{b}})]{Shenavar:2016xcp}
Shenavar, H.
\newblock {Motion of particles in solar and galactic systems by using Neumann
  boundary condition}.
\newblock {\em Astrophys. Space Sci.} {\bf 2016}, {\em 361},~378,
 doi:10.1007/s10509-016-2964-0.

\bibitem[van Putten(2017{\natexlab{a}})]{vanPutten:2017lik}
van Putten, M.H.P.M.
\newblock {Anomalous Galactic Dynamics by Collusion of Rindler and Cosmological
  Horizons}.
\newblock {\em Astrophys. J.} {\bf 2017}, {\em 837},~22, doi:10.3847/1538-4357/aa5da9.

\bibitem[van Putten(2017{\natexlab{b}})]{vanPutten:2017bqf}
van Putten, M.H.P.M.
\newblock {Evidence for galaxy dynamics tracing background cosmology below the
  de Sitter scale of acceleration}.
\newblock {\em Astrophys. J.} {\bf 2017}, {\em 848},~28,
 doi:10.3847/1538-4357/aa88cc.

\bibitem[Milgrom(1983{\natexlab{a}})]{Milgrom:1983ca}
Milgrom, M.
\newblock {A Modification of the Newtonian dynamics as a possible alternative
  to the hidden mass hypothesis}.
\newblock {\em Astrophys. J.} {\bf 1983}, {\em 270},~365--370, doi:10.1086/161130.

\bibitem[Milgrom(1983{\natexlab{b}})]{Milgrom:1983pn}
Milgrom, M.
\newblock {A Modification of the Newtonian dynamics: Implications for
  galaxies}.
\newblock {\em Astrophys. J.} {\bf 1983}, {\em 270},~371--383, doi:10.1086/161131.

\bibitem[Milgrom(1983{\natexlab{c}})]{Milgrom:1983zz}
Milgrom, M.
\newblock {A modification of the Newtonian dynamics: Implications for galaxy
  systems}.
\newblock {\em Astrophys. J.} {\bf 1983}, {\em 270},~384--389, doi:10.1086/161132.

\bibitem[Famaey and McGaugh(2012)]{Famaey:2011kh}
Famaey, B.; McGaugh, S.
\newblock {Modified Newtonian Dynamics (MOND): Observational Phenomenology and
  Relativistic Extensions}.
\newblock {\em Living Rev. Rel.} {\bf 2012}, {\em 15},~10,
doi:10.12942/lrr-2012-10.

\bibitem[Kroupa(2012)]{Kroupa:2012qj}
Kroupa, P.
\newblock {The dark matter crisis: Falsification of the current standard model
  of cosmology}.
\newblock {\em Publ. Astron. Soc. Austral.} {\bf 2012}, {\em 29},~395--433,
 doi:10.1071/AS12005.

\bibitem[Kroupa \em{et~al.}(2012)Kroupa, Pawlowski, and Milgrom]{Kroupa:2013yd}
Kroupa, P.; Pawlowski, M.; Milgrom, M.
\newblock {The failures of the standard model of cosmology require a new
  paradigm}.
\newblock {\em Int. J. Mod. Phys.} {\bf 2012}, {\em D21},~1230003.

\bibitem[Vagnozzi(2017)]{Vagnozzi:2017ilo}
Vagnozzi, S.
\newblock {Recovering a MOND-like acceleration law in mimetic gravity}.
\newblock {\em Class. Quant. Grav.} {\bf 2017}, {\em 34},~185006,
doi:10.1088/1361-6382/aa838b.

\bibitem[Rasanen(2004)]{Rasanen:2003fy}
Rasanen, S.
\newblock {Dark energy from backreaction}.
\newblock {\em JCAP} {\bf 2004}, {\em 0402},~003,
doi:10.1088/1475-7516/2004/02/003.

\bibitem[Hirata and Seljak(2005)]{Hirata:2005ei}
Hirata, C.M.; Seljak, U.
\newblock {Can superhorizon cosmological perturbations explain the acceleration
  of the Universe?}
\newblock {\em Phys. Rev.} {\bf 2005}, {\em D72},~083501,
 doi:10.1103/PhysRevD.72.083501.

\bibitem[Kolb \em{et~al.}(2005)Kolb, Matarrese, Notari, and
  Riotto]{Kolb:2004am}
Kolb, E.W.; Matarrese, S.; Notari, A.; Riotto, A.
\newblock {The Effect of inhomogeneities on the expansion rate of the
  universe}.
\newblock {\em Phys. Rev.} {\bf 2005}, {\em D71},~023524,
   doi:10.1103/PhysRevD.71.023524.

\bibitem[Martineau and Brandenberger(2005)]{Martineau:2005aa}
Martineau, P.; Brandenberger, R.H.
\newblock {The Effects of gravitational back-reaction on cosmological
  perturbations}.
\newblock {\em Phys. Rev.} {\bf 2005}, {\em D72},~023507,
 doi:10.1103/PhysRevD.72.023507.

\bibitem[Maldacena(2011)]{Maldacena:2011mk}
Maldacena, J.
\newblock {Einstein Gravity from Conformal Gravity}.
\newblock {\em arXiv} {\bf 2011}, arXiv:1105.5632.


\bibitem[Anastasiou and Olea(2016)]{Anastasiou:2016jix}
Anastasiou, G.; Olea, R.
\newblock {From conformal to Einstein Gravity}.
\newblock {\em Phys. Rev.} {\bf 2016}, {\em D94},~086008,
doi:10.1103/PhysRevD.94.086008.

\bibitem[Krishnan and Raju(2017)]{Krishnan:2016mcj}
Krishnan, C.; Raju, A.
\newblock {A Neumann Boundary Term for Gravity}.
\newblock {\em Mod. Phys. Lett.} {\bf 2017}, {\em A32},~1750077,
 doi:10.1142/S0217732317500778.

\bibitem[Ehlers \em{et~al.}(2012)Ehlers, Pirani, and
  Schild]{ehlers2012republication}
Ehlers, J.; Pirani, F.A.; Schild, A.
\newblock Republication of: The geometry of free fall and light propagation.
\newblock {\em Gen. Relativ. Gravit.} {\bf 2012}, {\em
  44},~1587--1609.

\bibitem[Cai \em{et~al.}(2011)Cai, Tuo, Zhang, and Su]{Cai:2010uf}
Cai, R.G.; Tuo, Z.L.; Zhang, H.B.; Su, Q.
\newblock {Notes on Ghost Dark Energy}.
\newblock {\em Phys. Rev.} {\bf 2011}, {\em D84},~123501,
   doi:10.1103/PhysRevD.84.123501.

\bibitem[Cai \em{et~al.}(2012)Cai, Tuo, Wu, and Zhao]{Cai:2012fq}
Cai, R.G.; Tuo, Z.L.; Wu, Y.B.; Zhao, Y.Y.
\newblock {More on QCD Ghost Dark Energy}.
\newblock {\em Phys. Rev.} {\bf 2012}, {\em D86},~023511,
   doi:10.1103/PhysRevD.86.023511.

\bibitem[Bento \em{et~al.}(2002)Bento, Bertolami, and Sen]{Bento:2002ps}
Bento, M.C.; Bertolami, O.; Sen, A.A.
\newblock {Generalized Chaplygin gas, accelerated expansion and dark energy
  matter unification}.
\newblock {\em Phys. Rev.} {\bf 2002}, {\em D66},~043507,
  doi:10.1103/PhysRevD.66.043507.

\bibitem[Scherrer(2004)]{Scherrer:2004au}
Scherrer, R.J.
\newblock {Purely kinetic k-essence as unified dark matter}.
\newblock {\em Phys. Rev. Lett.} {\bf 2004}, {\em 93},~011301,
   doi:10.1103/PhysRevLett.93.011301.

\bibitem[Fukuyama \em{et~al.}(2008)Fukuyama, Morikawa, and
  Tatekawa]{Fukuyama:2007sx}
Fukuyama, T.; Morikawa, M.; Tatekawa, T.
\newblock {Cosmic structures via Bose Einstein condensation and its collapse}.
\newblock {\em JCAP} {\bf 2008}, {\em 0806},~033,
   doi:10.1088/1475-7516/2008/06/033.

\bibitem[Moffat(2006)]{Moffat:2005si}
Moffat, J.W.
\newblock {Scalar-tensor-vector gravity theory}.
\newblock {\em JCAP} {\bf 2006}, {\em 0603},~004,
   doi:10.1088/1475-7516/2006/03/004.

\bibitem[Ellis and Uzan(2015)]{Ellis:2015wdi}
Ellis, G.F.R.; Uzan, J.P.
\newblock {Causal structures in inflation}.
\newblock {\em Comptes Rendus Phys.} {\bf 2015}, {\em 16},~928--947,
 doi:10.1016/j.crhy.2015.07.005.

\bibitem[Khoury and Weltman(2004)]{Khoury:2003rn}
Khoury, J.; Weltman, A.
\newblock {Chameleon cosmology}.
\newblock {\em Phys. Rev.} {\bf 2004}, {\em D69},~044026,
   doi:10.1103/PhysRevD.69.044026.

\bibitem[Di~Valentino \em{et~al.}(2016)Di~Valentino, Melchiorri, and
  Silk]{DiValentino:2015bja}
Di~Valentino, E.; Melchiorri, A.; Silk, J.
\newblock {Cosmological hints of modified gravity?}
\newblock {\em Phys. Rev.} {\bf 2016}, {\em D93},~023513,
 doi:10.1103/PhysRevD.93.023513.

\bibitem[Shenavar and Ghafourian(2018)]{shenavar2018local}
Shenavar, H.; Ghafourian, N.
\newblock Local stability of galactic discs in modified dynamics.
\newblock {\em Mon. Not. R. Astron. Soc.} {\bf 2018},
  {\em 475},~5603--5617.

\bibitem[Pyne and Birkinshaw(1996)]{Pyne:1995ng}
Pyne, T.; Birkinshaw, M.
\newblock {Beyond the thin lens approximation}.
\newblock {\em Astrophys. J.} {\bf 1996}, {\em 458},~46,
  doi:10.1086/176791.

\bibitem[Pyne and Carroll(1996)]{Pyne:1995bs}
Pyne, T.; Carroll, S.M.
\newblock {Higher order gravitational perturbations of the cosmic microwave
  background}.
\newblock {\em Phys. Rev.} {\bf 1996}, {\em D53},~2920--2929,
   doi:10.1103/PhysRevD.53.2920.

\bibitem[Ohanian(1976)]{ohaniangravitation}
Ohanian, H.
\newblock {\em {Gravitation and Spacetime}};  W. W. Norton, New York, USA, 1976. 


\bibitem[Jacobs \em{et~al.}(1992)Jacobs, Linder, and
  Wagoner]{jacobs1992obtaining}
Jacobs, M.W.; Linder, E.V.; Wagoner, R.V.
\newblock Obtaining the metric of our Universe.
\newblock {\em Phys. Rev. D} {\bf 1992}, {\em 45},~R3292.

\bibitem[Zalaletdinov(1992)]{Zalaletdinov:1992cg}
Zalaletdinov, R.M.
\newblock {Averaging out the Einstein equations and macroscopic space-time
  geometry}.
\newblock {\em Gen. Rel. Grav.} {\bf 1992}, {\em 24},~1015--1031, doi:10.1007/BF00756944.

\bibitem[Zalaletdinov(1993)]{Zalaletdinov:1992cf}
Zalaletdinov, R.
\newblock {Towards a theory of macroscopic gravity}.
\newblock {\em Gen. Rel. Grav.} {\bf 1993}, {\em 25},~673--695, doi:10.1007/BF00756937.

\bibitem[Bagheri and Schwarz(2014)]{Bagheri:2014gwa}
Bagheri, S.; Schwarz, D.J.
\newblock {Light propagation in the averaged universe}.
\newblock {\em J. Cosmol. Astropart. Phys.} {\bf 2014}, {\em 1410},~073,
  doi:10.1088/1475-7516/2014/10/073.

\bibitem[Overduin and Cooperstock(1998)]{Overduin:1998zv}
Overduin, J.M.; Cooperstock, F.I.
\newblock {Evolution of the scale factor with a variable cosmological term}.
\newblock {\em Phys. Rev.} {\bf 1998}, {\em D58},~043506,
  doi:10.1103/PhysRevD.58.043506.

\bibitem[Caldwell \em{et~al.}(2003)Caldwell, Kamionkowski, and
  Weinberg]{Caldwell:2003vq}
Caldwell, R.R.; Kamionkowski, M.; Weinberg, N.N.
\newblock {Phantom energy and cosmic doomsday}.
\newblock {\em Phys. Rev. Lett.} {\bf 2003}, {\em 91},~071301,
  doi:10.1103/PhysRevLett.91.071301.

\bibitem[Vagnozzi \em{et~al.}(2018)Vagnozzi, Dhawan, Gerbino, Freese, Goobar,
  and Mena]{Vagnozzi:2018jhn}
Vagnozzi, S.; Dhawan, S.; Gerbino, M.; Freese, K.; Goobar, A.; Mena, O.
\newblock {Constraints on the sum of the neutrino masses in dynamical dark
  energy models with $w(z) \geq -1$ are tighter than those obtained in
  $\Lambda$CDM}.
\newblock {\em Phys. Rev.} {\bf 2018}, {\em D98},~083501,
  doi:10.1103/PhysRevD.98.083501.

\bibitem[Amendola \em{et~al.}(2007)Amendola, Gannouji, Polarski, and
  Tsujikawa]{Amendola:2006we}
Amendola, L.; Gannouji, R.; Polarski, D.; Tsujikawa, S.
\newblock {Conditions for the cosmological viability of f(R) dark energy
  models}.
\newblock {\em Phys. Rev.} {\bf 2007}, {\em D75},~083504,
   doi:10.1103/PhysRevD.75.083504.

\bibitem[Jamali and Roshan(2016)]{Jamali:2016zww}
Jamali, S.; Roshan, M.
\newblock {The phase space analysis of modified gravity (MOG)}.
\newblock {\em Eur. Phys. J.} {\bf 2016}, {\em C76},~490,
  doi:10.1140/epjc/s10052-016-4336-x.

\bibitem[Jamali \em{et~al.}(2018)Jamali, Roshan, and Amendola]{Jamali:2017zrh}
Jamali, S.; Roshan, M.; Amendola, L.
\newblock {On the cosmology of scalar-tensor-vector gravity theory}.
\newblock {\em J. Cosmol. Astropart. Phys.} {\bf 2018}, {\em 1801},~048,
   doi:10.1088/1475-7516/2018/01/048.

\bibitem[{Balakrishna Subramani} \em{et~al.}(2019){Balakrishna Subramani},
  {Kroupa}, {Shenavar}, and {Muralidhara}]{2019MNRAS.488.3876B}
{Balakrishna Subramani}, V.; {Kroupa}, P.; {Shenavar}, H.; {Muralidhara}, V.
\newblock {Pseudo-evolution of galaxies in {\ensuremath{\Lambda}} CDM
  cosmology}.
\newblock {\em Mon. Not. R. Astron. Soc.} {\bf 2019},
  {\em 488},~3876--3883,
doi:10.1093/mnras/stz2027.

\bibitem[Gurvits \em{et~al.}(1999)Gurvits, Kellermann, and
  Frey]{Gurvits:1998hs}
Gurvits, L.I.; Kellermann, K.I.; Frey, S.
\newblock {The ``angular size---Redshift'' relation for compact radio
  structures in quasars and radio galaxies}.
\newblock {\em Astron. Astrophys.} {\bf 1999}, {\em 342},~378.

\bibitem[Realdi and Peruzzi(2009)]{Realdi:2009zz}
Realdi, M.; Peruzzi, G.
\newblock {Einstein, de Sitter and the beginning of relativistic cosmology in
  1917}.
\newblock {\em Gen. Rel. Grav.} {\bf 2009}, {\em 41},~225--247, doi:10.1007/s10714-008-0664-y.

\bibitem[York(1972)]{York:1972sj}
York, J.W., Jr.
\newblock {Role of conformal three geometry in the dynamics of gravitation}.
\newblock {\em Phys. Rev. Lett.} {\bf 1972}, {\em 28},~1082--1085, doi:10.1103/PhysRevLett.28.1082.

\bibitem[Gibbons and Hawking(1977)]{Gibbons:1976ue}
Gibbons, G.W.; Hawking, S.W.
\newblock {Action Integrals and Partition Functions in Quantum Gravity}.
\newblock {\em Phys. Rev.} {\bf 1977}, {\em D15},~2752--2756, doi:10.1103/PhysRevD.15.2752.

\bibitem[Charap and Nelson(1983)]{Charap:1982kn}
Charap, J.M.; Nelson, J.E.
\newblock {Surface Integrals and the Gravitational Action}.
\newblock {\em J. Phys.} {\bf 1983}, {\em A16},~1661, doi:10.1088/0305-4470/16/8/013.

\bibitem[Peebles \em{et~al.}(2011)Peebles, Tully, and Shaya]{Peebles:2011kv}
Peebles, P.J.E.; Tully, R.B.; Shaya, E.J.
\newblock {A Dynamical Model of the Local Group}.
\newblock {\em arXiv} {\bf 2011}, arXiv:1105.5596.


\bibitem[Peebles(2017)]{peebles2017dynamics}
Peebles, P.
\newblock Dynamics of the Local Group: The Dwarf Galaxies.
\newblock {\em arXiv} {\bf 2017}, arXiv:1705.10683.

\bibitem[Gourgoulhon(2012)]{gourgoulhon20123+}
Gourgoulhon, E.
\newblock {\em 3+ 1 Formalism in General Relativity: Bases of Numerical
  Relativity}; Springer Science \& Business Media:  Berlin/Heidelberg, Germany,   2012;  Volume 846.

\bibitem[Arnowitt \em{et~al.}(2008)Arnowitt, Deser, and
  Misner]{Arnowitt:1962hi}
Arnowitt, R.L.; Deser, S.; Misner, C.W.
\newblock {The Dynamics of general relativity}.
\newblock {\em Gen. Rel. Grav.} {\bf 2008}, {\em 40},~1997--2027,
  doi:10.1007/s10714-008-0661-1.

\bibitem[Kidder \em{et~al.}(2005)Kidder, Lindblom, Scheel, Buchman, and
  Pfeiffer]{Kidder:2004rw}
Kidder, L.E.; Lindblom, L.; Scheel, M.A.; Buchman, L.T.; Pfeiffer, H.P.
\newblock {Boundary conditions for the Einstein evolution system}.
\newblock {\em Phys. Rev.} {\bf 2005}, {\em D71},~064020,
 doi:10.1103/PhysRevD.71.064020.

\bibitem[Szilagyi \em{et~al.}(2000)Szilagyi, Gomez, Bishop, and
  Winicour]{Szilagyi:1999nu}
Szilagyi, B.; Gomez, R.; Bishop, N.T.; Winicour, J.
\newblock {Cauchy boundaries in linearized gravitational theory}.
\newblock {\em Phys. Rev.} {\bf 2000}, {\em D62},~104006,
   doi:10.1103/PhysRevD.62.104006.

\bibitem[{Brill}(1973)]{1973ASSL...38..127B}
{Brill}, D.R.
\newblock {Observational contacts of general relativity.}
\newblock In  \emph{Relativity, Astrophysics and Cosmology}; {Israel}, W., Ed.,  1973;
  Volume~38: {Astrophysics and Space Science Library}, pp. 127--152, doi:10.1007/978-94-010-2639-0\_2.

\bibitem[Jackson(1998)]{Jackson:1998nia}
Jackson, J.D.
\newblock {\em {Classical Electrodynamics}}; Wiley:  Hoboken, NJ, USA, 
  1998.

\bibitem[Weinberg(2003)]{Weinberg:2003sw}
Weinberg, S.
\newblock {Adiabatic modes in cosmology}.
\newblock {\em Phys. Rev.} {\bf 2003}, {\em D67},~123504,
   doi:10.1103/PhysRevD.67.123504.

\bibitem[Akhshik \em{et~al.}(2015)Akhshik, Firouzjahi, and
  Jazayeri]{Akhshik:2015rwa}
Akhshik, M.; Firouzjahi, H.; Jazayeri, S.
\newblock {Cosmological Perturbations and the Weinberg Theorem}.
\newblock {\em J. Cosmol. Astropart. Phys.} {\bf 2015}, {\em 1512},~027,
  doi:10.1088/1475-7516/2015/12/027.

\bibitem[Gasperini \em{et~al.}(2011)Gasperini, Marozzi, Nugier, and
  Veneziano]{Gasperini:2011us}
Gasperini, M.; Marozzi, G.; Nugier, F.; Veneziano, G.
\newblock {Light-cone averaging in cosmology: Formalism and applications}.
\newblock {\em J. Cosmol. Astropart. Phys.} {\bf 2011}, {\em 1107},~008,
   doi:10.1088/1475-7516/2011/07/008.

\bibitem[Misner \em{et~al.}(1973)Misner, Thorne, and Wheeler]{Misner:1974qy}
Misner, C.W.; Thorne, K.S.; Wheeler, J.A.
\newblock {\em {Gravitation}}; W. H. Freeman: San Francisco, CA, USA, 1973.

\bibitem[Navarro \em{et~al.}(1996)Navarro, Frenk, and White]{Navarro:1995iw}
Navarro, J.F.; Frenk, C.S.; White, S.D.M.
\newblock {The Structure of cold dark matter halos}.
\newblock {\em Astrophys. J.} {\bf 1996}, {\em 462},~563--575,
   doi:10.1086/177173.

\bibitem[Dicke and Peebles(1964)]{dicke1964evolution}
Dicke, R.H.; Peebles, P.J.E.
\newblock Evolution of the Solar System and the Expansion of the Universe.
\newblock {\em Phys. Rev. Lett.} {\bf 1964}, {\em 12},~435.

\bibitem[Pachner(1963)]{Pachner:1963zz}
Pachner, J.
\newblock {Mach's Principle in Classical and Relativistic Physics}.
\newblock {\em Phys. Rev.} {\bf 1963}, {\em 132},~1837--1842, doi:10.1103/PhysRev.132.1837.

\bibitem[Pachner(1965)]{pachner1965problem}
Pachner, J.
\newblock Problem of Energy in an Expanding Universe.
\newblock {\em Phys. Rev.} {\bf 1965}, {\em 137},~B1379.

\bibitem[Marzke and Wheeler(1964)]{marzke1964gravitation}
Marzke, R.F.; Wheeler, J.A.
\newblock Gravitation as geometry. I: The geometry of space-time and the
  geometrodynamical standard meter.
In   {\em Gravitation and Relativity};  WA Benjamin, New York, USA,  {\bf 1964},  40--64. 

\bibitem[Ehlers(1973)]{ehlers1973survey}
Ehlers, J.
\newblock Survey of general relativity theory. In {\em Relativity, Astrophysics
  and Cosmology}; Springer:  Berlin/Heidelberg, Germany,  
  1973; pp. 1--125.

\bibitem[Kundt and Hoffmann(1962)]{kundt1962determination}
Kundt, W.; Hoffmann, B.
\newblock Determination of gravitational standard time.
In   {\em Recent Developments in  General Relativity}; Pergamon Press, New York, USA,  {\bf 1962},  303. 

\bibitem[Desloge(1989)]{desloge1989simple}
Desloge, E.A.
\newblock A simple variation of the Marzke-Wheeler clock.
\newblock {\em Gen. Relativ. Gravit.} {\bf 1989}, {\em
  21},~677--681.

\bibitem[{Trautman}(2012)]{2012GReGr..44.1581T}
{Trautman}, A.
\newblock {Editorial note to: J. Ehlers, F. A. E. Pirani and A. Schild, The
  geometry of free fall and light propagation}.
\newblock {\em Gen. Relativ. Gravit.} {\bf 2012}, {\em
  44},~1581--1586, doi:10.1007/s10714-012-1352-5.

\bibitem[Dodelson(2003)]{Dodelson:2003ft}
Dodelson, S.
\newblock {\em {Modern Cosmology}}; Academic Press: Amsterdam, The Netherlands, 2003.

\bibitem[{Sofue}(2018)]{2018PASJ...70...31S}
{Sofue}, Y.
\newblock {Radial distributions of surface mass density and mass-to-luminosity
  ratio in spiral galaxies}.
\newblock {\em Publ. Astron. Soc. Jpn.} {\bf 2018}, {\em 70},~31,
   doi:10.1093/pasj/psy014.

\bibitem[{Sofue}(2016)]{2016PASJ...68....2S}
{Sofue}, Y.
\newblock {Rotation curve decomposition for size-mass relations of bulge, disk,
  and dark halo components in spiral galaxies}.
\newblock {\em Publ. Astron. Soc. Jpn.} {\bf 2016}, {\em 68},~2,
  doi:10.1093/pasj/psv103.

\bibitem[{Pilyugin} \em{et~al.}(2014){Pilyugin}, {Grebel}, {Zinchenko}, and
  {Kniazev}]{2014AJ....148..134P}
{Pilyugin}, L.S.; {Grebel}, E.K.; {Zinchenko}, I.A.; {Kniazev}, A.Y.
\newblock {The Abundance Properties of Nearby Late-Type Galaxies. II. the
  Relation between Abundance Distributions and Surface Brightness Profiles}.
\newblock {\em Astron. J.} {\bf 2014}, {\em 148},~134,
   doi:10.1088/0004-6256/148/6/134.

\bibitem[{Courteau}(1996)]{1996ApJS..103..363C}
{Courteau}, S.
\newblock {Deep r-Band Photometry for Northern Spiral Galaxies}.
\newblock {\em Astrophys. J. Suppl. Ser.} {\bf 1996}, {\em 103},~363, doi:10.1086/192281.

\bibitem[{Courteau}(1997)]{1997AJ....114.2402C}
{Courteau}, S.
\newblock {Optical Rotation Curves and Linewidths for Tully-Fisher
  Applications}.
\newblock {\em Astron. J.} {\bf 1997}, {\em 114},~2402,
 doi:10.1086/118656.

\bibitem[Dirac(1937)]{Dirac:1937ti}
Dirac, P.A.M.
\newblock {The Cosmological constants}.
\newblock {\em Nature} {\bf 1937}, {\em 139},~323, doi:10.1038/139323a0.

\bibitem[Dirac(1938)]{Dirac:1938mt}
Dirac, P.A.M.
\newblock {New basis for cosmology}.
\newblock {\em Proc. Roy. Soc. Lond.} {\bf 1938}, {\em A165},~199--208, doi:10.1098/rspa.1938.0053.

\bibitem[Weinberg(1972)]{Weinberg:1972kfs}
Weinberg, S.
\newblock {\em {Gravitation and Cosmology}}; John Wiley and Sons: New York,
  NY, USA, 1972.

\bibitem[Hassani(2013)]{hassani2013mathematical}
Hassani, S.
\newblock {\em Mathematical Physics: A Modern Introduction to Its Foundations};
  Springer Science \& Business Media:  Berlin/Heidelberg, Germany,  
  2013.

\bibitem[{Sch{\"u}cking}(1967)]{1967rta1.book..218S}
{Sch{\"u}cking}, E.L., {Cosmology}.
\newblock In {\em Relativity Theory and Astrophysics.~Vol.1: Relativity and
  Cosmology}; {Ehlers}, J., Ed.; Providence, Rhode Island, USA, {\bf 1967}; p. 218.

\bibitem[Benisty and Guendelman(2019)]{Benisty:2019wpm}
Benisty, D.; Guendelman, E.I.
\newblock {Homogeneity of the universe emerging from the Equivalence Principle
  and Poisson equation: A comparison between Newtonian and MONDian cosmology}.
\newblock {\em arXiv} {\bf 2019}, arXiv:1902.06511.

\end{thebibliography}
\end{document}